\documentclass[preprint,prd,aps,showpacs,preprintnumbers,amsmath,amssymb]{revtex4-1}
\usepackage{graphics,epsfig,subfigure}
\usepackage{diagbox}
\usepackage[usenames]{color}
\usepackage{color}
\usepackage{graphicx}
\usepackage{amsfonts}
\usepackage{indentfirst}
\usepackage{booktabs}
\usepackage{hyperref}
\hypersetup{hypertex=ture,backref=true,colorlinks=ture,linkcolor=blue,anchorcolor=blue,citecolor=blue}
\usepackage{float}
\usepackage{multirow}
\usepackage[section]{placeins}
\usepackage{array}
\usepackage{amsmath}

\begin{document}
\title{\Large \bf Emergency of black holes from wormholes}
\author{Chen-Hao Hao}
\author{Shi-Xian Sun}
\author{Long-Xing Huang}
\author{Rong Zhang}
\author{Xin Su}
\author{Yong-Qiang Wang\footnote{yqwang@lzu.edu.cn, corresponding author}}
\affiliation{ $^{1}$Lanzhou Center for Theoretical Physics, Key Laboratory of Theoretical Physics of Gansu Province,
	School of Physical Science and Technology, Lanzhou University, Lanzhou 730000, China\\
	$^{2}$Institute of Theoretical Physics $\&$ Research Center of Gravitation, Lanzhou University, Lanzhou 730000, China}

\begin{abstract}
In this paper, we study the spherically symmetric Dirac star model in the presence of a phantom field, obtaining a traversable wormhole solution in non-trivial topological spacetime. This solution exhibits asymmetry in both the field configuration and the metric and possesses a finite ADM mass $M$ and Noether charge $Q$. Furthermore, we find that due to the presence of a wormhole at the center, this solution exhibits many differences from the Dirac star under trivial spacetime. Notably, when the wormhole throat size is small, our numerical calculations indicate the emergence of an extremely approximate black hole solution on one side of the wormhole spacetime, a phenomenon unexplored. At this time, the Kretschmann scalar near the throat tends to infinity, indicating the wormhole becomes untraversable.
\end{abstract}

\maketitle

\section{INTRODUCTION}\label{Sec1}

Wormhole is a hypothetical structure that can connect two separate points in spacetime. This concept was first proposed by Austrian physicist L. Flamm in 1916, only one year after A. Einstein published his theory of gravity. After that, until 1935, A. Einstein and his collaborator N. Rosen formally developed the wormhole theory, which was known as the "Einstein-Rosen bridge" \cite{Einstein:1935tc}. However, any matter and even light cannot pass through the Einstein-Rosen bridge, which means this kind of wormhole cannot pass through \cite{Kruskal:1959vx,Fuller:1962zza}. After that, the field fell silent for more than twenty years. Interest in this work was rekindled by J. A. Wheeler in the 1950s, he and C. W. Misner mentioned the word ``wormhole'' for the first time in their 1957 paper \cite{Misner:1957mt}. In 1973, a traversable wormhole was independently proposed by H. G. Ellis and K. A. Bronnikov. At the same time, the same type of traversable wormhole solutions were proposed \cite{Ellis:1973yv,Ellis:1979bh,Bronnikov:1973fh,Kodama:1978dw}. Later, in 1988, the excellent work done by M. Morris and K. Thorne conducted a clear study on this type of traversable wormhole \cite{Morris:1988cz}, so people later called this traversable wormhole the Morris-Thorne Wormhole. A more comprehensive introduction is in \cite{Visser:1995cc}. Constructing such traversable wormholes requires exotic matter with negative energy density, violating the Null Energy Condition (NEC) \cite{Visser:1989kh}. Phantom field can then be used as a candidate for exotic matter to construct this type of wormhole, with the kinetic energy term having the opposite sign. Furthermore, people are also finding traversable wormholes without exotic matter. Some of these works need modified gravitational theory \cite{Bronnikov:2002rn,McFadden:2004ni,Kanti:2011jz,Maldacena:2020sxe,Tello-Ortiz:2021kxg}, and some are based on Einstein-Dirac-Maxwell theory \cite{Blazquez-Salcedo:2020czn,Konoplya:2021hsm,Bolokhov:2021fil,Wang:2022aze,Kain:2023ann,Kain:2023pvp}. Recently, researchers constructed a new type of traversable wormhole by using three-dimensional spacetime defects \cite{Klinkhamer:2022rsj,Klinkhamer:2023avf,Wang:2023rfq}. In addition, the theory of ``black bounce" also links traversable wormholes with regular black holes \cite{Simpson:2018tsi,Simpson:2019oft,Lobo:2020ffi,Bronnikov:2022bud,Rodrigues:2023vtm}.

In the 1950s, J. A. Wheeler proposed a solution of the classical electromagnetic waves coupled with general relativity, a concept he termed ``geons'' \cite{Wheeler:1955zz,Power:1957zz}. Based on this idea, people found that the scalar field matter with spin of 0 coupled with the gravitational field will obtain the so-called Boson stars \cite{Kaup:1968zz,Ruffini:1969qy}. Similarly, the coupling of a Proca field with a spin of 1 and the gravitational field gives rise to Proca stars \cite{Brito:2015pxa}. In 1999, Shing-Tung Yau and his collaborators pioneered a solution for coupling the spinor field with spin 1/2 and gravitational field. By introducing a pair of fermions with opposite spins, they constructed a spherically symmetric system. This is the first study on Dirac stars \cite{Finster:1998ws}. Subsequently, research led to the construction of charged Dirac stars through the coupling of the spinor field and the electromagnetic field \cite{Finster:1998ux}. The study of Dirac stars with self-interactions was undertaken by V. Dzhunushaliev and V. Folomeev \cite{Dzhunushaliev:2018jhj}. Furthermore, Brito et al. proposed the first non-spherically symmetric configuration for a rotating Dirac star \cite{Herdeiro:2019mbz}. These studies predominantly focus on the ground state of the Dirac field, and for excited states with higher energy levels, the construction of excited state Dirac stars has been explored \cite{Herdeiro:2017fhv,Leith:2021urf}. Additionally, numerous intriguing studies have been conducted on the Einstein-Dirac theoretical model \cite{Liang:2023ywv,Liang:2022mjo,Kain:2023jgu}.

Previously, most of the research on Boson star-related models was based on the topologically trivial spacetime. In \cite{Dzhunushaliev:2014bya,Hoffmann:2017jfs,Yue:2023ela,Ding:2023syj}, the researchers introduced wormholes in the center of the Bose stars. The bosonic matter that comprises these systems can be composed of real or complex boson fields with self-interaction, which could exhibit symmetric or asymmetrical distribution across the two asymptotically flat regions. As a result, the wormhole spacetime could be either symmetric or asymmetric. Furthermore, if a rotating scalar field is considered, the rotating symmetric or asymmetric wormhole spacetime solution can also be obtained \cite{Hoffmann:2019jva,Hoffmann:2018oml}. However, the solutions of Proca stars and Dirac stars in non-trivial spacetime topology have not been studied so far. In this work, we investigate a spherically symmetric Dirac star model in the presence of a phantom field, obtaining a traversable wormhole which is asymmetric. Particularly, when the wormhole throat size is small, an extremely approximate black hole emerges on the side of the wormhole.

The paper is organized as follows. In Sec.~\ref{sec2}, we present the model four-dimensional
Einstein gravity coupled to a phantom field and two Dirac fields. In Sec.~\ref{sec3}, the boundary conditions are studied. We compared this situation with the Boson star model. The numerical results of the three different cases are discussed in Sec.~\ref{sec4}. We conclude in Sec.~\ref{sec5} with a summary and illustrate the range for future work.

\section{THE MODEL}\label{sec2}

\subsection{Action}

We consider the Einstein-Hilbert action including the Lagrangian for two massive Dirac fields and the phantom scalar field, the action is given by
\begin{equation}\label{action}
 S=\int\sqrt{-g}d^4x\left(\frac{R}{2\kappa}+\mathcal{L}_{p}+\mathcal{L}_{1}+\mathcal{L}_{2}\right),
\end{equation}
where $R$ is the Ricci scalar.  The term
$\mathcal{L}_{p}$ and $\mathcal{L}_{1,2}$ are  the Lagrangians defined by with

\begin{eqnarray}
\mathcal{L}_{1}  &= & \left( \frac{\mathrm i}{2} \overline{\Psi}_{1} \gamma^\nu \hat{D}_\nu \Psi_{1} - \frac{\mathrm i}{2} \hat{D}_\nu \overline{\Psi}_{1}  \gamma^\nu \Psi_{1}  - \mu \overline{\Psi}_{1}  \Psi_{1}  \right),  \\
\mathcal{L}_{2}  &= & \left( \frac{\mathrm i}{2} \overline{\Psi}_{2} \gamma^\nu \hat{D}_\nu \Psi_{2} - \frac{\mathrm i}{2} \hat{D}_\nu \overline{\Psi}_{2}  \gamma^\nu \Psi_{2}  - \mu \overline{\Psi}_{2}  \Psi_{2}  \right),  \\
\mathcal{L}_{p}  & =   &   \nabla_a\Phi\nabla^a\Phi \ .
\end{eqnarray}

Here $\Psi$ and $\Phi$ represent the Dirac field and the phantom field, respectively.
By varying the action (\ref{action}) with respect to the metric, we can obtain the Einstein equations 

\begin{equation}
  \label{eq:EKG1}
R_{\mu\nu}-\frac{1}{2}g_{\mu\nu}R-\kappa T_{\mu\nu}=0 \ ,
\end{equation}
with stress-energy tensor
\begin{equation}
T_{\mu\nu} = g_{\mu\nu}({{\cal L}}_1+{{\cal L}}_2+{{\cal L}}_p)
-2 \frac{\partial ({{\cal L}}_1+{{\cal L}}_2+{{\cal L}}_p)}{\partial g^{\mu\nu}} \ ,
\end{equation}

and the matter field equations by varying with respect to the phantom field and Dirac fields.
\begin{eqnarray}
\Box\Phi=0,  \\
(\gamma^\nu \hat{D}_\nu-\mu) \Psi_{\varepsilon}=0  \     (\varepsilon = 1,2).
\end{eqnarray}
 $\gamma^\nu$ are the gamma matrices in curved spacetime, and $\mu$ is the mass of Dirac field $\Psi$. $\hat{D}_\nu = \partial_{ \nu} + \Gamma_\nu$ is the spinor covariant derivative,  where $\Gamma_\nu$ are the spinor connection matrices.

In order to study the Dirac equation in curved spacetime, we need to introduce the vierbein $e_a^{\alpha}$, defined by
\begin{equation}
g_{\alpha\beta}e_a^{\alpha}e_b^{\beta} = \eta_{ab}\,,
\end{equation}
where $g_{\alpha\beta}$ is the curved space metric and $\eta_{ab} = (-1,+1,+1,+1)$. We use Greek letters $\alpha,\beta,\cdots$ for curved spacetime indices and the Latin letters $a,b,\cdots$ for flat spacetime indices. Greek indices are raised and lowered with $g_{\alpha\beta}$ and Latin indices are raised and lowered with $\eta_{ab}$. 

\begin{equation}\label{equ27}
\gamma^{\alpha} = e_a^{\alpha}\gamma^a,\qquad \gamma^a = e^a_{\alpha}\gamma^{\alpha}.
\end{equation}
$\gamma^a$ and $\gamma^{\alpha}$ satisfying the anticommutation relations:
\begin{equation}
\{\gamma^{\alpha},\gamma^{\beta}\} = 2g^{\alpha \beta},\qquad \{\gamma^a,\gamma^b\} = 2\eta^{ab}.
\end{equation}
$\gamma$-matrices indices are raised and lowered in the following ways:
\begin{equation}
\gamma_a = \eta_{ab} \gamma^b,\qquad \gamma_{\alpha} = g_{\alpha\beta}\gamma^{\beta}.
\end{equation}

For $\gamma^a$ we use the Weyl representation:
\begin{equation}
\setlength{\arraycolsep}{8pt}
\gamma^0 = i\begin{pmatrix} 0&1\\1&0 \end{pmatrix},\qquad \gamma^k = i\begin{pmatrix} 0&\sigma_k\\-\sigma_k &0 \end{pmatrix},\qquad k = 1,2,3,
\end{equation}
where $\sigma_k$ are the Pauli matrices:
\begin{equation}
\setlength{\arraycolsep}{8pt}
\sigma_1 = \begin{pmatrix} 0&1\\1&0 \end{pmatrix},\qquad \sigma_2 = \begin{pmatrix} 0&-i\\i&0 \end{pmatrix},\qquad \sigma_3 = \begin{pmatrix} 1&0\\0&-1 \end{pmatrix}.
\end{equation}

\subsection{Ansatze}

We consider the  general static spherically symmetric solution with a wormhole,
and adopt the Ans\"atzes as follows, see {\em e.g.}~\cite{Hoffmann:2017jfs}:
\begin{equation}  \label{line_element1}
 ds^2 = -e^{A}  dt^2 +B e^{-A}   \left[ d r^2 + h (d \theta^2+\sin^2 \theta d\varphi^2)   \right]\,,
\end{equation}
here $A$ and $B$ are functions of  radial coordinate $r$,  $h=r^2+r_0^2$ with  the throat parameter  $r_0$.
 and $r$  ranges from positive infinity to negative infinity.
It should be emphasized that the two limits $r\rightarrow \pm\infty$ correspond to two distinct asymptotically flat spacetime.
Furthermore, we assume Dirac fields and phantom field in the form
\begin{equation}
\Phi=\phi(r),
\end{equation}
\begin{equation}
 \Psi_{1} = \begin{pmatrix}\cos(\frac{\theta}{2})[(1 + i)f(r) - (1 - i)g(r)]\\ i\sin(\frac{\theta}{2})[(1 - i)f(r) - (1 + i)g(r)]\\-i\cos(\frac{\theta}{2})[(1 - i)f(r) - (1 + i)g(r)]\\ -\sin(\frac{\theta}{2})[(1 + i)f(r) - (1 - i)g(r)] \end{pmatrix}e^{i\frac{\varphi}{2} - i\omega t}\,,
\end{equation}

\begin{equation}
 \Psi_{2} = \begin{pmatrix}i\sin(\frac{\theta}{2})[(1 + i)f(r) - (1 - i)g(r)]\\ \cos(\frac{\theta}{2})[(1 - i)f(r) - (1 + i)g(r)]\\ \sin(\frac{\theta}{2})[(1 - i)f(r) - (1 + i)g(r)]\\ i\cos(\frac{\theta}{2})[(1 + i)f(r) - (1 - i)g(r)] \end{pmatrix}e^{-i\frac{\varphi}{2} - i\omega t}\,,
\end{equation}
there $f(r)$ and $G(r)$ is the real function of the radial coordinate $r$, the phantom field $\Phi$ is also a real function and is independent of the time coordinate $t$, and the constant $\omega$ is referred to as the synchronization frequency. From the equations, we can see the  $f(r)$ and $G(r)$ include information of the Dirac field.

Substituting the above Ansatze into the Einstein equations leads to the following field equations
\begin{equation}
\frac{32 \kappa \omega B}{e^{\frac{3 A}{2}}}(F^2+G^2) - 3 \frac{B'^2}{B^2} + \frac{4 r_0^2}{h^2} - \frac{8 r A'}{h} + A'^2 - 2\kappa \psi'- 4 A'' = 0,
\end{equation}
\begin{equation}
- \frac{4 r_0^2}{h^2} - A'^2 + 2 \kappa (16 \sqrt{e^{- A} B}(- G F' +F G') + \psi'^2) + \frac{4 r B'}{hB} + \frac{B'^2}{B^2} = 0,
\end{equation}
\begin{equation}
\frac{4 r_0^2}{h^2} - \frac{32 \kappa F G \sqrt{e^{-A} h B}}{h} + A'^2 - 2 \kappa \psi'^2 - 2 \frac{B'^2}{B^2} + \frac{\frac{2 r B'^2}{h} + 2 B'}{B} = 0,
\end{equation}

By variation of the action with respect to the
Dirac fields and to the phantom field
leads to the equations

\begin{equation}
\begin{split}
&4 h (\sqrt{e^A} \mu - \omega) G B^2 + e^{\frac{3 A}{2}} \\ &(4 h B \sqrt{e^{-A} B} F' + 4 B F r \sqrt{e^{-A} B} + 4 B F \sqrt{e^{-A} h B} - h B F \sqrt{e^{-A} B} A' + 2 h F \sqrt{e^{-A} B} B') = 0,
\end{split}
\end{equation}

\begin{equation}
- 4 h B(\sqrt{B} F \mu + \sqrt{e^{-A} B} F \omega + G' \sqrt{e^{A}}) + \frac{4 B \sqrt{h B} G}{\sqrt{e^{-A} B}} + B G \sqrt{e^{A}} (-4 r + h A') - G \sqrt{e^{A}} 2 h B' = 0,
\end{equation}

\begin{equation}
(h \sqrt{B} \phi')' =0,
\end{equation}

Solving these OED equations numerically, we can get all information about metric functions $A$ and $B$, field $F$,$G$, and $\phi$.
Furthermore, we can transform the last expression into the following form

\begin{equation}
\phi' = \frac{\sqrt{\cal D}}{h \sqrt{B}}\ .
\end{equation}

the $\cal D$ is a constant represents the scalar charge of the phantom field and can be used to check the accuracy of numerical calculations. Its value as a function of frequency $\omega$ should be the same at different locations while fixing $r_0$.
We give the expression of scalar charge $\cal D$.

\begin{eqnarray}
{\cal D}
  = -\frac{B^2 [h^2 A'^2 +4 (r_0^2 +8 \kappa h^2 G \sqrt{e^{-A} B} F' - 8 \kappa h^2 F \sqrt{e^{-A} B} G')] - 4 r h B B' - h^2 B'^2}{2 \kappa B}.
\label{eqD2}
\end{eqnarray}

Before going any further, we want to make two points; Firstly, when the Dirac fields vanish, one can derive the solution for an Ellis wormhole. Secondly, as the throat size $r_0$ approaches zero, the asymptotically flat spacetime on both sides of the wormhole does not have the well-known Dirac star solution (Although some similarities emerge in some cases).

\subsection{Mass and Charge}

The ADM mass $M$ is the key quantities we are interested in,
which is encoded in the asymptotic expansion of metric components
\begin{eqnarray}
g_{tt}= -1+\frac{2 M}{r}+\cdots \ .
\end{eqnarray}
The action of the dirac field is invariant under the $U(1)$ transformation $\Psi^{(1,2)}\rightarrow e^{i\alpha}\Psi^{(1,2)}$, $\alpha$ is a constant. According to Noether's theorem, there is a conserved current corresponding to the Dirac field:

\begin{equation}
\qquad J_D^{\nu} = \overline{\Psi}\gamma^\nu\Psi\,.
\end{equation}

Integrating the timelike component of the above conserved currents on a spacelike hypersurface $\cal{S}$, ones could obtain Noether charge:
\begin{eqnarray}
Q  &=& \int_{\cal S}J_D^t \nonumber \\
&= &- \int J^t \left| g \right|^{1/2} dr d\Omega_{2}.
\end{eqnarray}
The Noether charge reflects the number of particles in the Dirac field, which is an essential reason determines mass.

\section{BOUNDARY CONDITIONS}\label{sec3}

Before numerically solving the differential equations instead of seeking the analytical
solutions, we should obtain the asymptotic behaviors of the four functions $f$, $g$, $A$, $B$,
which is equivalent to giving the boundary conditions we need.

It should be emphasized that for the Boson star model of non-trivial spacetime, because it often has symmetric or antisymmetric properties, only necessary to numerically solve half of the area, and then obtain the solution of the entire spacetime by applying specific boundary conditions (this usually means it needs to be restricted at $x=0$). 

For our work, the solution is asymmetric and continuous at the origin, and we can get the solution in the whole spacetime at once, which means there is no need to limit the boundary conditions at the origin. Meanwhile, in order to make spacetime satisfy asymptotic flatness, it is only necessary to require:
\begin{eqnarray}
 f = g =A=0, \hspace{5pt}   B=1,\hspace{5pt}    \partial_r B= 0
\end{eqnarray}
at infinity ($r \rightarrow \infty$).

\section{NUMERICAL RESULTS}\label{sec4}

In this work, all the numbers are dimensionless as follows
\begin{eqnarray}
r \rightarrow r\mu \hspace{5pt}, \hspace{5pt} \phi \rightarrow \phi \kappa^{-1/2}\hspace{5pt}, \hspace{5pt} \omega \rightarrow \omega/\mu \hspace{5pt}.
\end{eqnarray}
without loss of generality, we can fix the specific parameters as $\mu_0 = 1$ and $\kappa= 2$.
In order to facilitate numerical calculations, we transform the radial coordinates by the following equation
\begin{eqnarray}
\label{transform}
x= \frac{2}{\pi}\arctan(r) \;,
\end{eqnarray}
map the infinite region ($-\infty$,$+\infty$) to the finite region (-1,1).
This allows the ordinary differential equations to be approximated by algebraic equations. The grid with 2000 points covers the integration region and the relative errors are less $10^{-5}$.

At first, we give a brief introduction to the excited states of Dirac fields. As quantum mechanics tells us, there are $n$ intersection points (we call them nodes) between the image of the field and its zero-value axis, and we call the field the $n$-th excited state. According to the excited state configurations of different solutions in this work, we divide them into three cases.

\textbf{Case 1: $F$ has one node and $G$ has no nodes.}

\begin{figure}
\begin{center}
\subfigure{\includegraphics[width=0.49\textwidth]{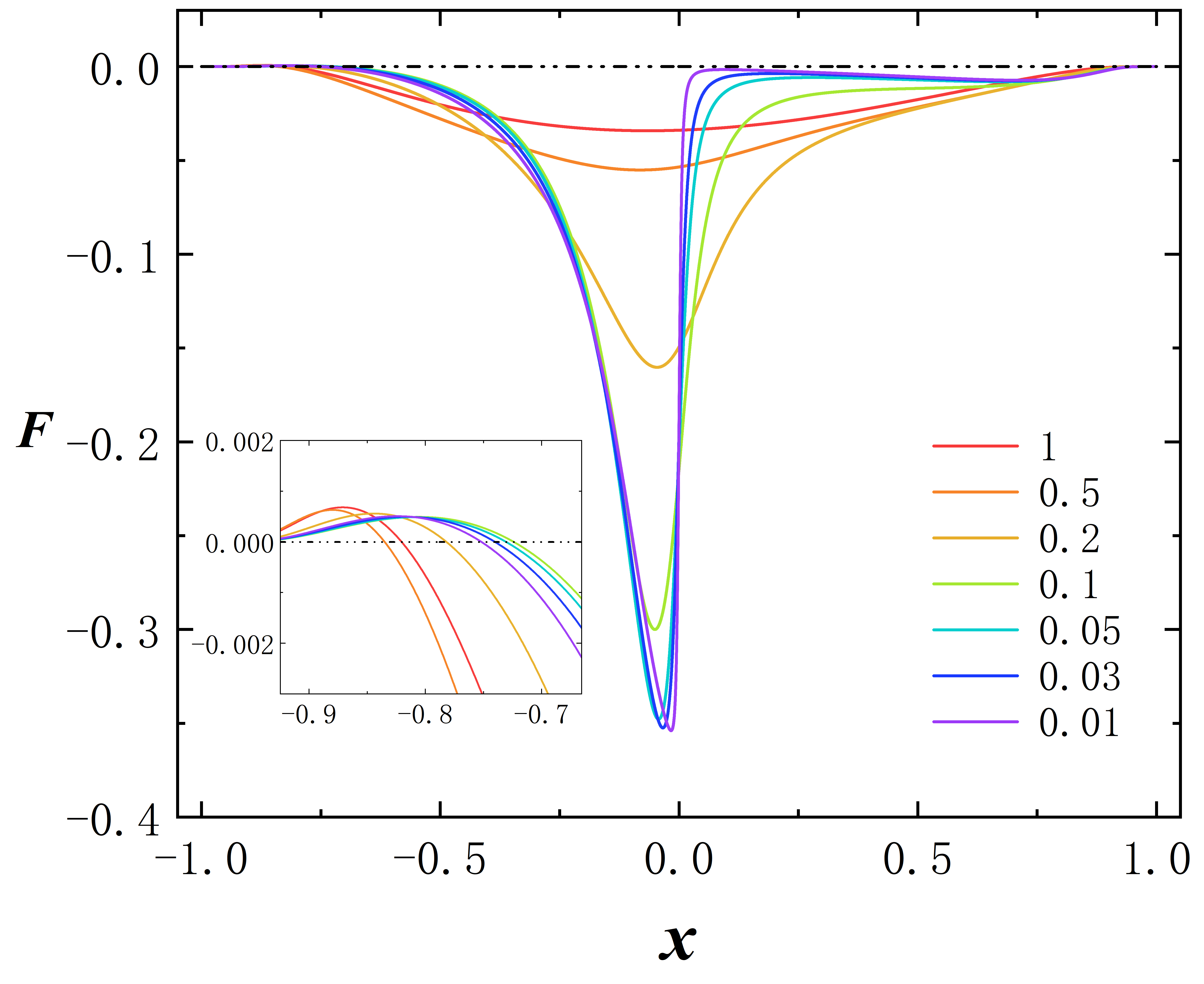}}
\subfigure{\includegraphics[width=0.49\textwidth]{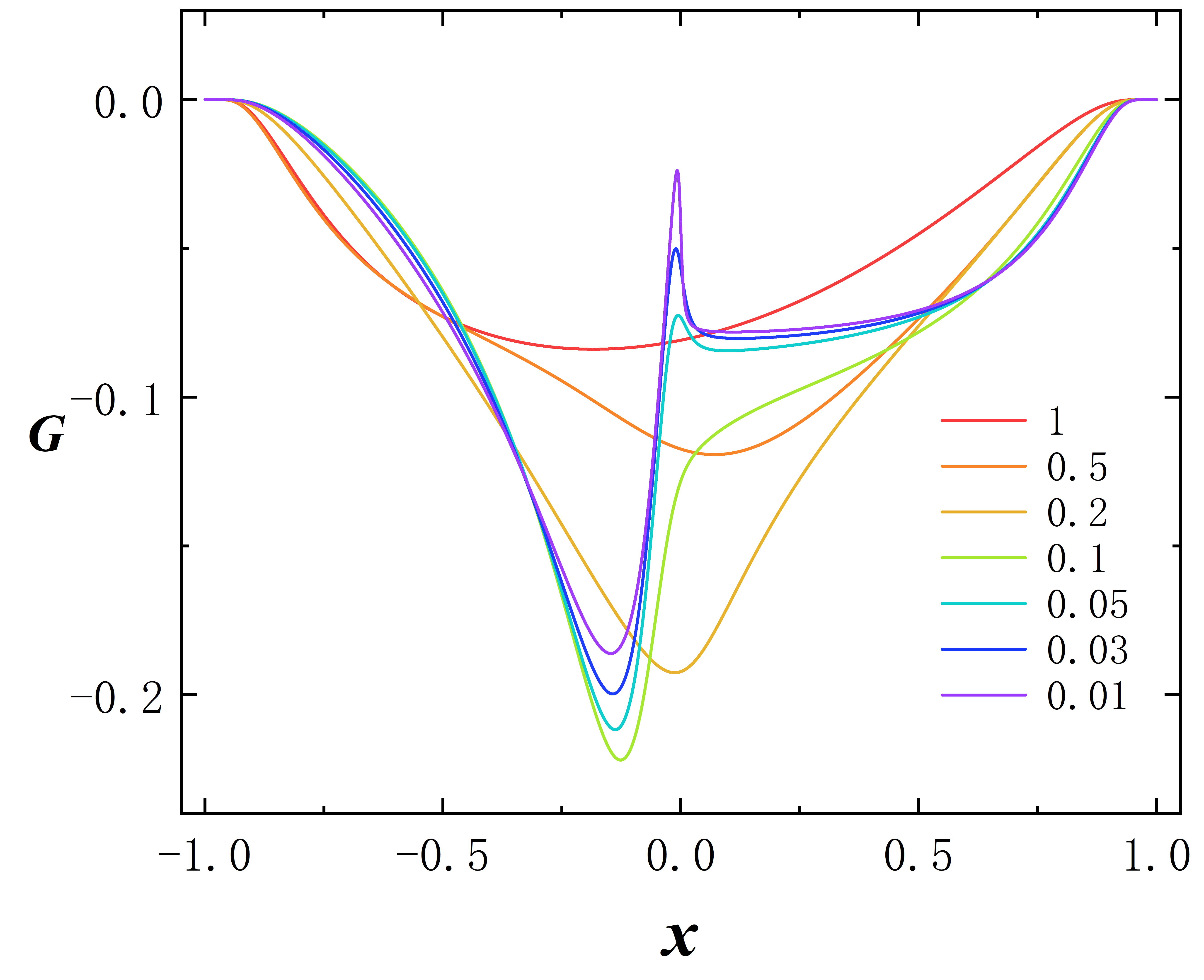}}
\end{center}
\caption{The radial distribution of the dirac fields $F$ and $G$ with several values of $r_0$ for frequency $\omega$ = 0.87.}
\label{phase1}
\end{figure}

We plot in Fig. \ref{phase1} the numerical results for the Dirac fields $F$ and $G$ as a function of $x$, choose various values of the throat size $r_0$ and fix the frequency $\omega$ = 0.87. The left panel displays the intersection of the $F$ field with the black dotted line representing $F = 0$, which is further magnified in a subgraph. The $F$ field exhibits one node between $x= -0.9$ and $x= -0.7$ and the asymmetry of the fields is evident from the picture.

\begin{figure}
\begin{center}
\subfigure{\includegraphics[width=0.4\textwidth]{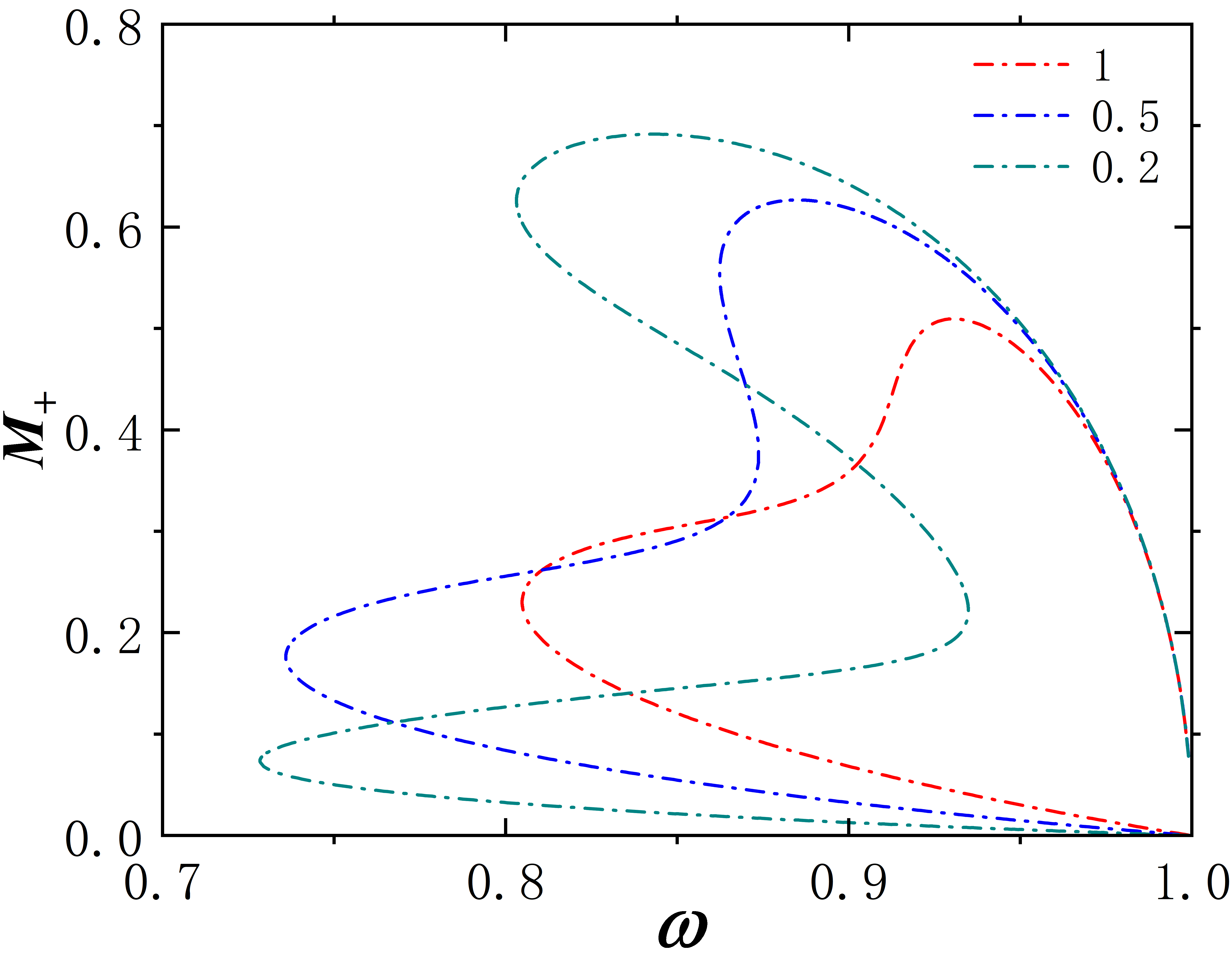}}
\subfigure{\includegraphics[width=0.4\textwidth]{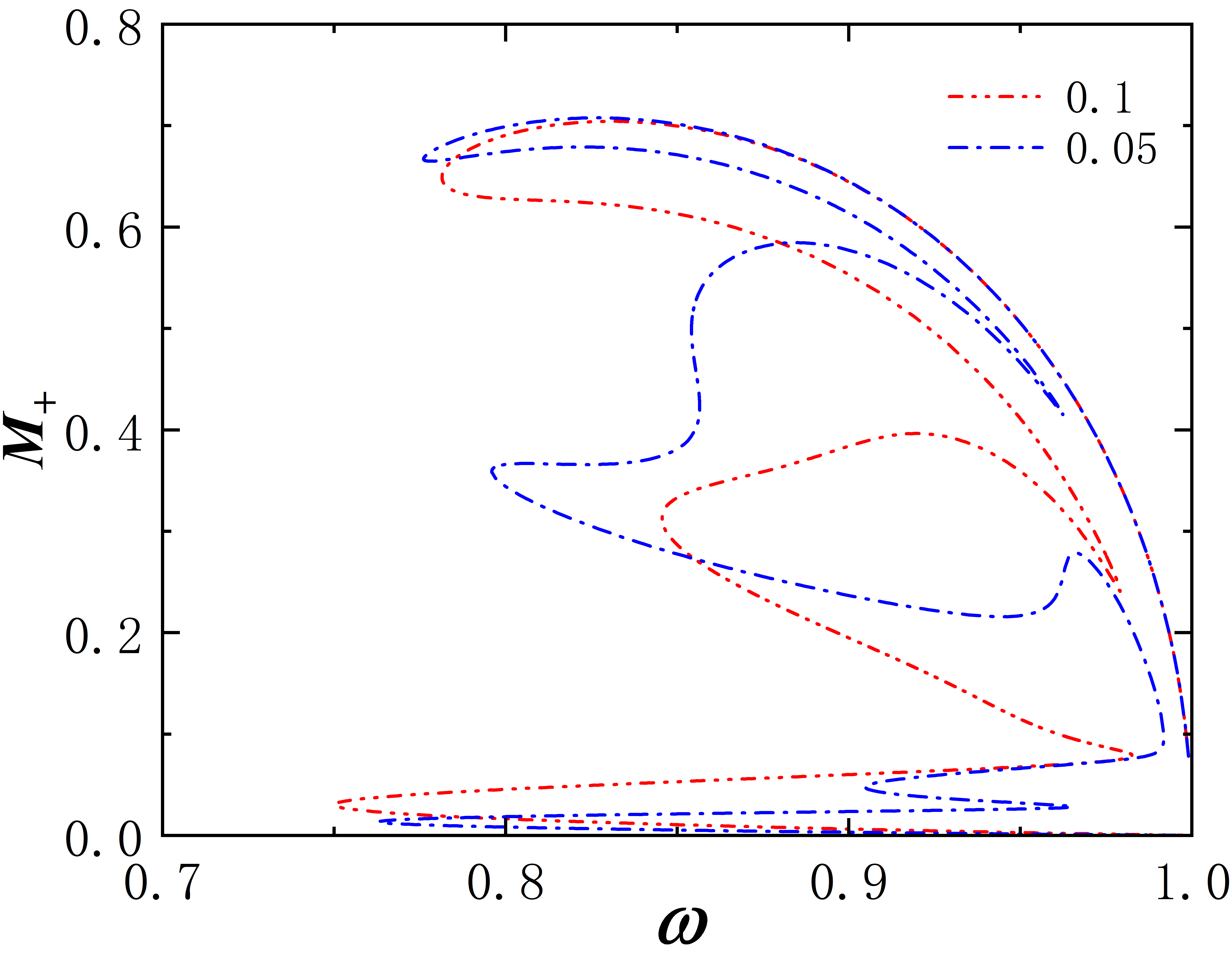}}
\subfigure{\includegraphics[width=0.4\textwidth]{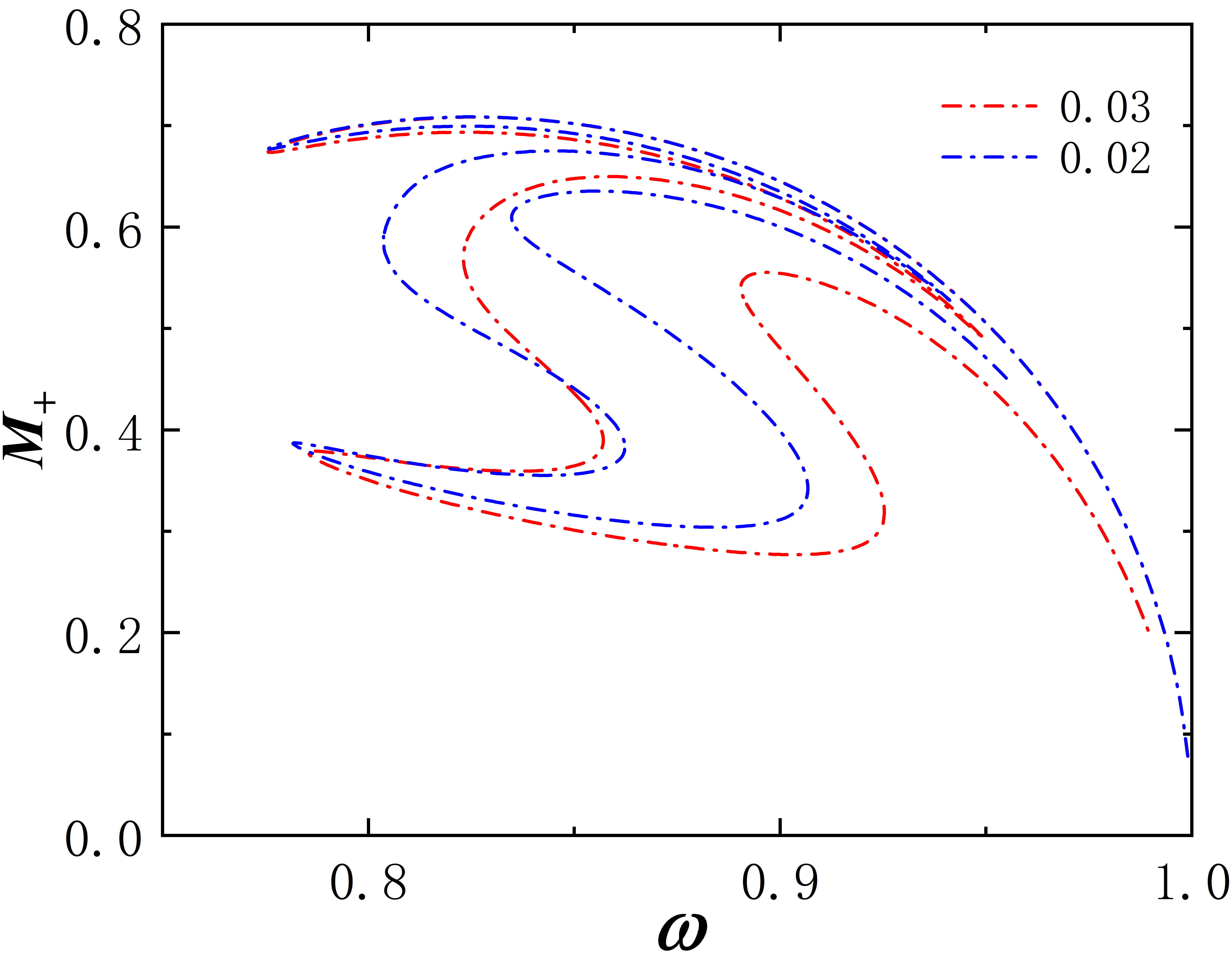}}
\subfigure{\includegraphics[width=0.4\textwidth]{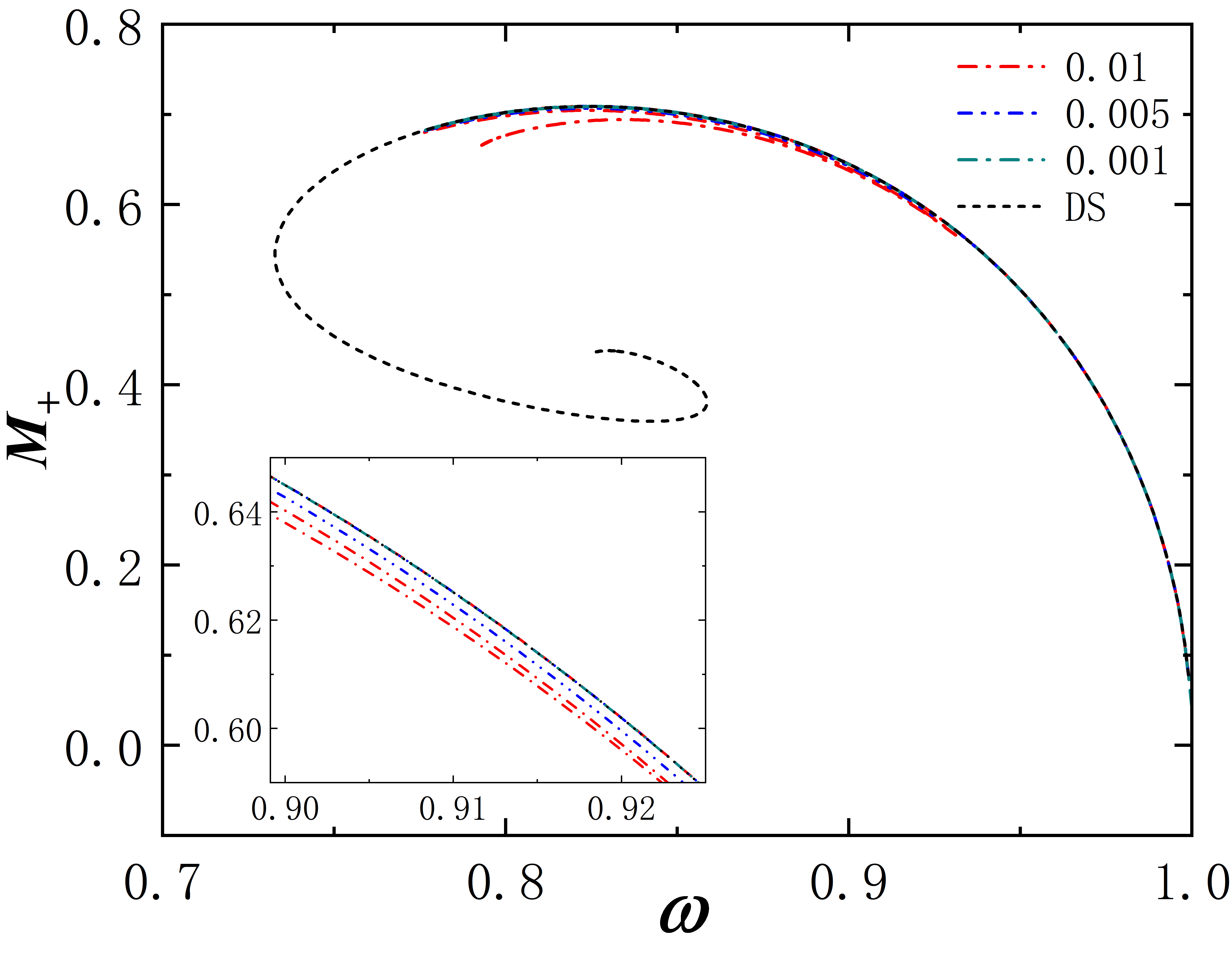}}
\subfigure{\includegraphics[width=0.4\textwidth]{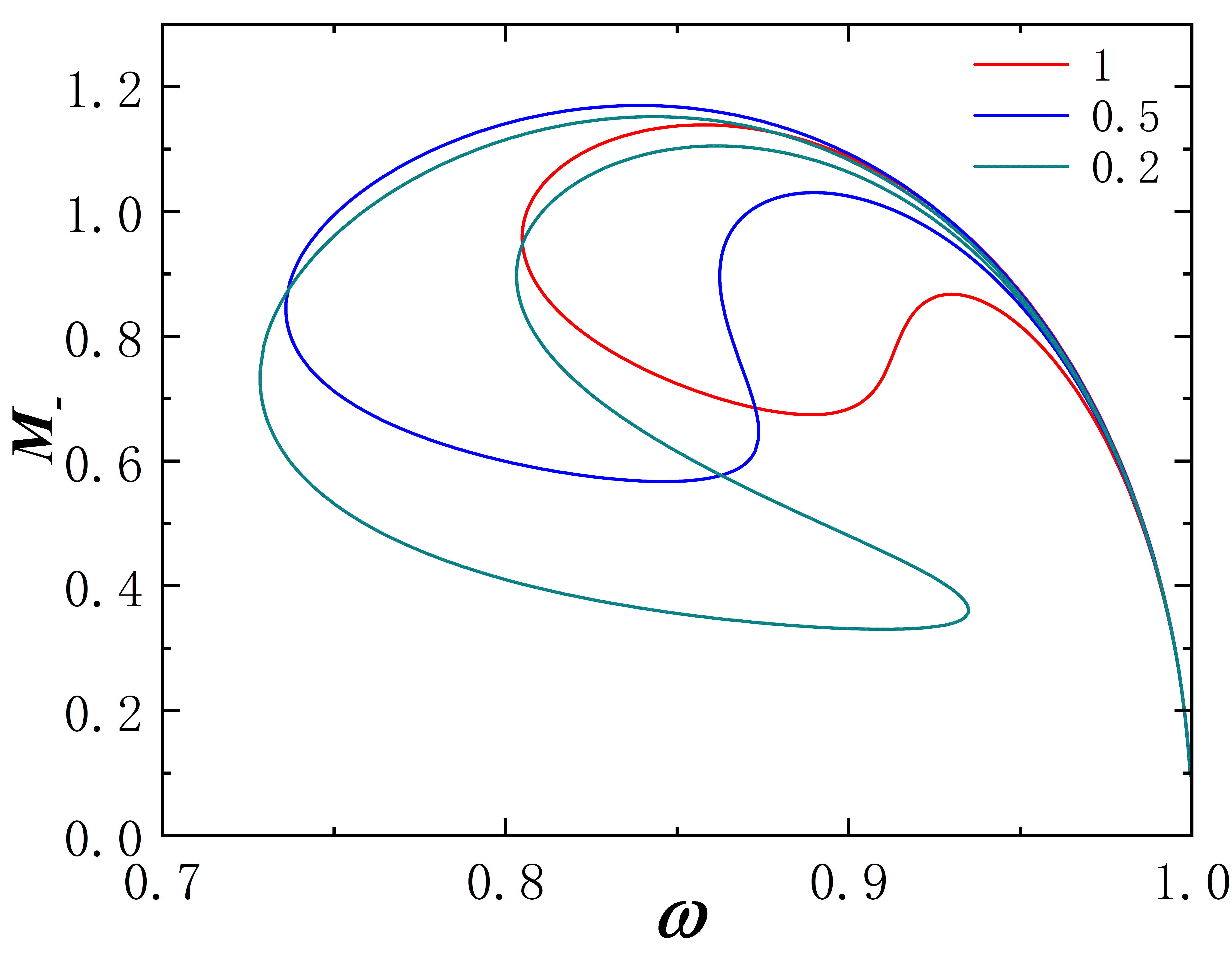}}
\subfigure{\includegraphics[width=0.4\textwidth]{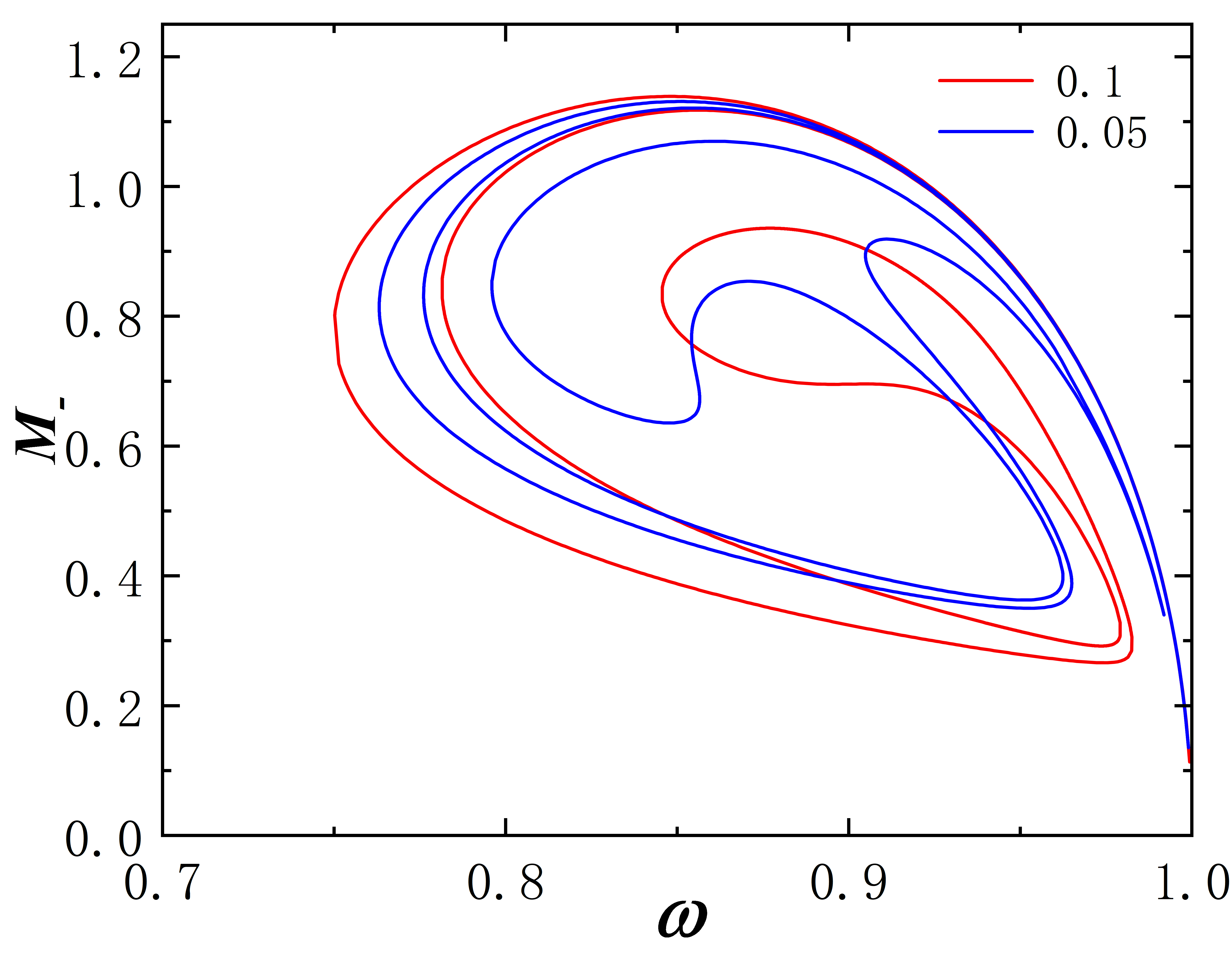}}
\subfigure{\includegraphics[width=0.4\textwidth]{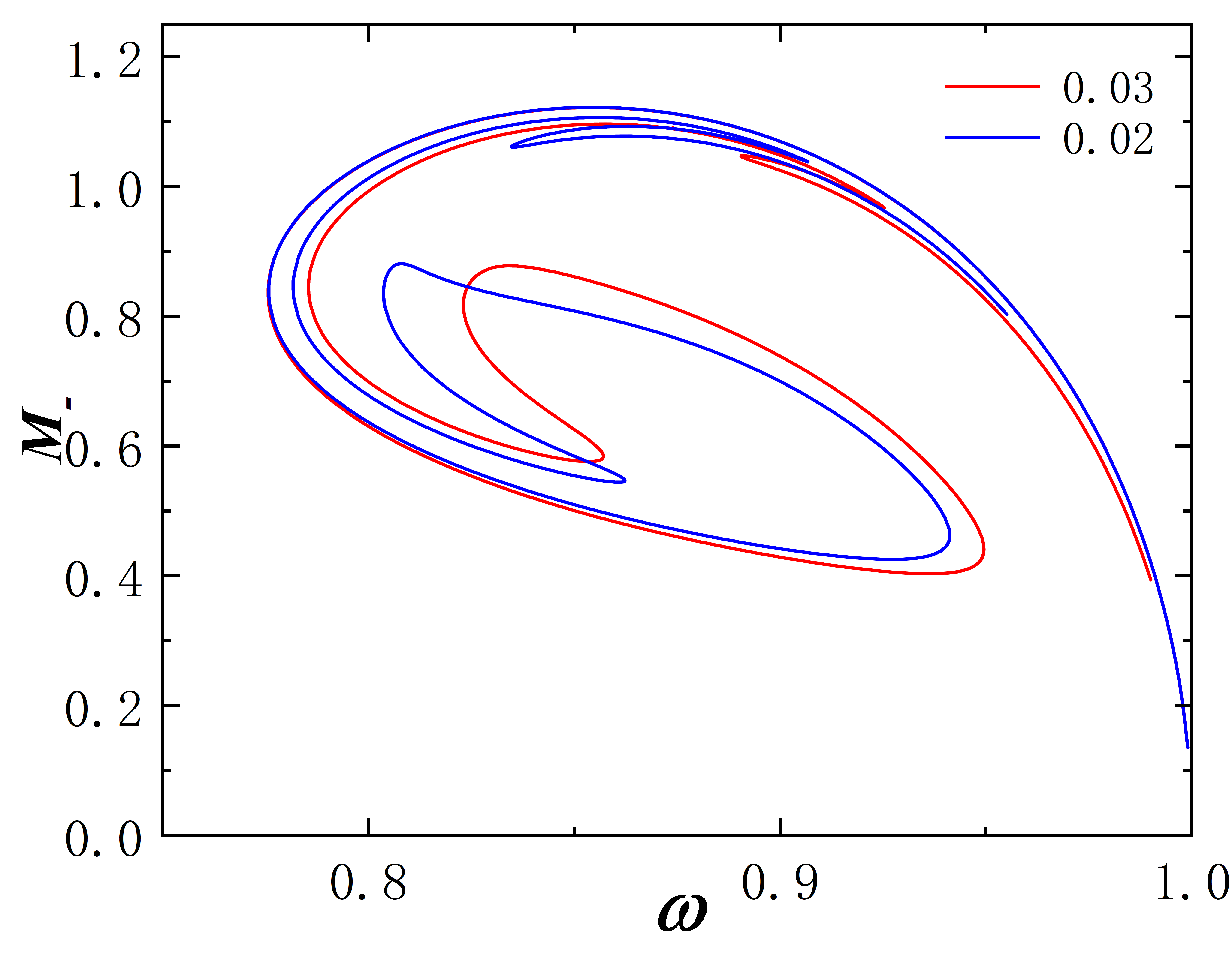}}
\subfigure{\includegraphics[width=0.4\textwidth]{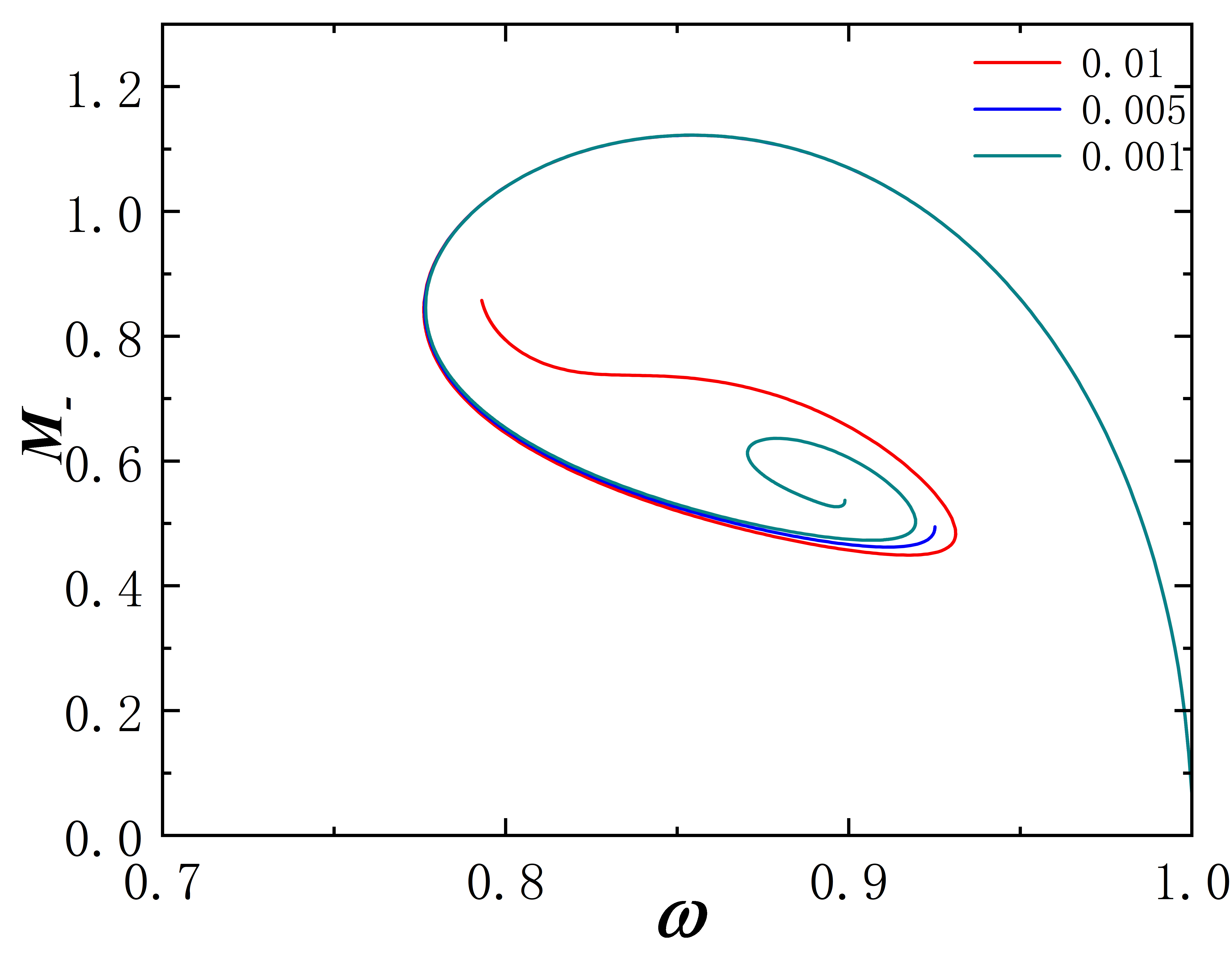}}
\end{center}
\caption{The mass M as the function of frequency $\omega$ for some values of $r_0$. The dotted line represents $M_+$, the solid line represents $M_-$.}
\label{phase2}
\end{figure}

\begin{figure}
\begin{center}
\subfigure{\includegraphics[width=0.49\textwidth]{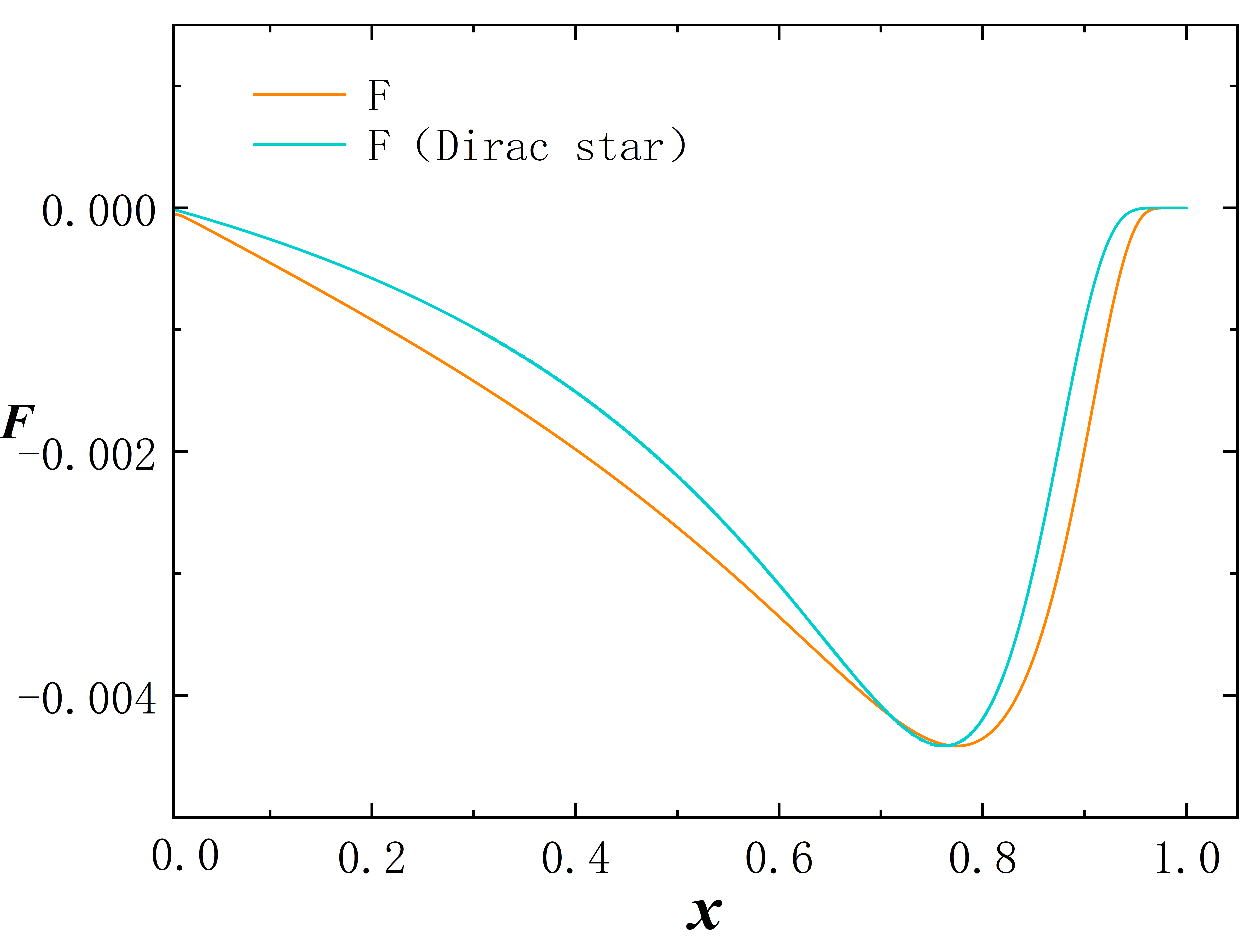}}
\subfigure{\includegraphics[width=0.49\textwidth]{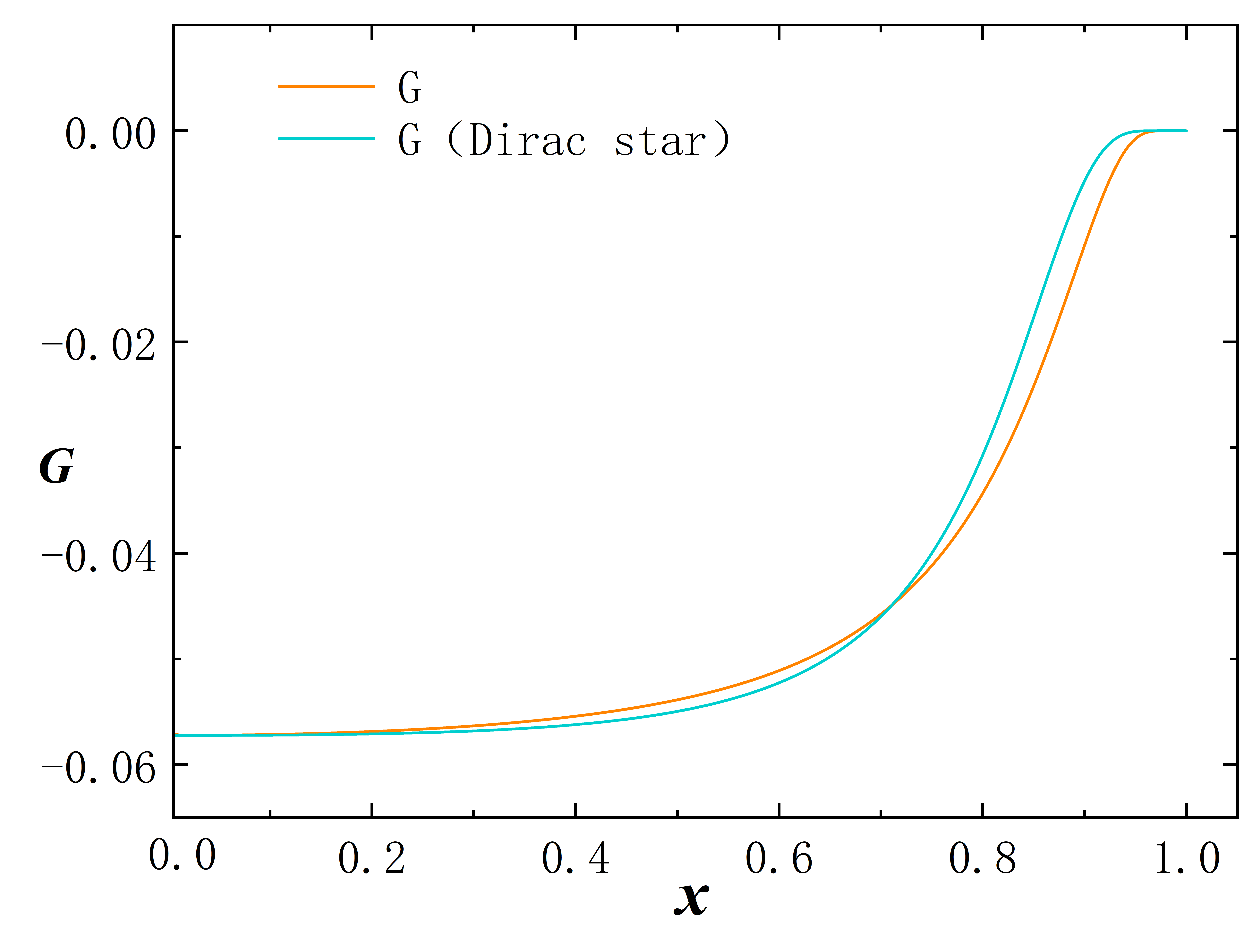}}
\end{center}
\caption{The orange line represents the fields in the region of $M_+$, blue line reflect trivial spacetime Dirac star.}
\label{phase3}
\end{figure}

\begin{figure}
\begin{center}
\subfigure{\includegraphics[width=0.4\textwidth]{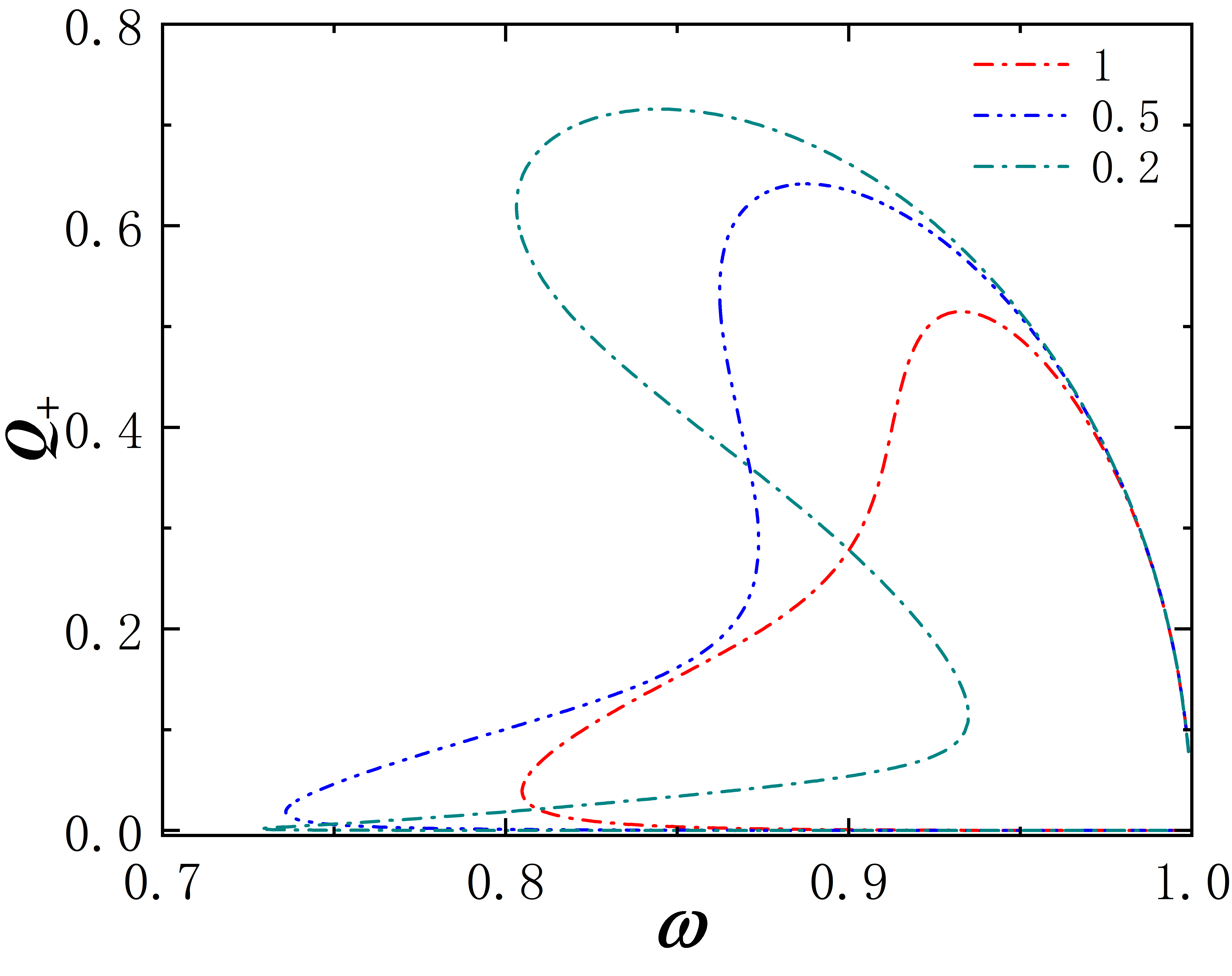}}
\subfigure{\includegraphics[width=0.4\textwidth]{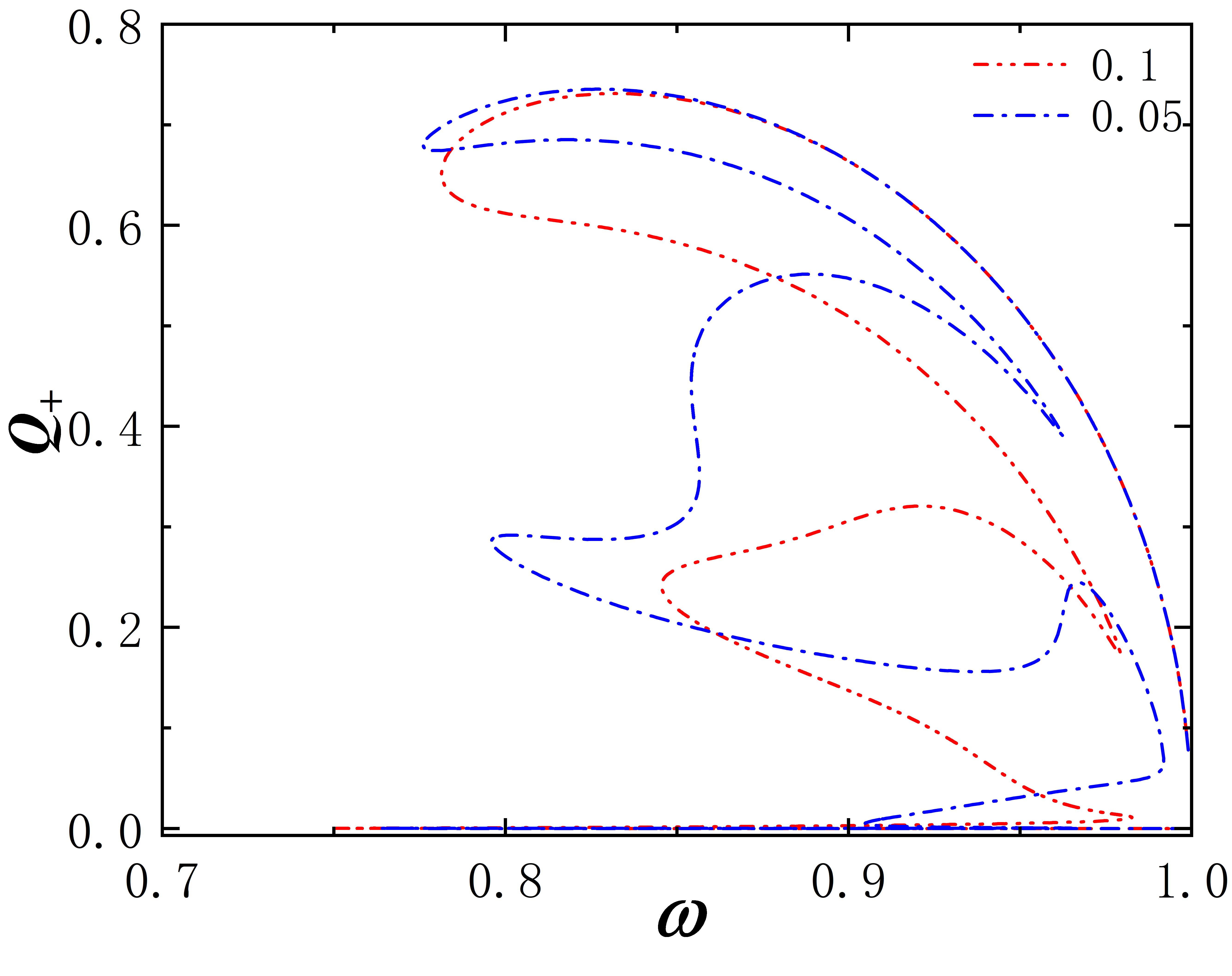}}
\subfigure{\includegraphics[width=0.4\textwidth]{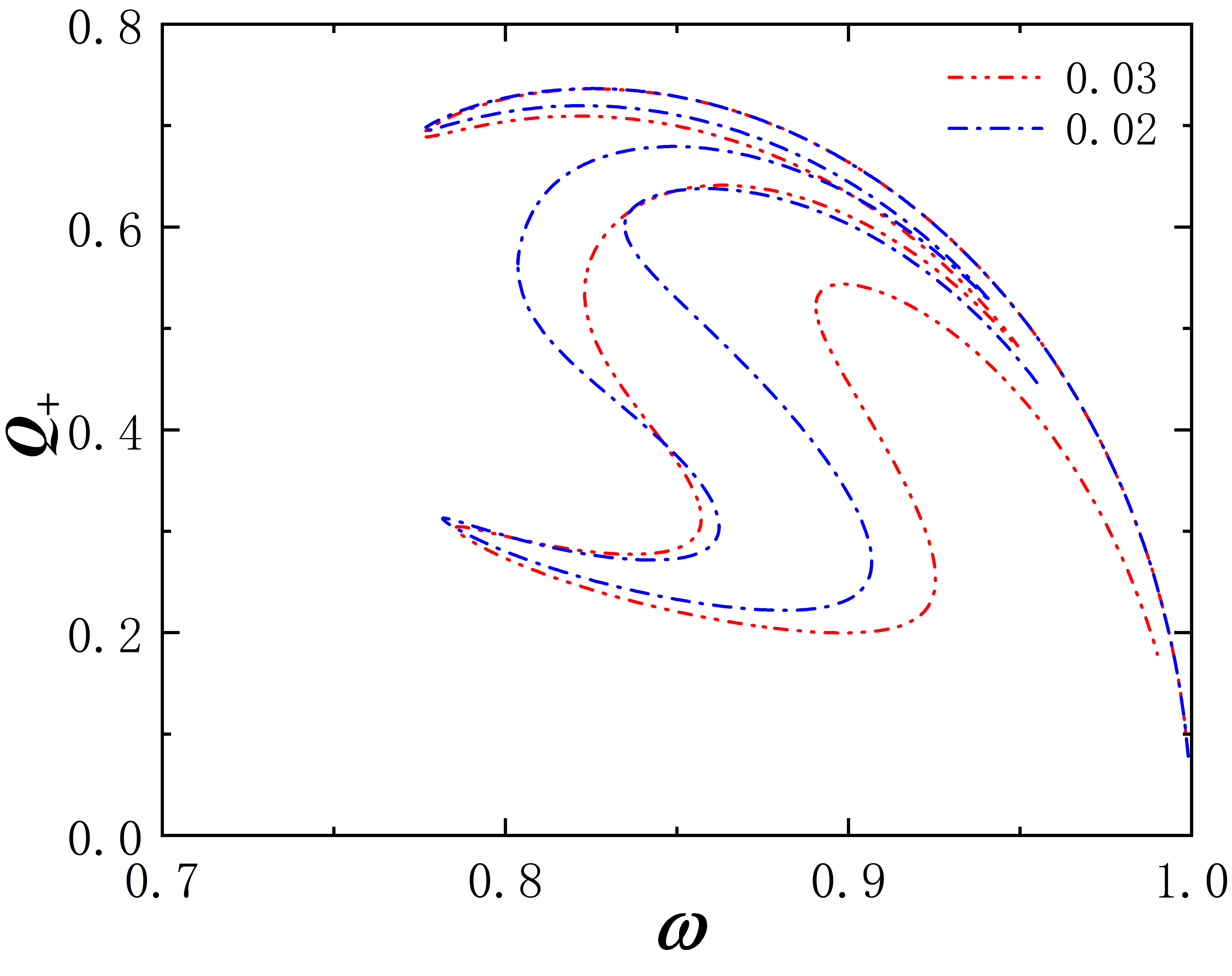}}
\subfigure{\includegraphics[width=0.4\textwidth]{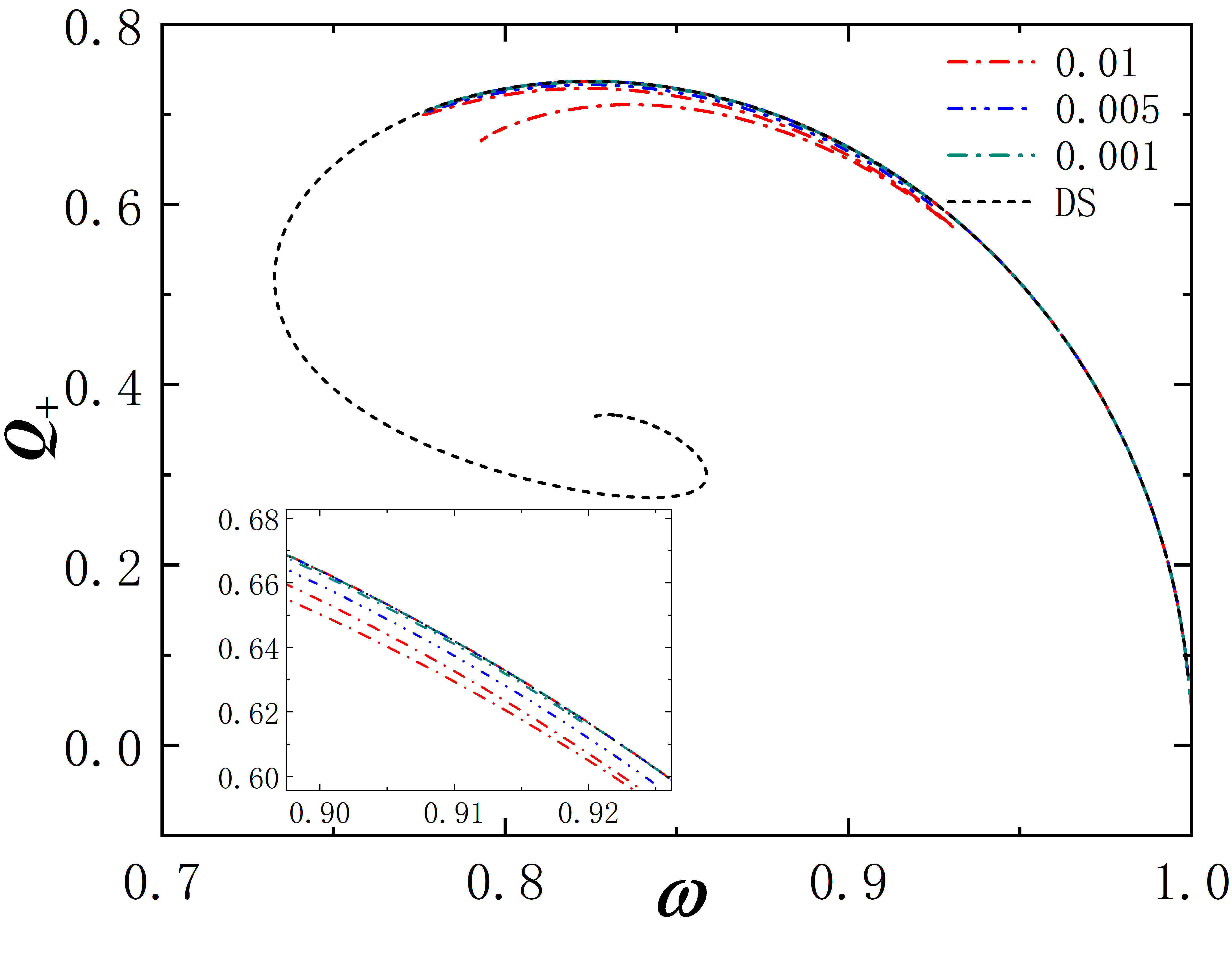}}
\subfigure{\includegraphics[width=0.4\textwidth]{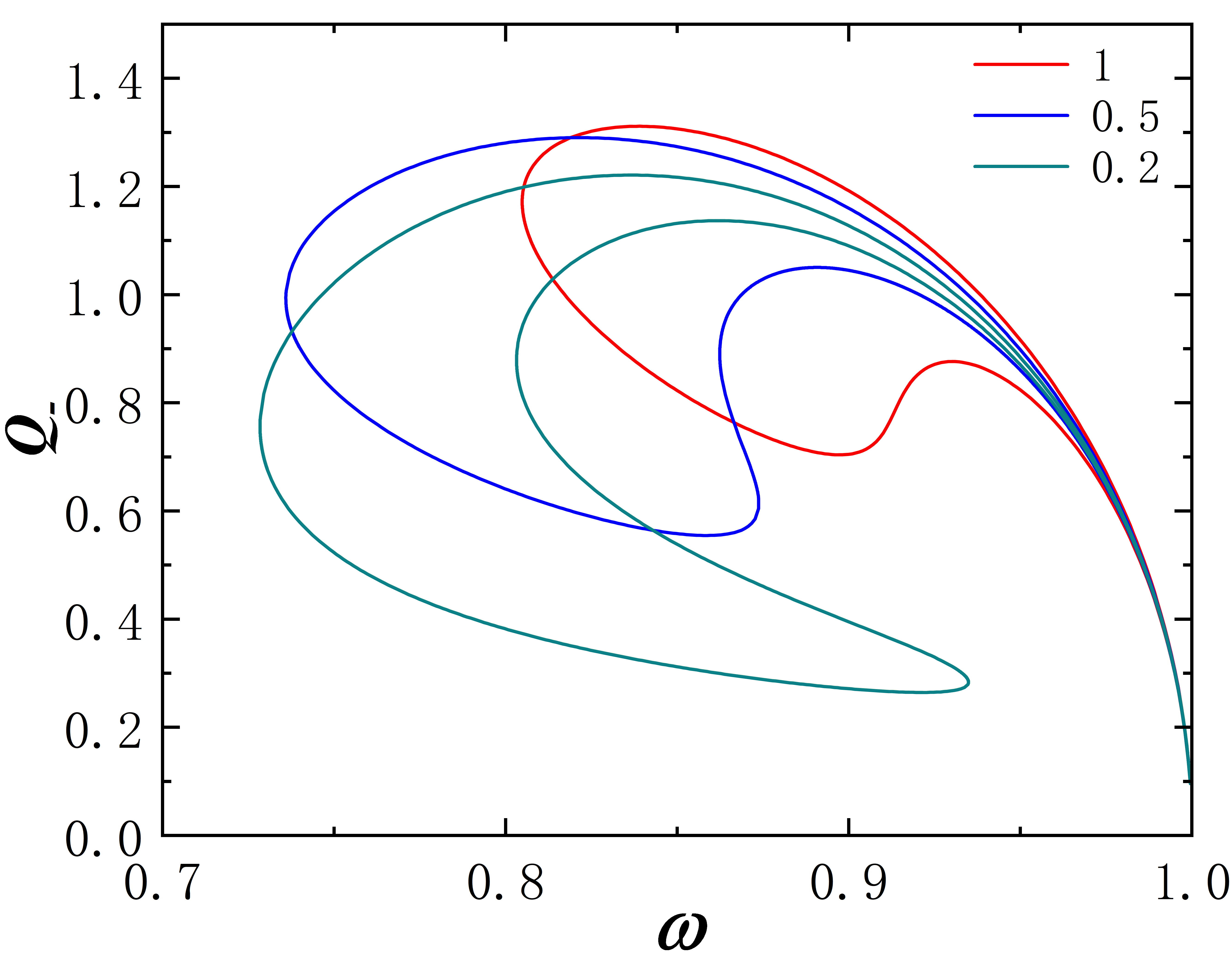}}
\subfigure{\includegraphics[width=0.4\textwidth]{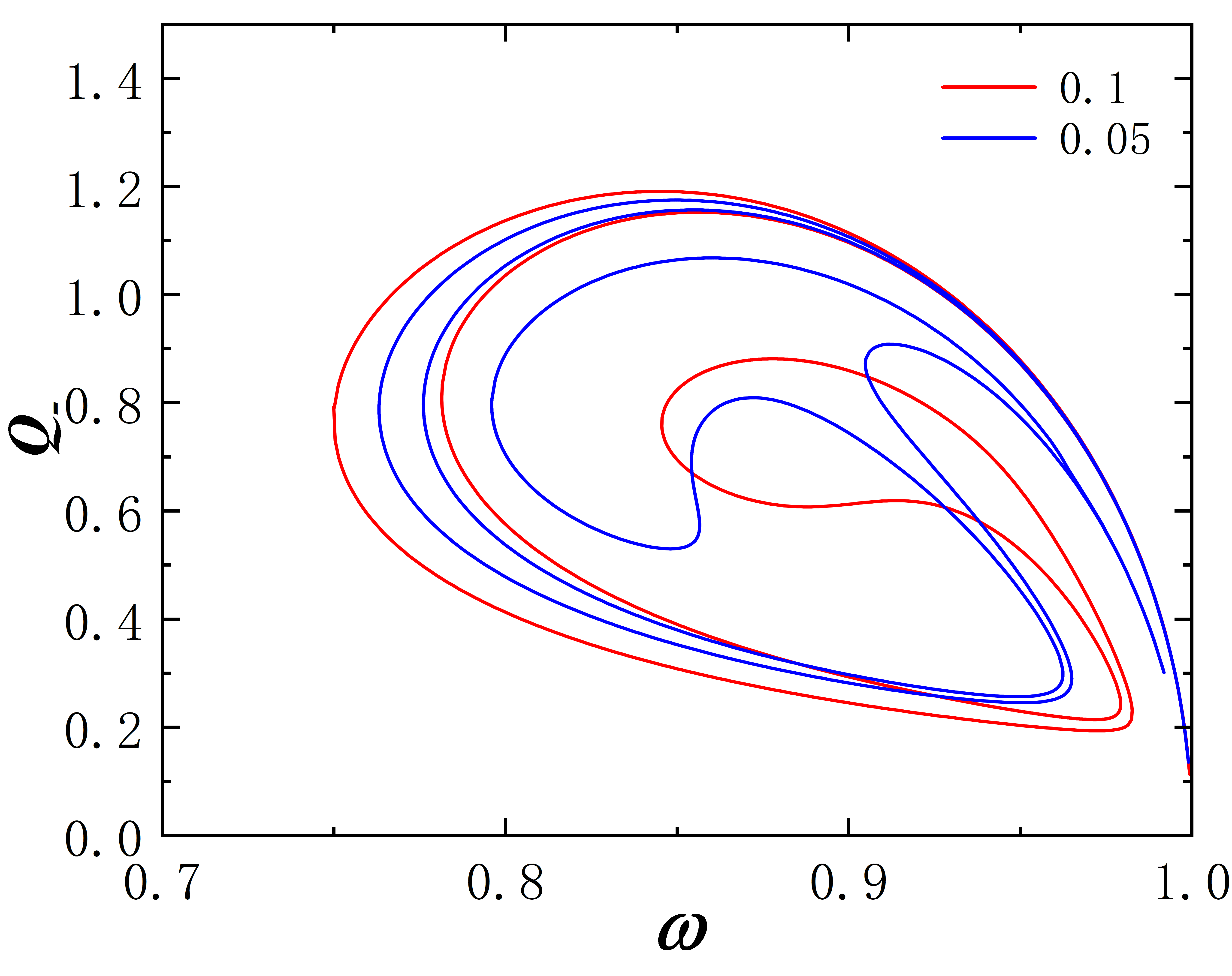}}
\subfigure{\includegraphics[width=0.4\textwidth]{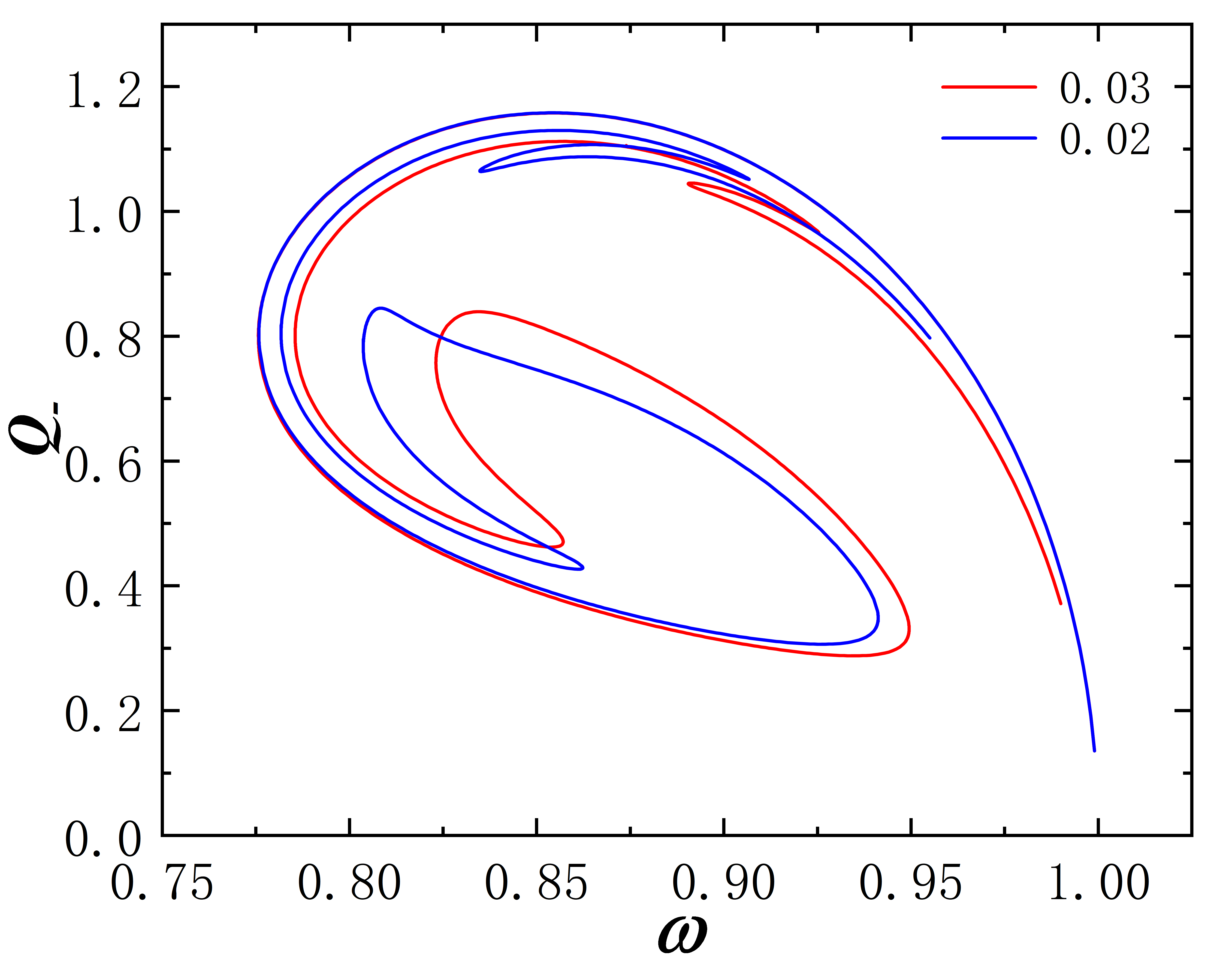}}
\subfigure{\includegraphics[width=0.4\textwidth]{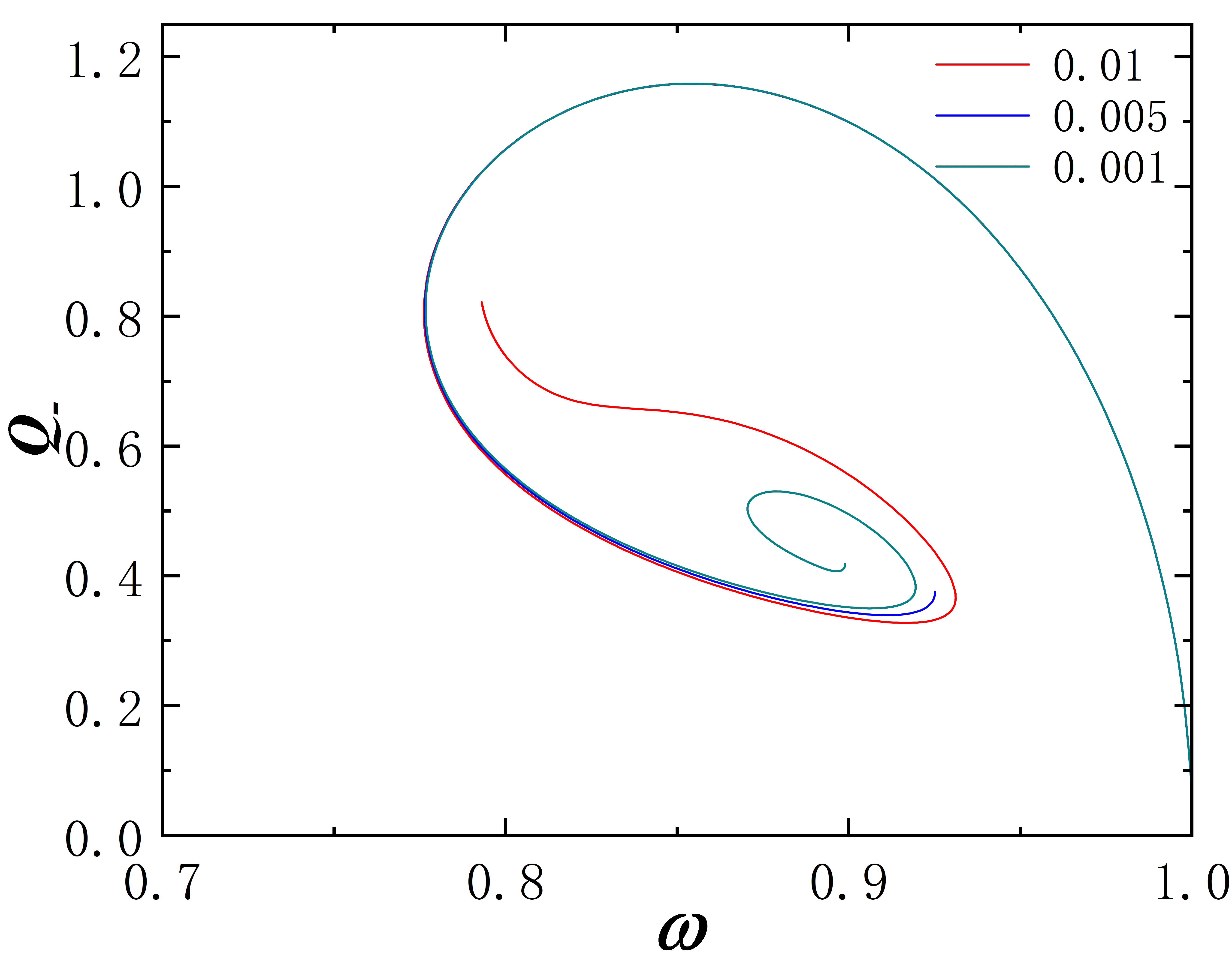}}
\end{center}
\caption{The Noether charge Q as the function of frequency $\omega$ for some values of $r_0$. The dotted line represents $Q_+$, the solid line represents $Q_-$.}
\label{phase4}
\end{figure}

The ADM mass is shown in Fig. \ref{phase2}, $M$ is the function of frequency $\omega$. In particular, Due to the presence of a wormhole at the center, two asymptotically flat spacetime regions exist, resulting in an asymmetric solution with two distinct mass distributions on either side of the wormhole, denoted as $M_+$ and $M_-$. These are represented by dotted and solid lines respectively. To clearly illustrate the mass for different throat sizes, we categorize the results into four groups based on the values of $r_0$. For larger values of $r_0$, such as 1, 0.5, and 0.2, the mass exhibits fewer branches. Starting from the vacuum solution at the first branch's starting point (where $\omega$ is close to 1), it returns to the last branch's endpoint (where $\omega$ is also close to 1), where the vacuum solution is still obtained. As $r_0$ decreases, as shown in the figure for $r_0$ = 0.1, 0.05, and 0.03, the mass branches become more numerous and complex. Interestingly, for $r_0 < 0.03$, the mass distribution starts from the vacuum solution but does not return to it at the end. As $r_0$ decreases further, the number of branches reduces, and the differences between the various branches of $M_+$ become so small that the overall shape appears as a single curve. In contrast, the distribution of $M_-$ assumes the spiral shape characteristic of the well-known Boson star model. In the last panel, by incorporating the Dirac star mass, the first branch of $M_+$ with $r_0$ = 0.001 nearly coincides with the first branch of the Dirac star mass in trivial spacetime. In Fig. \ref{phase3}, we find that the Dirac field in the region of $M_+$ and the field of Dirac star basically overlap. However, as all mass distributions do not completely overlap, the two solutions are not completely same. The Noether charge $Q$ reflects the number of matter field particles in spacetime and its distribution should be similar to ADM mass. We show it in Fig. \ref{phase4}.


\begin{figure}
  \begin{center}
\subfigure{\includegraphics[width=0.49\textwidth]{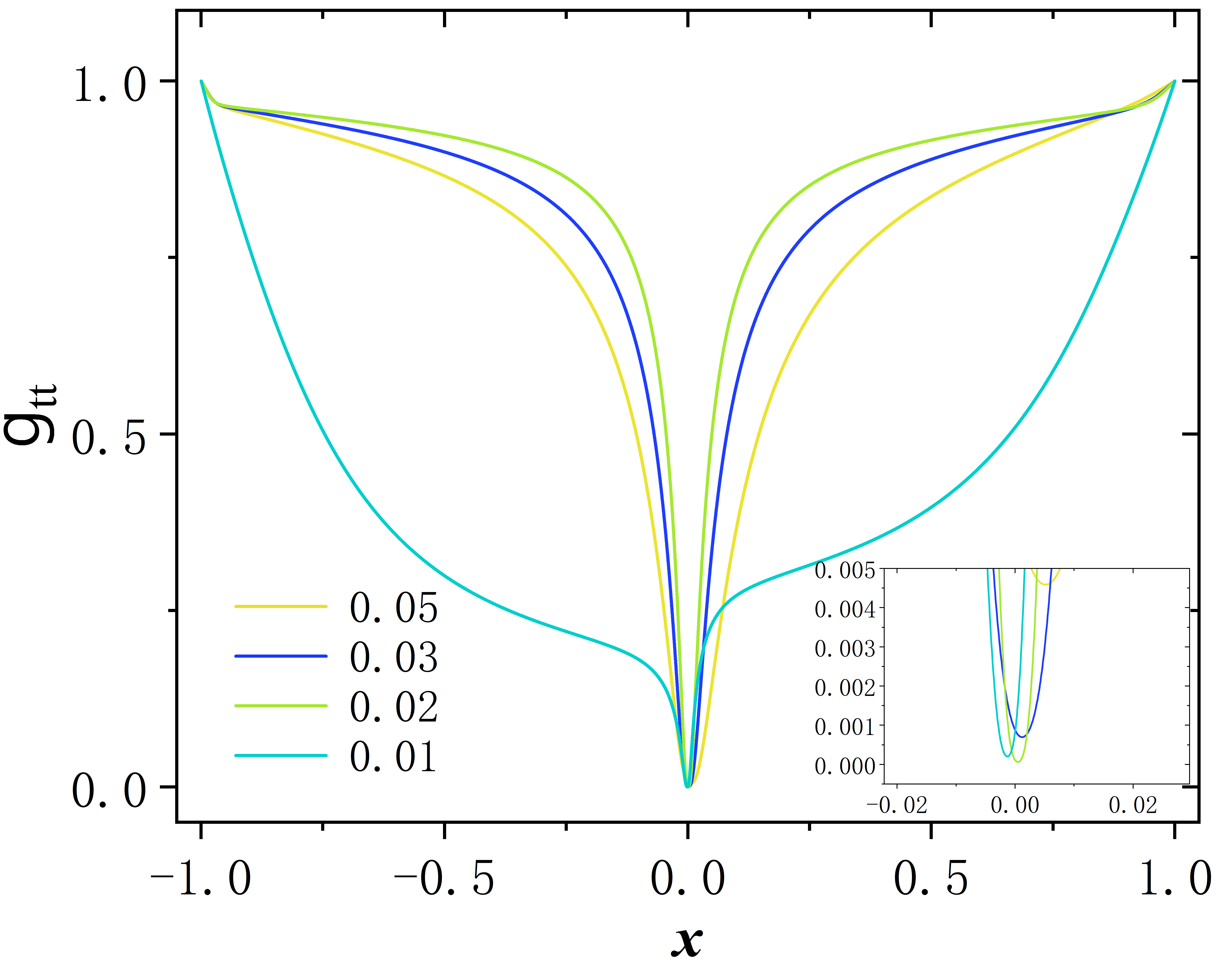}}
\subfigure{\includegraphics[width=0.49\textwidth]{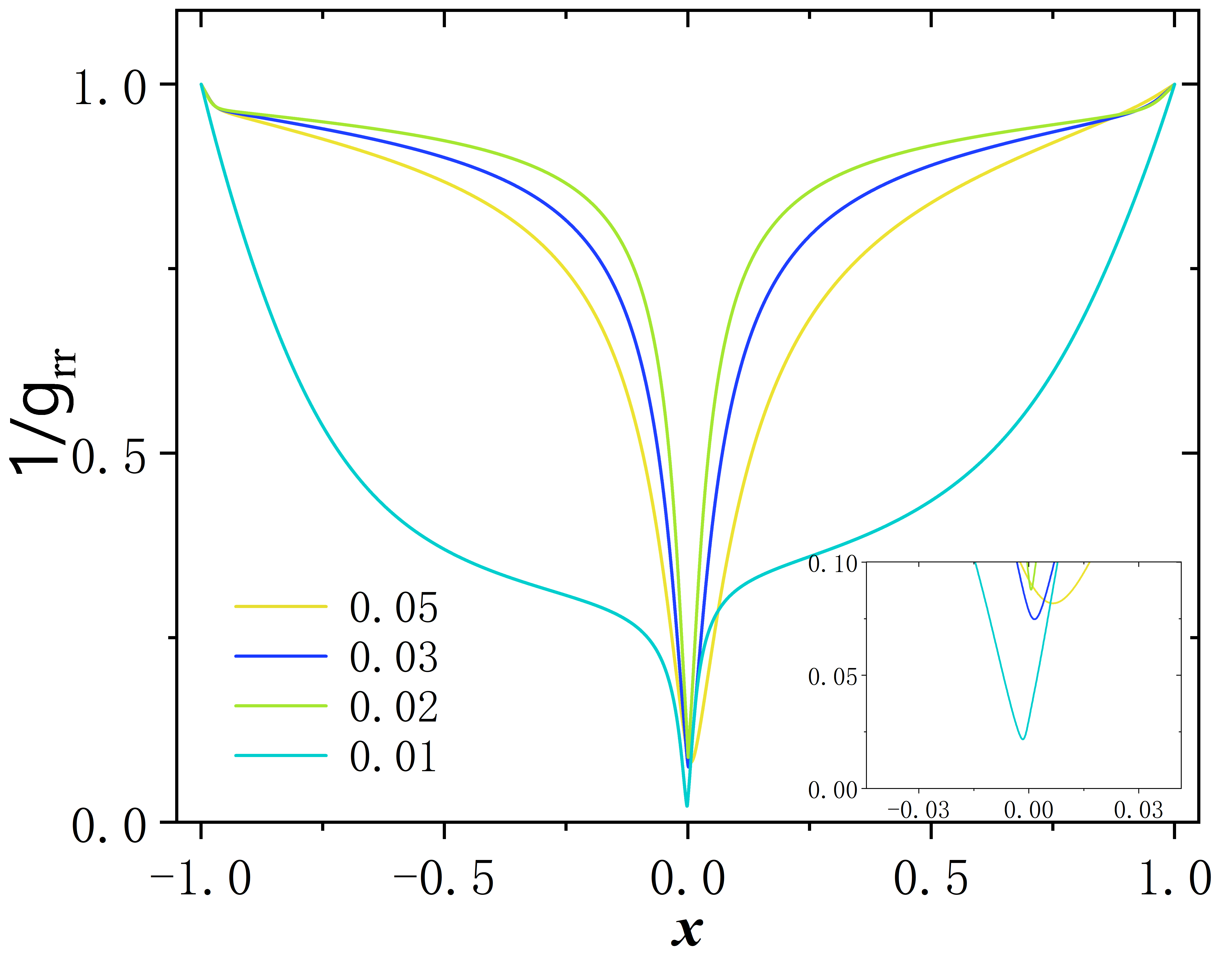}}
  \end{center}
\caption{The left panel is $g_{tt}$ and the right panel is $g_{rr}$, they are functions of $x$. For $r_0$ = $0.05$, $0.03$, $0.02$, the frequency $\omega$ fixed in 0.99. When $r_0$ = $0.01$, the $\omega$ = $0.794$. We use subfigures to show the trend of $g_{tt}$ tending to zero and $g_{rr}$ tending to infinity under four different values of $r_0$.}
\label{phase5}
\end{figure}

\begin{figure}
  \begin{center}
\subfigure{\includegraphics[width=0.49\textwidth]{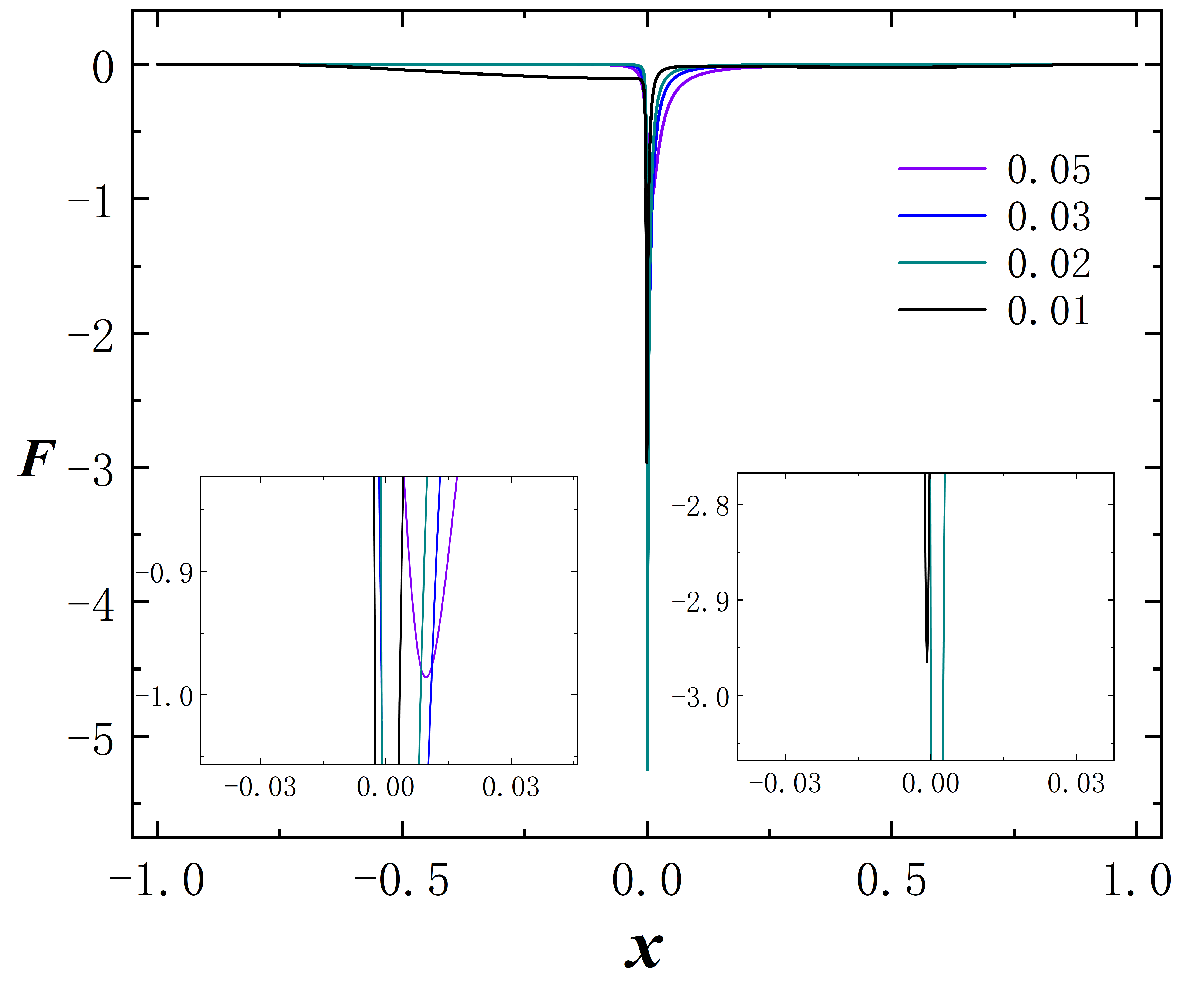}}
\subfigure{\includegraphics[width=0.49\textwidth]{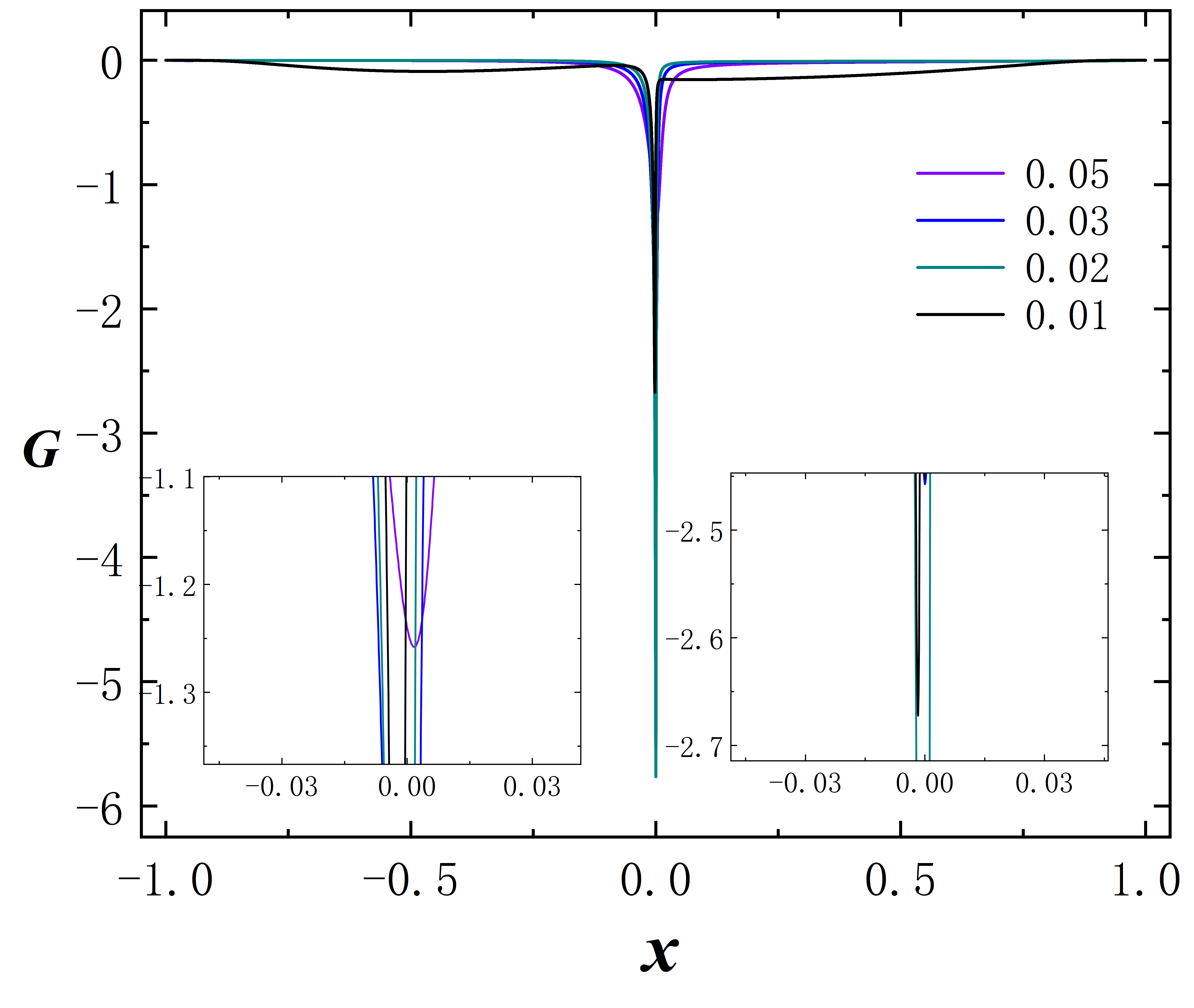}}
  \end{center}
\caption{The left panel is $F$ and the right panel is $G$, they are functions of $x$. For $r_0$ = $0.05$, $0.03$, $0.02$, the frequency $\omega$ fixed in 0.99. When $r_0$ = $0.01$, the $\omega$ = $0.794$. We use subfigures to show the convergence of the fields under four different values of $r_0$.}
\label{phase6}
\end{figure}

From the distribution of ADM mass $M$ and Nother charge $Q$, we observed that when $r_0 \leq 0.03$, the solution originates from the vacuum solution but does not return to it. This led us to discover the emergence of an extremely approximate black hole solution. Moreover, this phenomenon occurs near the end of the last branches, and as the frequency $\omega$ approaches the end point, the appearance of the extremely approximate black hole becomes more and more obvious. To further test this hypothesis, we calculated the values of the metric, and fields $F$ and $G$ at a specific frequency, Fig. \ref{phase5}, Fig. \ref{phase6}.

In the metric component, the $g_{tt} \rightarrow 0$ and $g_{rr} \rightarrow \infty$ are important indicators of black hole emergence. As the throat size $r_0$ decreases, the minimum value of $g_{tt}$ is getting closer and closer to 0, while the maximum value of $g_{rr}$ is constantly approaching infinity, which is represented by its reciprocal approaching 0 in the figure. We consider an extremely approximate black hole to have emerged when the value of $g_{tt}$ is less than $10^{-4}$. In fact, for all cases in which  $r_0 < 0.03$, the criterion is satisfied, Tab. \ref{tab:t1}. And we do not show the solutions of $r_0$ are too small because the numerical error is large at this time. When the $r_0$ = 0.01, the value of metric $g_{tt}$ is larger than $r_0$ = 0.02 for the same reason, due to limitations in numerical calculation accuracy, we cannot calculate the last branch to further $\omega$ when $r_0$ = 0.01. In addition, we also found that as $r_0$ decreases, the minimum value points of both images move to the left from greater than 0, and finally cross the zero point and reach the area less than 0. This shows that from the perspective of an external observer, when the size of the wormhole throat is small, all he can observe in the spacetime on his side is a black hole. As the size of the throat decreases, he will eventually be unable to observe it.

Under the non-vacuum black hole solution, we observed the distribution of the material field is concentrated near the event horizon. When $r_0$ is sufficiently small, the field distribution may even diverge at the limit. As the divergence of matter field, we numerically calculated the Kretschmann scalar when the extremely approximate black hole appeared, and compared it with the results when $r_0$ was larger. These contents will be explained in the last section of this chapter.

    	\begin{table}[!t] 
	\centering 
	\begin{tabular}{|c||c|c|c|}
\hline
		$r_0$ & 0.03($\omega=0.99$) & 0.02($\omega=0.99$) & 0.01($\omega=0.794$) \\
\hline
		$g_{tt}(min)$ & $0.00070$ & $0.00006$ & $0.00020$ \\
\hline
		$g_{rr}(min)$ & $0.0748$ & $0.0879$ & $0.0218$ \\
\hline
	\end{tabular}
 	\caption{Under different values of $r_0$, the minimum mertic $g_{tt}$ and $g_{rr}$ in the specific frequency.}
	\label{tab:t1}
\end{table}

\begin{figure}
  \begin{center}
\subfigure{\includegraphics[width=0.75\textwidth]{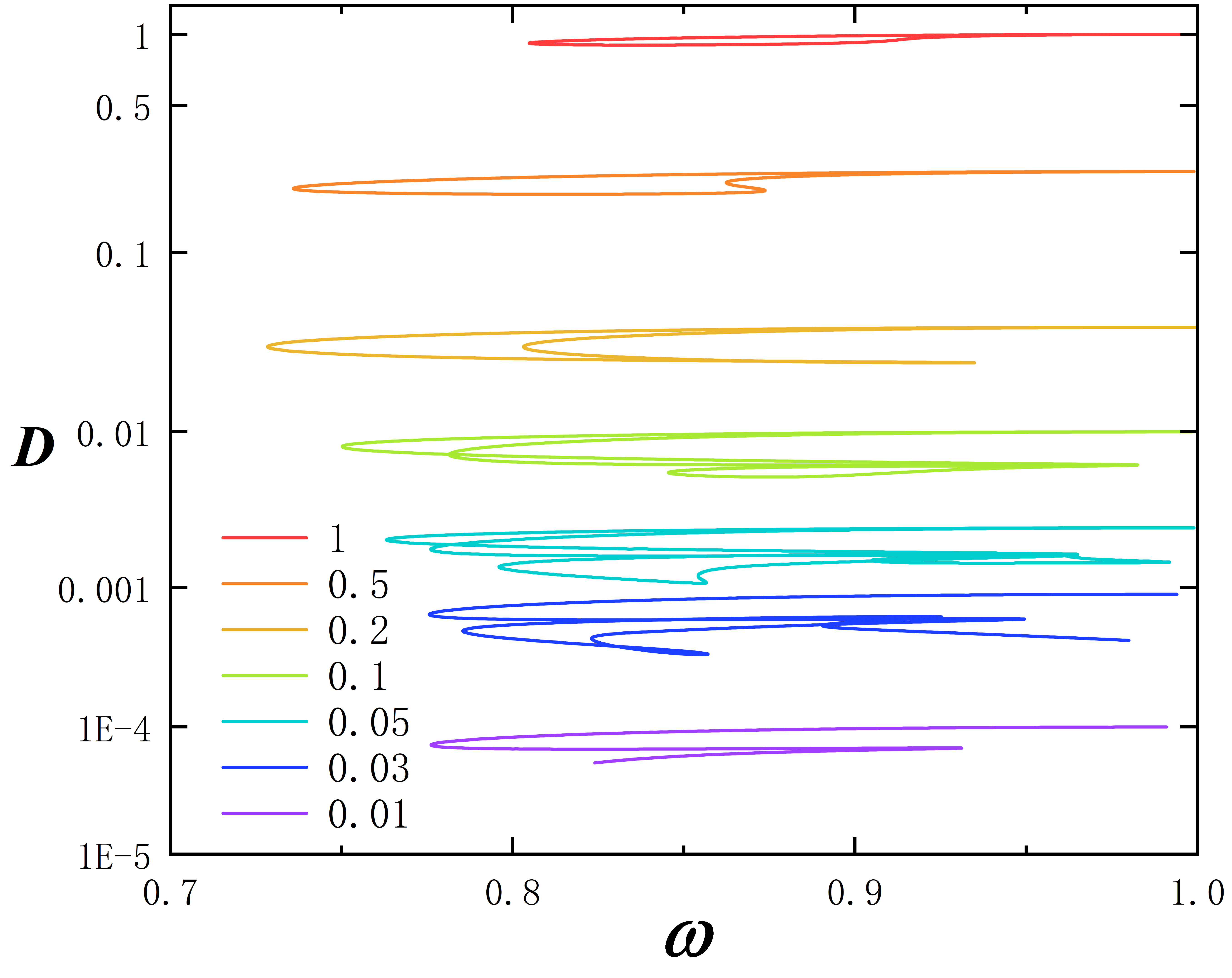}}
\end{center}
\caption{The scalar charge $\cal D$ of the phantom field as a function of the frequency $\omega$ with several values of the throat size $r_0$.}
\label{phase7}
\end{figure}

In our previous article, we defined the constant $\cal D$ to reflect the scalar charge of the phantom field and to test the accuracy of numerical calculations. Its value, fixed at $r_0$ as a function of frequency $\omega$, should be consistent at different positions, as shown in Fig. \ref{phase7}. The shrink of throat size $r_0$ indicates the phantom field that holds the throat open is decreasing, this is reflected in the fact that $\cal D$ decreases with $r_0$.

Finally, we study the geometric properties of the wormhole in this case. We can make use of a geometrical embedding diagram by fixing $t$ and $\theta$. The resulting two-dimensional spatial hypersurface of the wormhole spacetime can then be embedded in a three-dimensional Euclidean space, where the embedding diagram can be used to visualize the wormhole geometry. This technique allows us to better understand the topology and properties of the wormhole solution.

The specific method is: we begin by constructing the embeddings of planes with  $\theta = \pi/2$, and then use the cylindrical coordinates $(\rho,\varphi,z)$, the metric on this plane can be expressed by the following formula
\begin{align}
ds^2 &= B e^{-A}  d r^2 + B e^{-A} h   d\varphi^2 \, \\
&= d \rho^2 + dz^2 + \rho^2 d \varphi^2   \,.
\end{align}
Comparing the two equations above, we then obtain the expression for $\rho$ and $z$,
\begin{equation} \label{formula_embedding}
 \rho(r)= \sqrt{ B(r) e^{-A(r)} h(r) } ,\;\;\;\;\;\;\;\;\;\;   z(r) = \pm  \int  \sqrt{ B(r) e^{-A(r)}  -   \left( \frac{d \rho}{d r} \right)^2    }     d r \;.
\end{equation}
Here $\rho$ corresponds to the circumferential radius, which corresponds to the radius of a circle located in the equatorial plane and having a constant coordinate $r$. The function $\rho(r)$ has extreme points, where the first derivative is zero.
When the second derivative of the extreme point is greater than zero, we call this point a throat, which corresponds to a minimal surface. When the second derivative of the extreme point is less than zero, we call this point an equator, which corresponds to a maximal surface.

\begin{figure}
  \begin{center}
\subfigure{\includegraphics[width=0.425\textwidth]{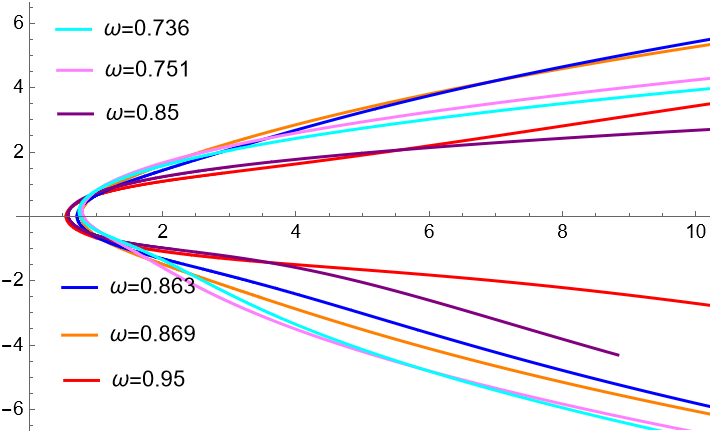}}
\subfigure{\includegraphics[width=0.425\textwidth]{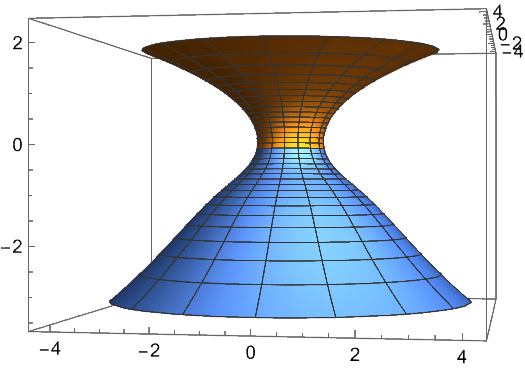}}
\end{center}
\caption{Geometric properties of throats: \textit{left}: Two-dimensional view of the isometric embedding of the equatorial plane. \textit{Right}: Isometric embeddings of the equatorial plane of this solution with throat parameter $r_0$ = 0.5 and  $\omega$ = 0.736.}
\label{phase8}
\end{figure}

\begin{figure}
  \begin{center}
\subfigure{\includegraphics[width=0.425\textwidth]{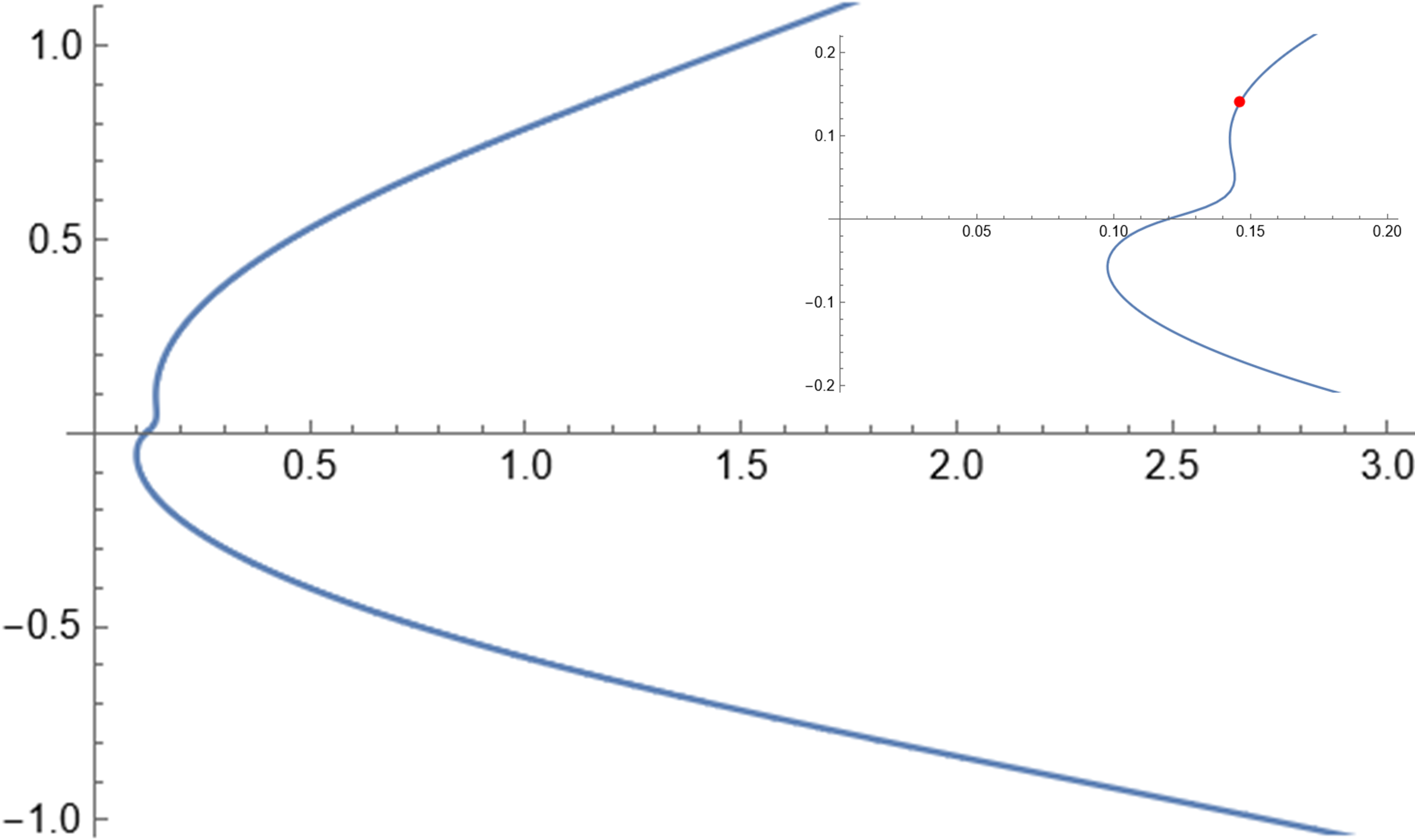}}
\subfigure{\includegraphics[width=0.425\textwidth]{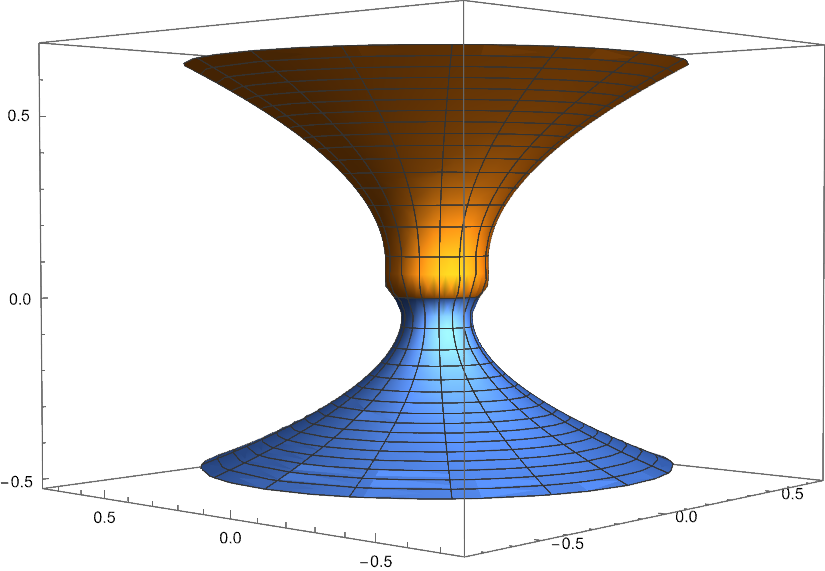}}
\end{center}
\caption{ \textit{left}: Two-dimensional view of the isometric embedding of the equatorial plane with $r_0$ = $0.03$, $\omega$ = $0.9$. \textit{Right}: The corresponding 3D embedded image. }
\label{phase9}
\end{figure}

In Fig. \ref{phase8}, we show a two-dimensional view of the isometric embedding of the equatorial plane with throat parameter $r_0$ = 0.5 and several values of the frequency $\omega$ at the left panel. The right panel shows the 3D plot corresponding to the frequency $\omega$ = 0.736. We can see that the geometry of the wormhole at any frequency in this case is asymmetric and has only one throat, no equatorial plane.
\FloatBarrier
It is worth noting that when the throat size $r_0$ is small and reaches the approximate black hole solution mentioned before, the situation changes. An equatorial plane appears in the wormhole, with a throat on both sides of the equatorial plane, Fig. \ref{phase9}. In the subfigure on the left, we zoom in on the throats and equatorial plane regions, and mark with red dot the location where an extremely approximate black hole appears at this time.
\FloatBarrier

\textbf{Case2: $F$ has two nodes and $G$ has one node.}

\begin{figure}
  \begin{center}
\subfigure{\includegraphics[width=0.49\textwidth]{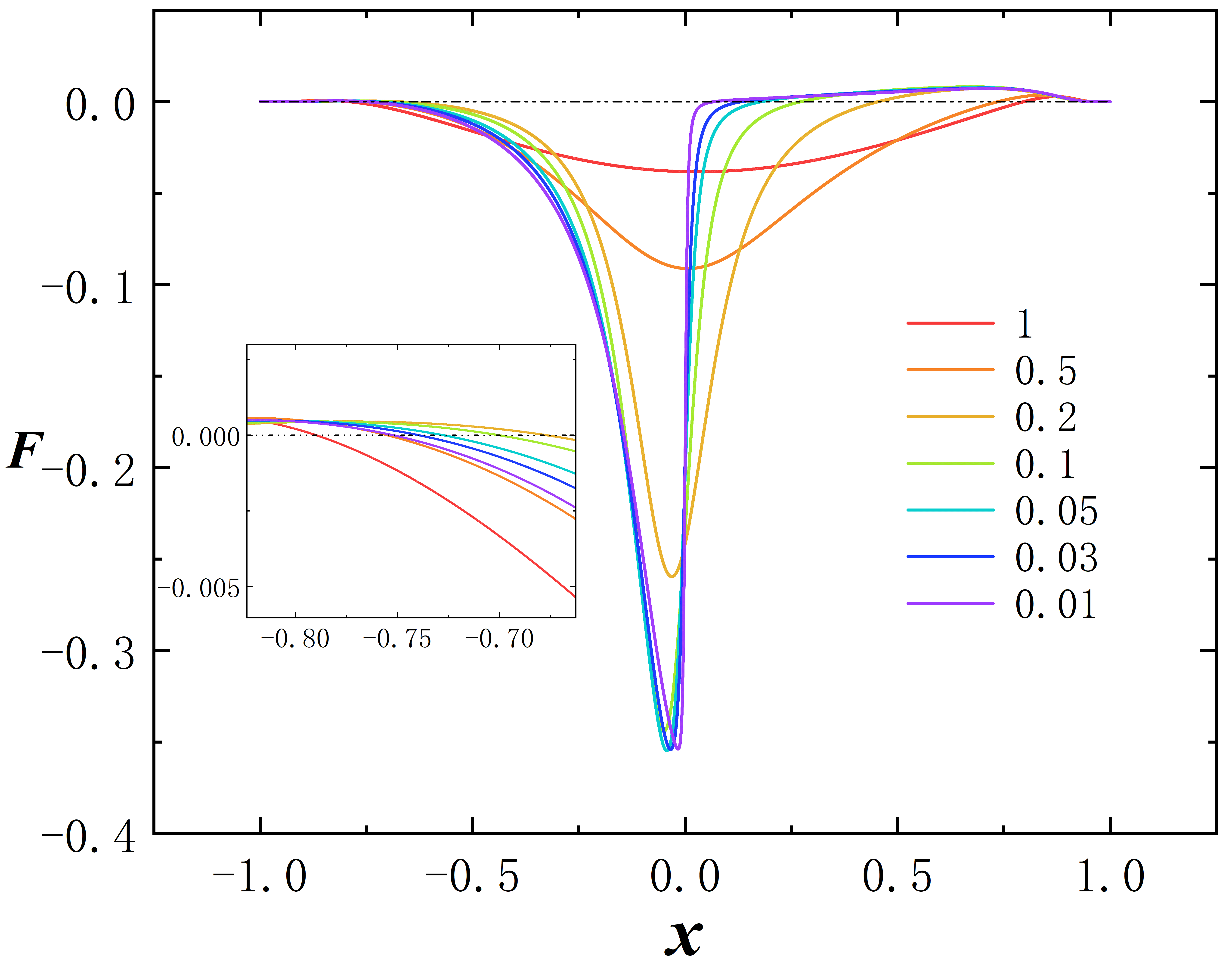}}
\subfigure{\includegraphics[width=0.49\textwidth]{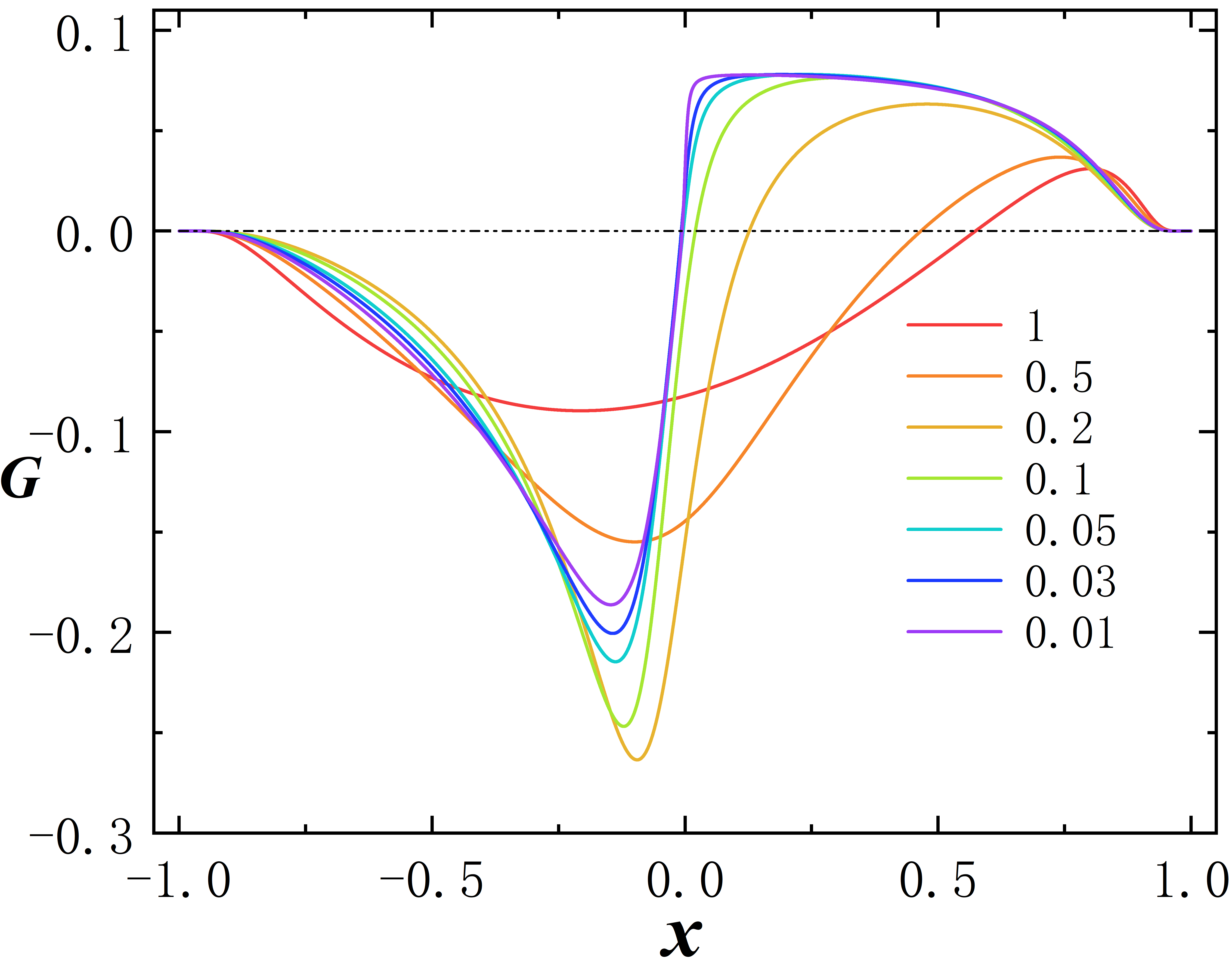}}
  \end{center}
\caption{The radial distribution of the dirac fields $F$ and $G$ with several values of $r_0$ for frequency $\omega$ = 0.87.}
\label{phase10}
\end{figure}

We obtained solutions for higher excited states with two nodes in the $F$ field and one node in the $G$ field, maintaining an asymmetric field configuration, Fig. \ref{phase10}. For comparison, we fixed the frequency $\omega$ at 0.87 and selected several values of throat size $r_0$ as before, and added black dotted lines at $F$ = 0 and $G$ = 0 to indicate node positions.

The ADM mass in this case differs significantly from that in the low excited state, but the mass distribution becomes similar for small $r_0$. The branches of $M_+$ and $M_-$ are less of branches for different throat sizes $r_0$. As $r_0$ decreases, the distribution of $M_+$ becomes monotonous and the distribution of $M_-$ becomes spiral as the first case. We once again compared the ADM masses of $M_+$ and Dirac stars when $r_0$ was very small. The distributions of them still almost coincided. This is also reflected in the field and will not be shown again.

\begin{figure}
\begin{center}
\subfigure{\includegraphics[width=0.32\textwidth]{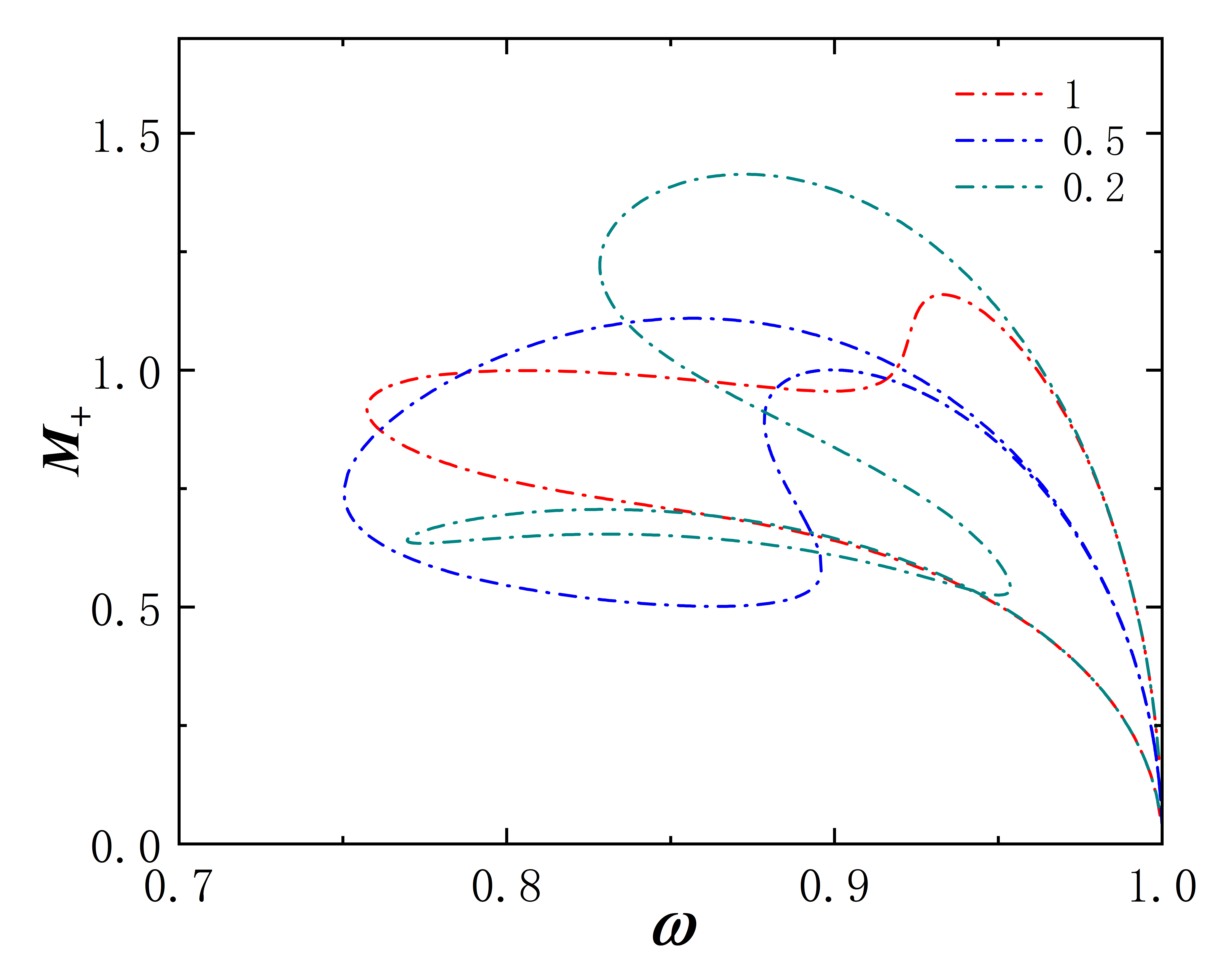}}
\subfigure{\includegraphics[width=0.32\textwidth]{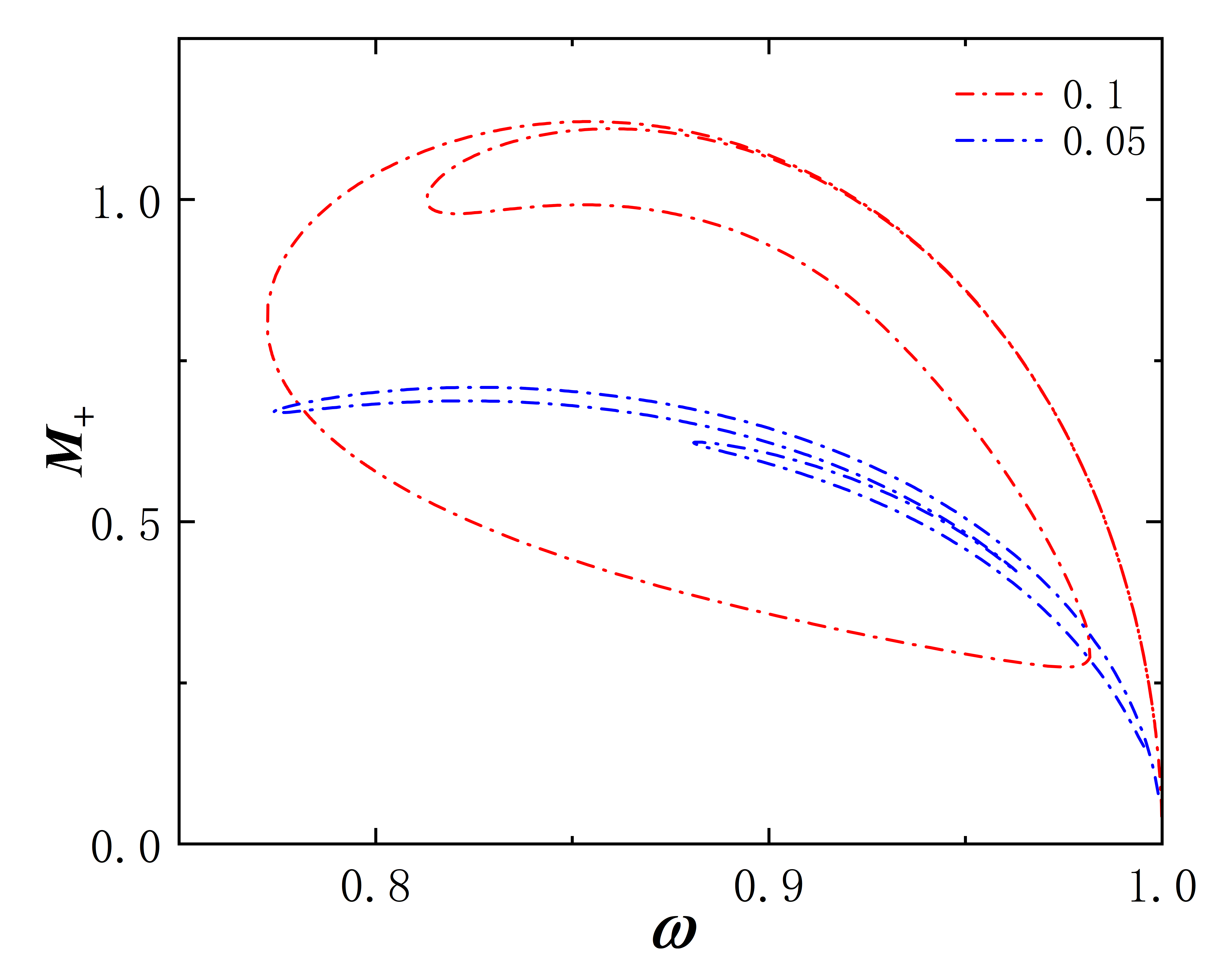}}
\subfigure{\includegraphics[width=0.32\textwidth]{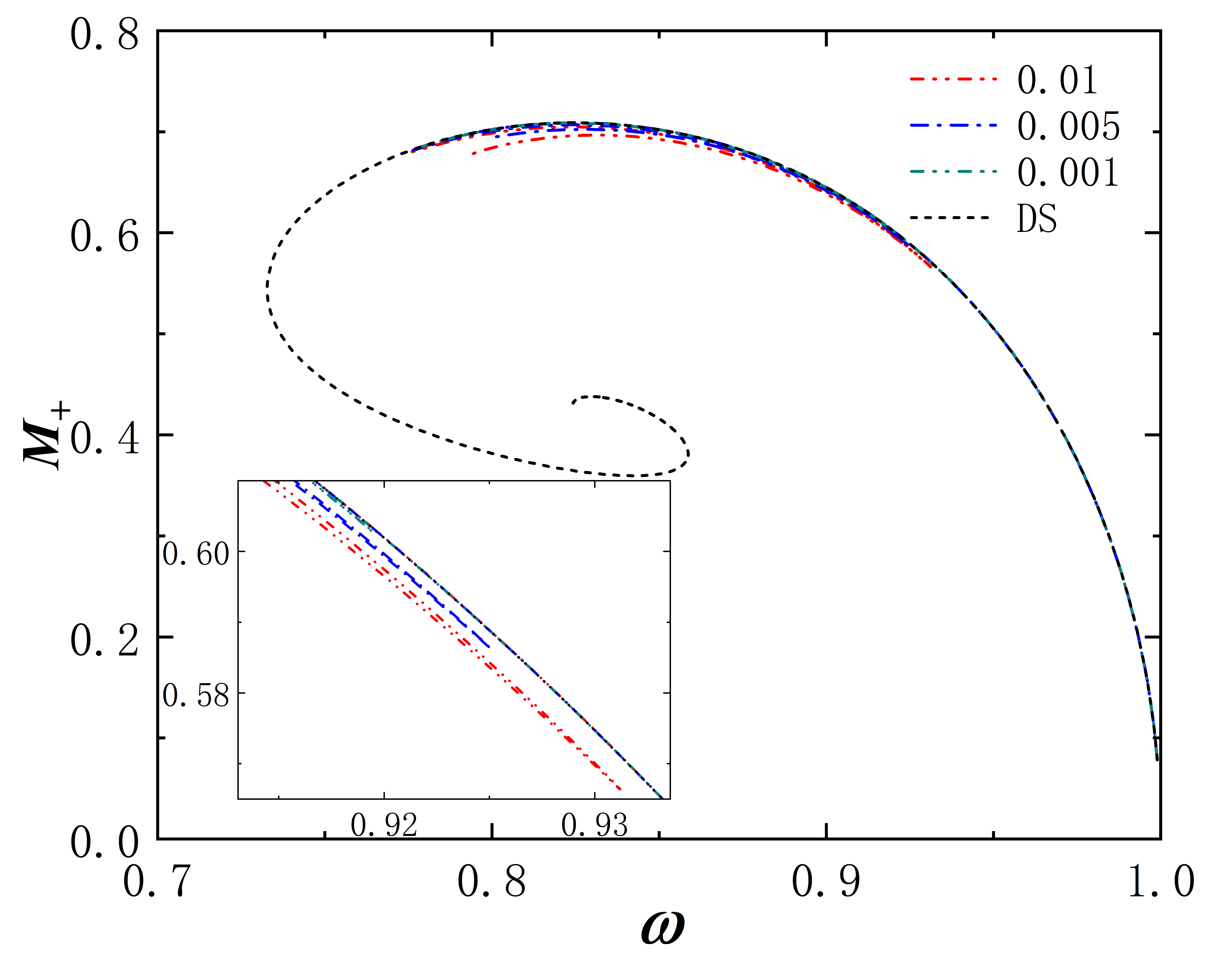}}
\subfigure{\includegraphics[width=0.32\textwidth]{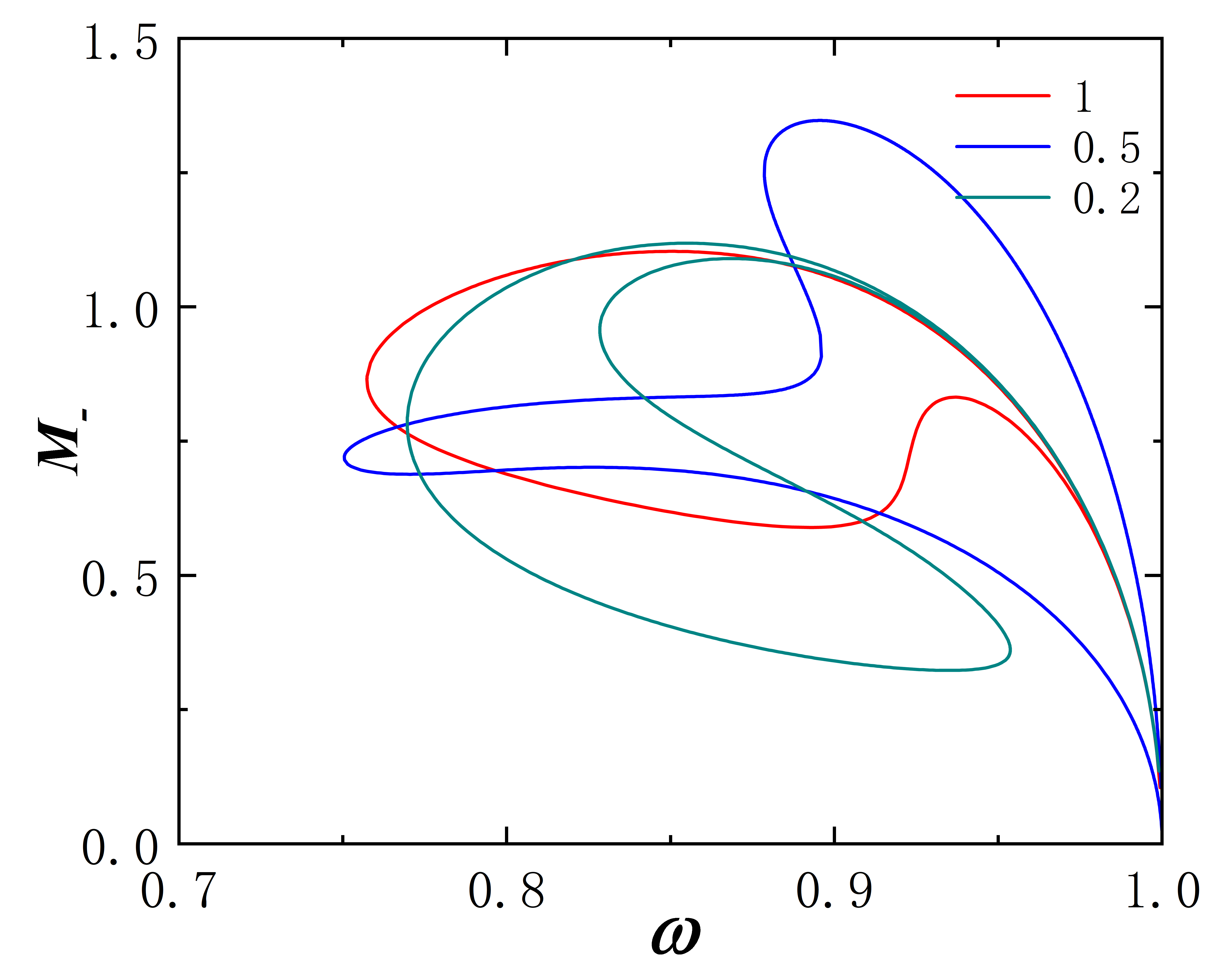}}
\subfigure{\includegraphics[width=0.32\textwidth]{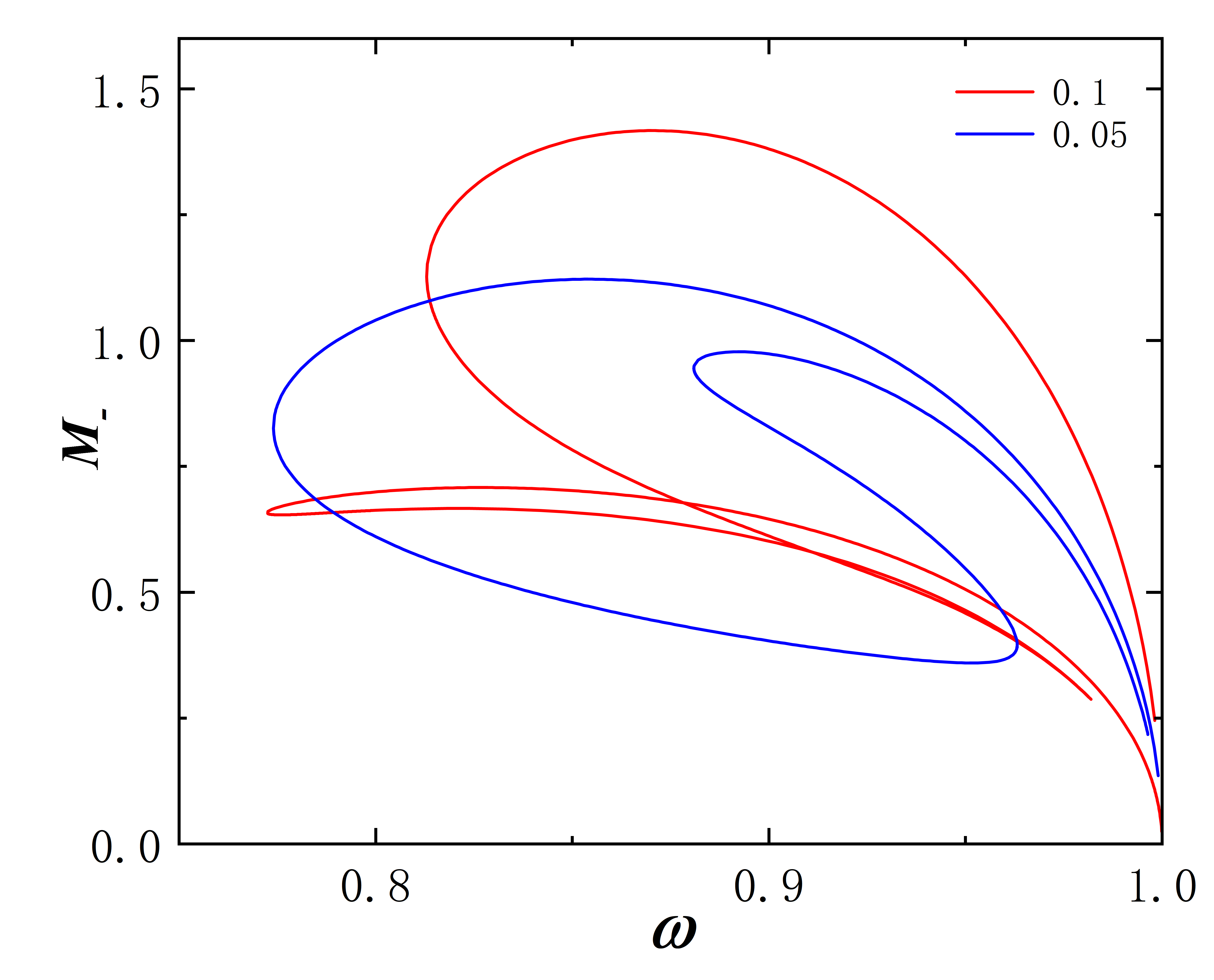}}
\subfigure{\includegraphics[width=0.32\textwidth]{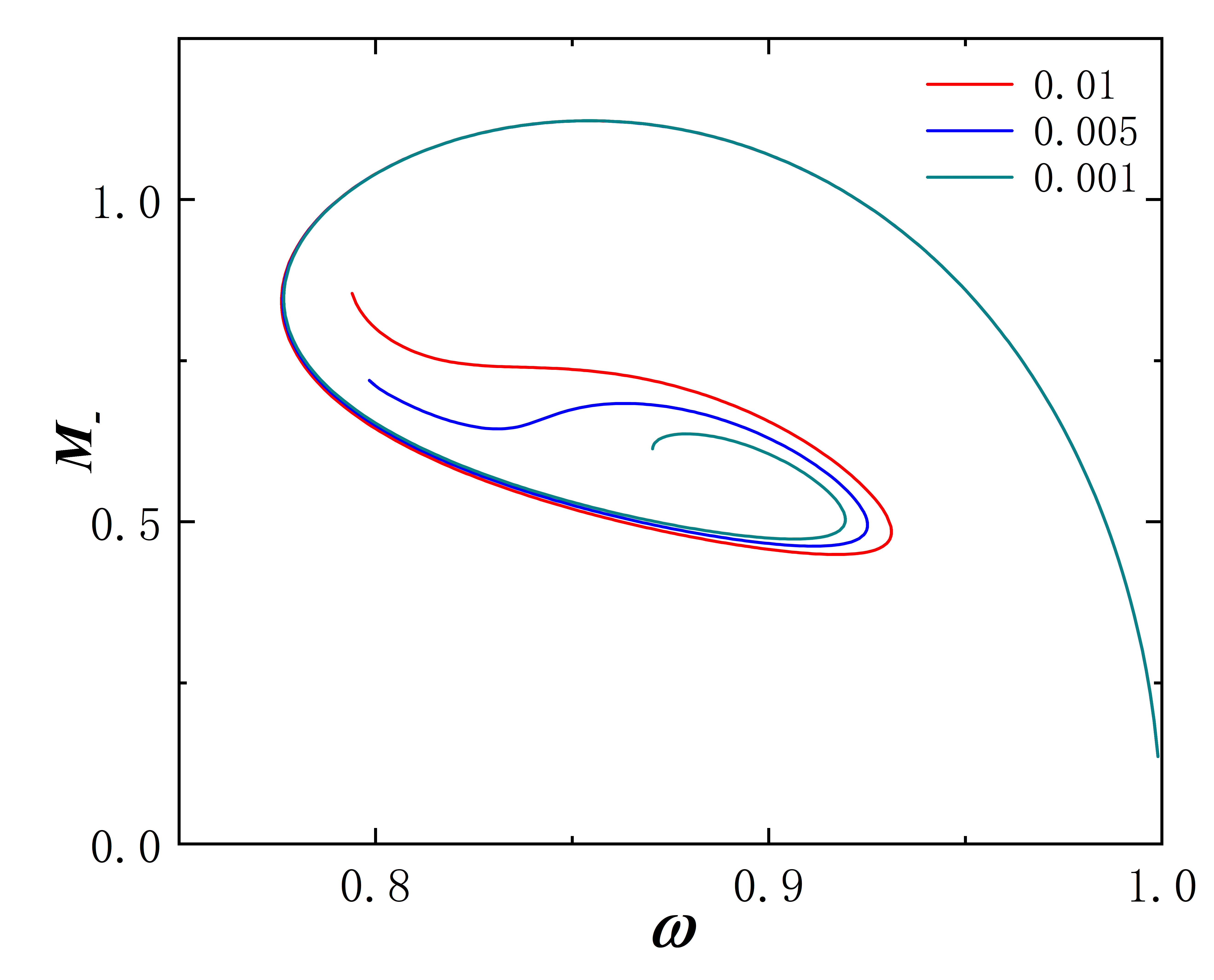}}
\end{center}
\caption{The mass M as the function of frequency $\omega$ for some values of $r_0$. The dotted line represents $M_+$, the solid line represents $M_-$.}
\label{phase11}
\end{figure}

\begin{figure}
\begin{center}
\subfigure{\includegraphics[width=0.32\textwidth]{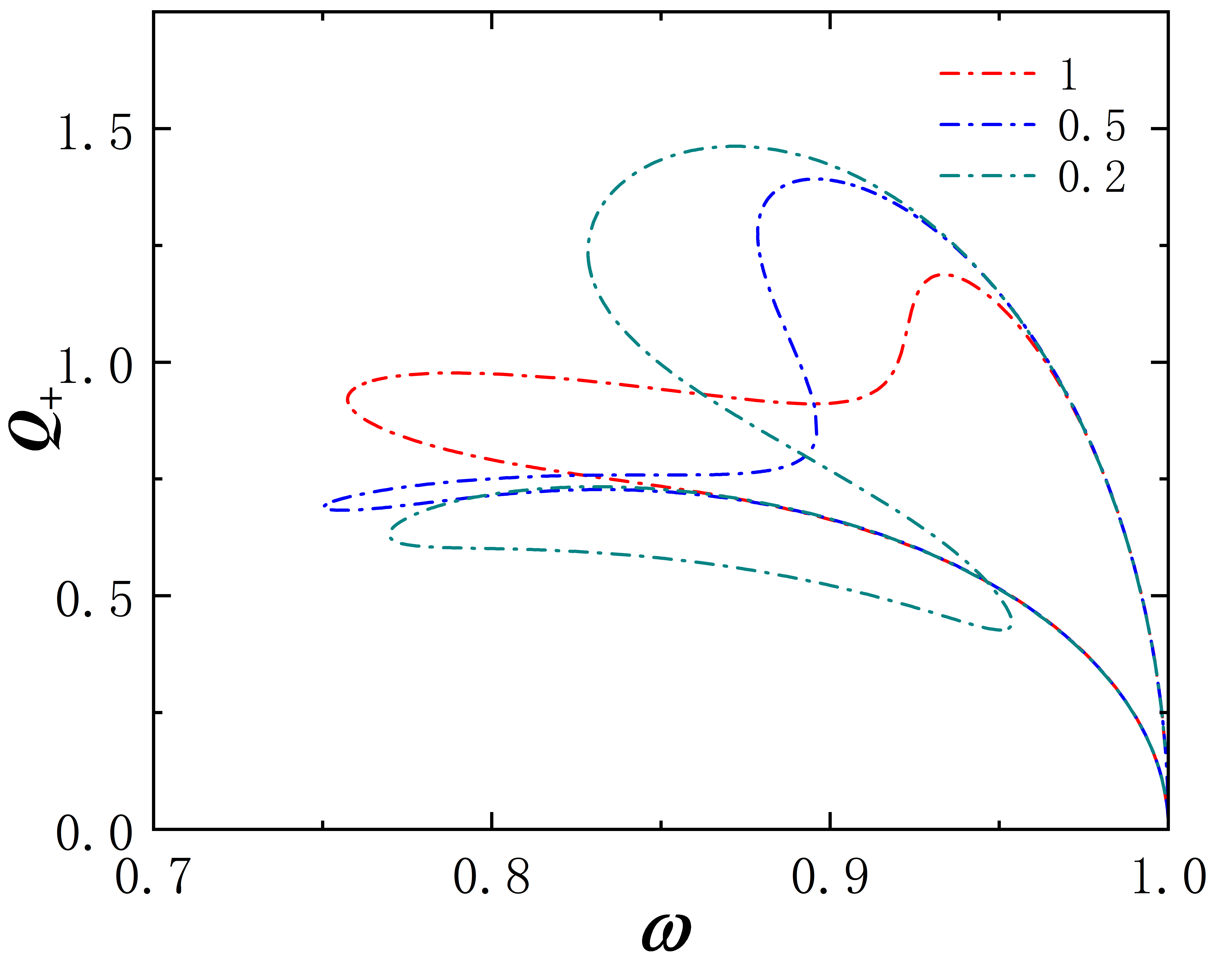}}
\subfigure{\includegraphics[width=0.32\textwidth]{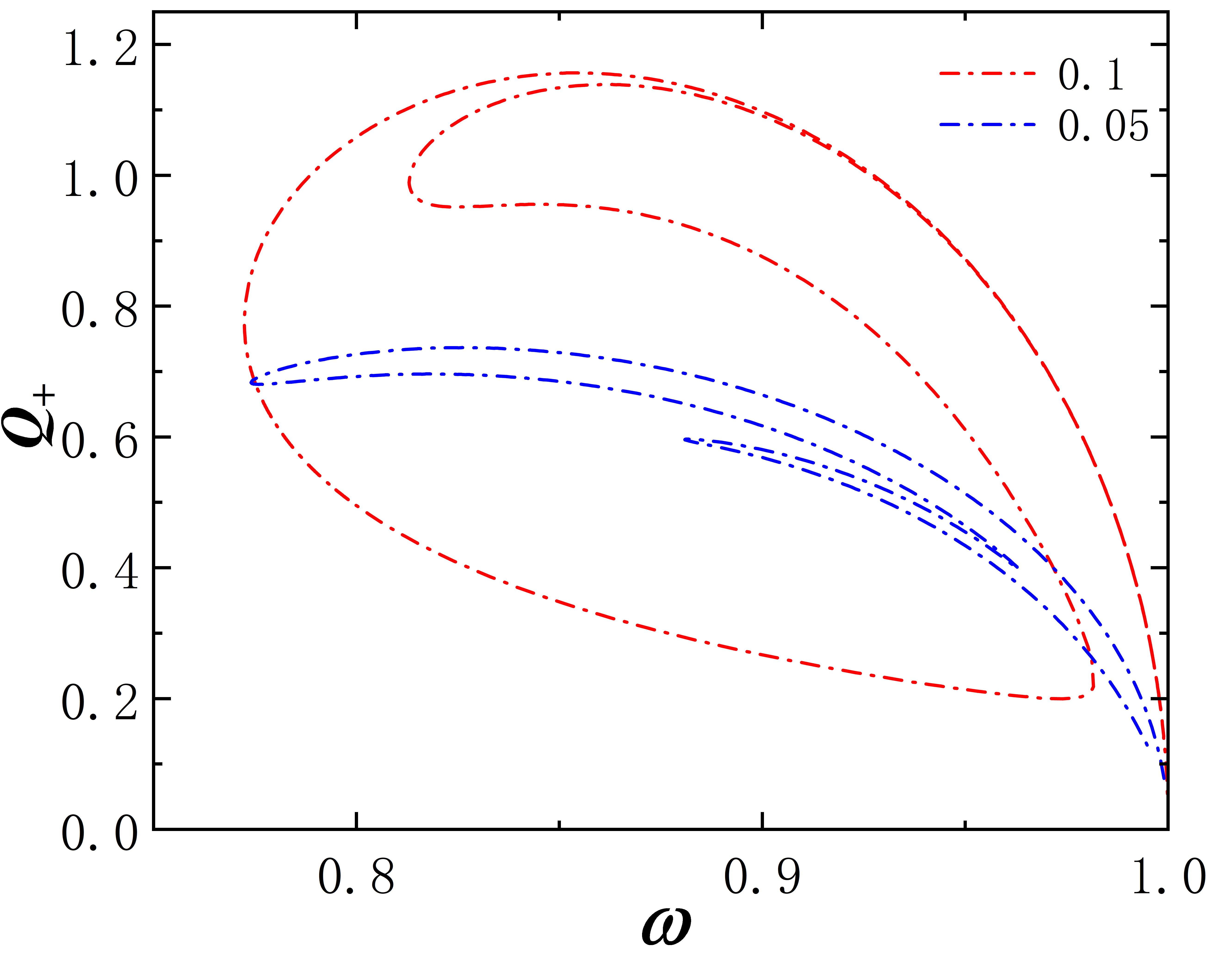}}
\subfigure{\includegraphics[width=0.32\textwidth]{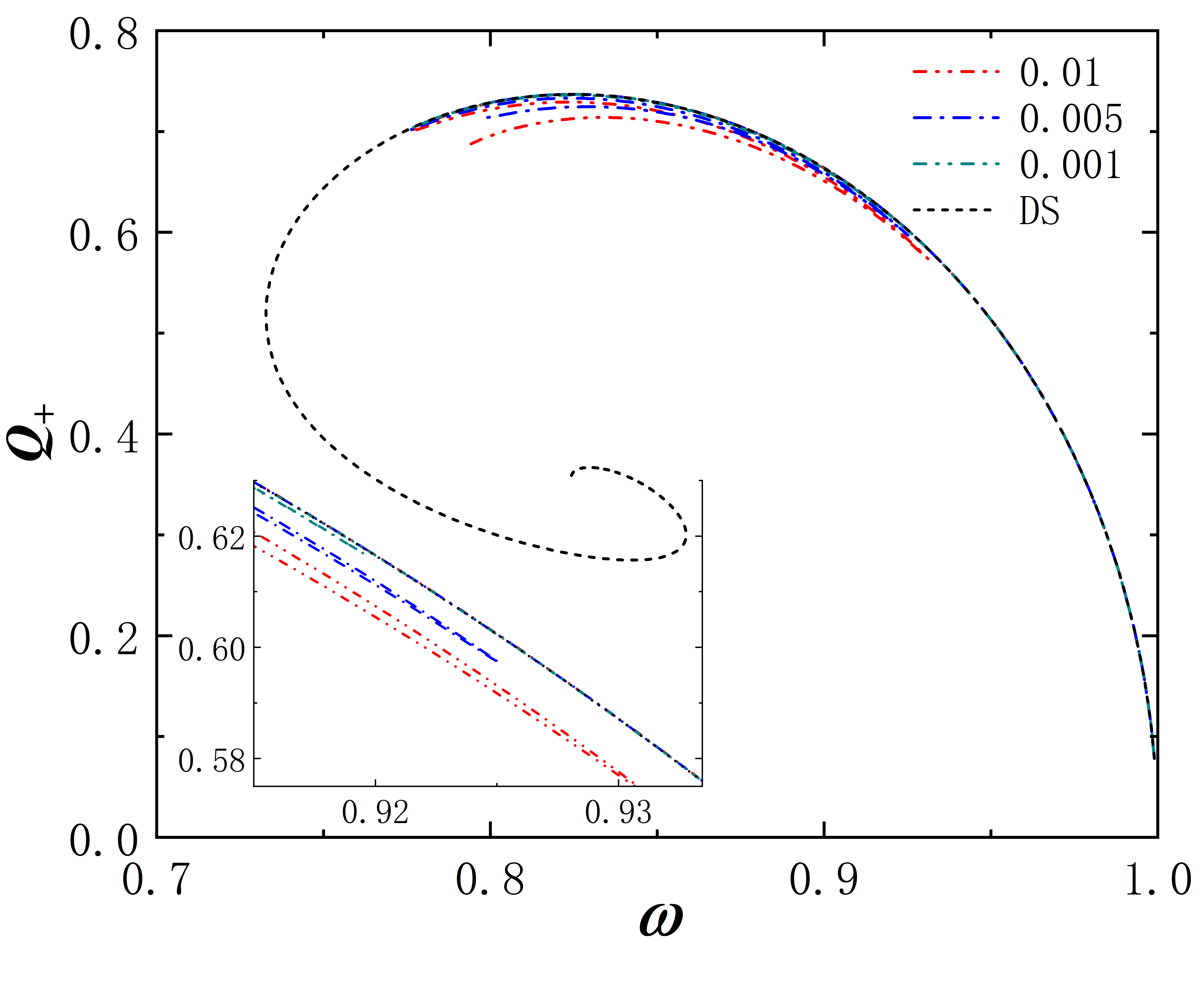}}
\subfigure{\includegraphics[width=0.32\textwidth]{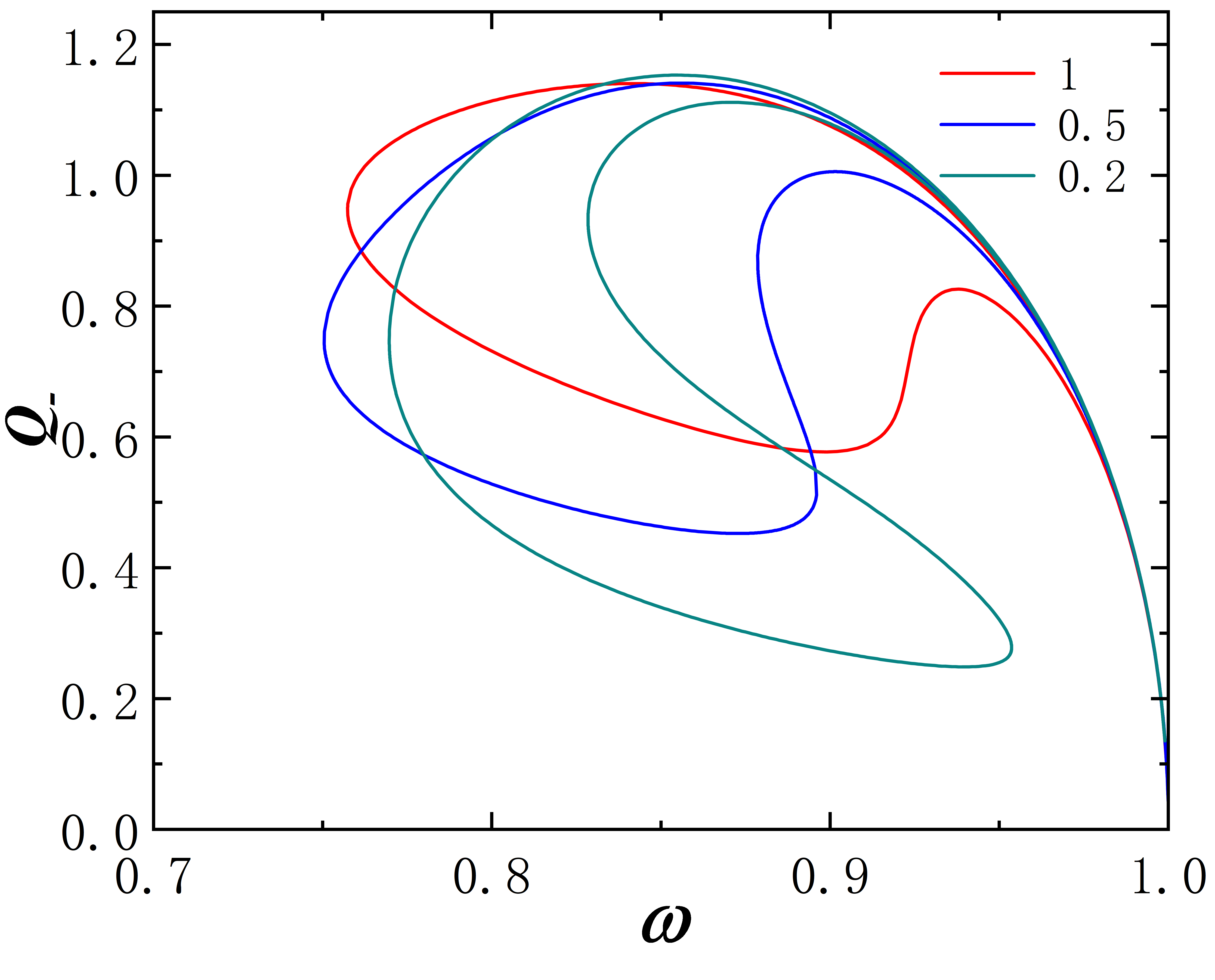}}
\subfigure{\includegraphics[width=0.32\textwidth]{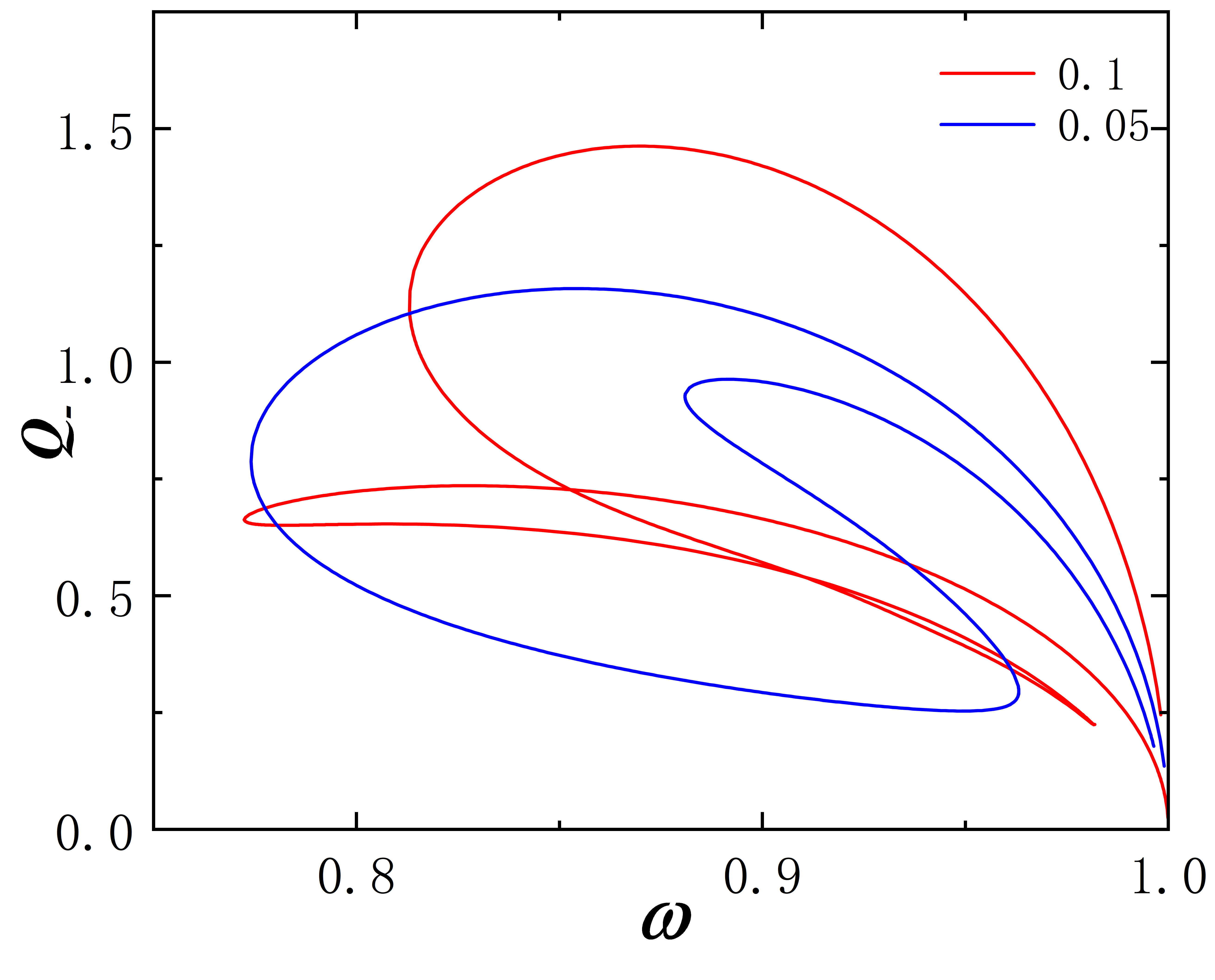}}
\subfigure{\includegraphics[width=0.32\textwidth]{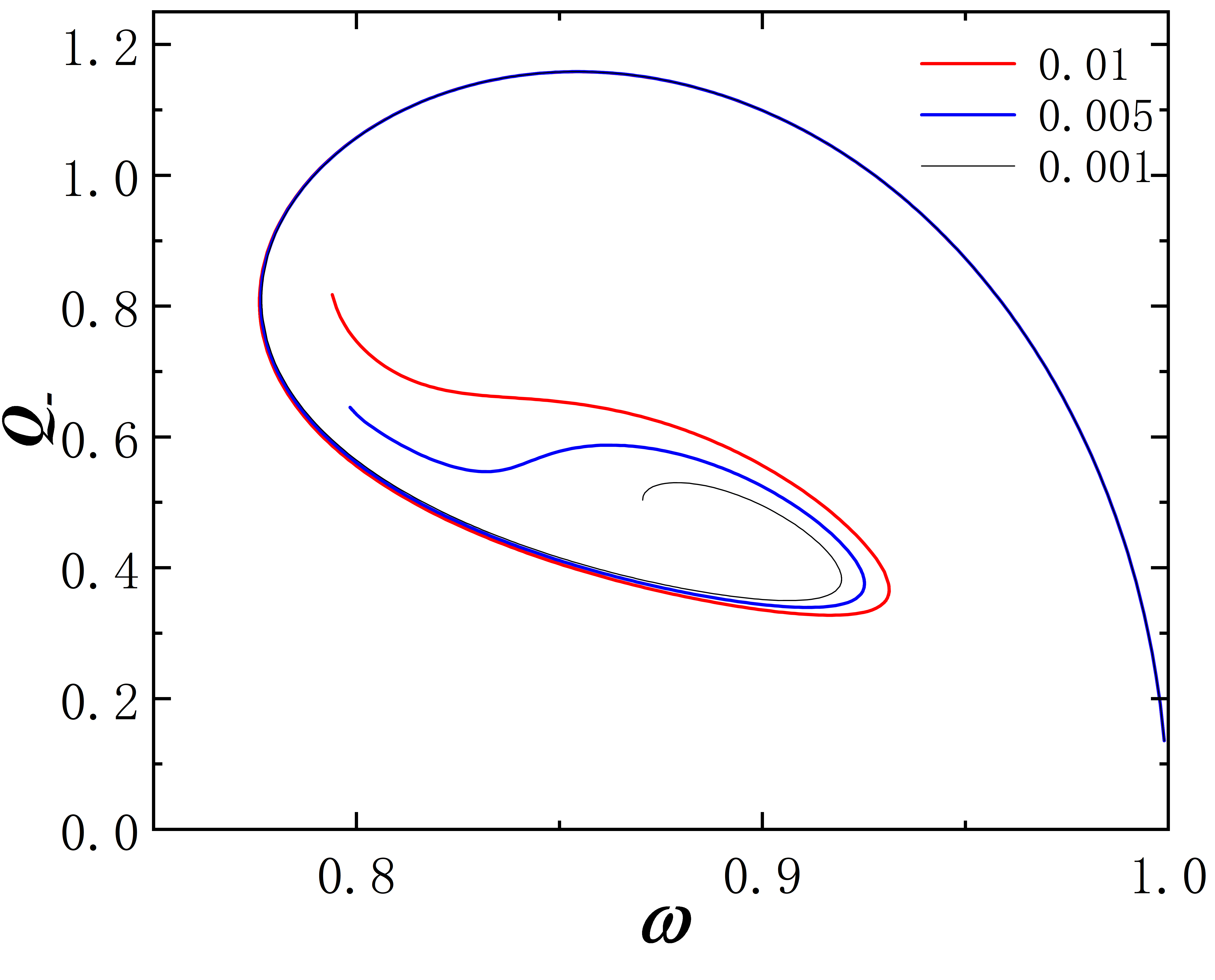}}
\end{center}
\caption{The Nother charge Q as the function of frequency $\omega$ for some values of $r_0$. The dotted line represents $Q_+$, the solid line represents $Q_-$.}
\label{phase12}
\end{figure}

\begin{figure}
  \begin{center}
\subfigure{\includegraphics[width=0.75\textwidth]{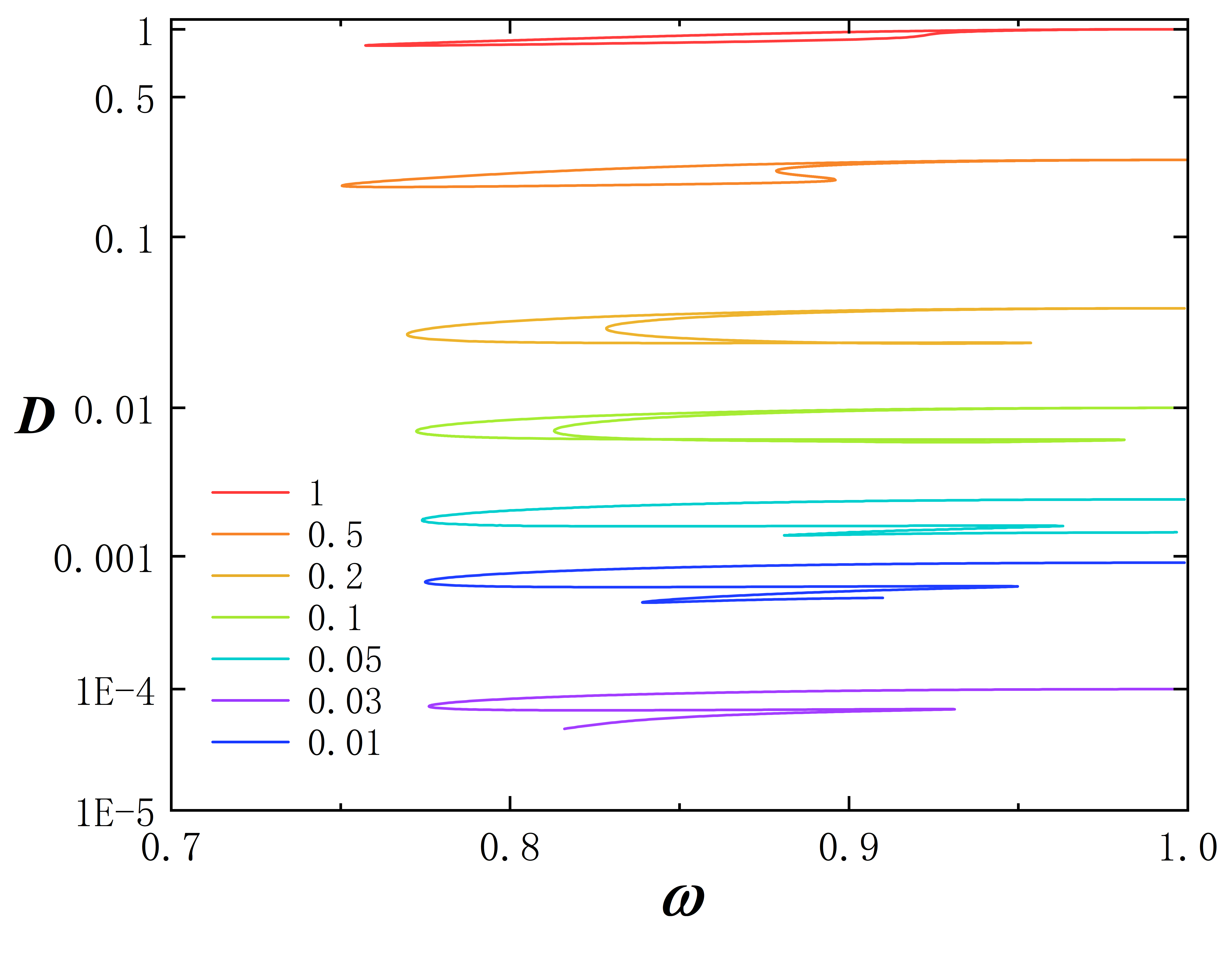}}
\end{center}
\caption{The scalar charge $\cal D$ of the phantom field as a function of the frequency $\omega$ with several values of the throat size $r_0$.}
\label{phase13}
\end{figure}

In particular, for $r_0 \leq 0.03$, the solution can start from the vacuum state but cannot return to it, indicating the emergence of an extremely approximate black hole solution, and we will explain in detail later, Fig. \ref{phase11}. In Fig. \ref{phase12}, We show the Noether charge $Q$ which also agrees well with the ADM mass distribution. The scalar charge of the phantom field $\cal D$ in Fig. \ref{phase13}, the branches of it are simple and its value also decreases as the throat size $r_0$ decreases.

\FloatBarrier

\begin{figure}
  \begin{center}
\subfigure{\includegraphics[width=0.49\textwidth]{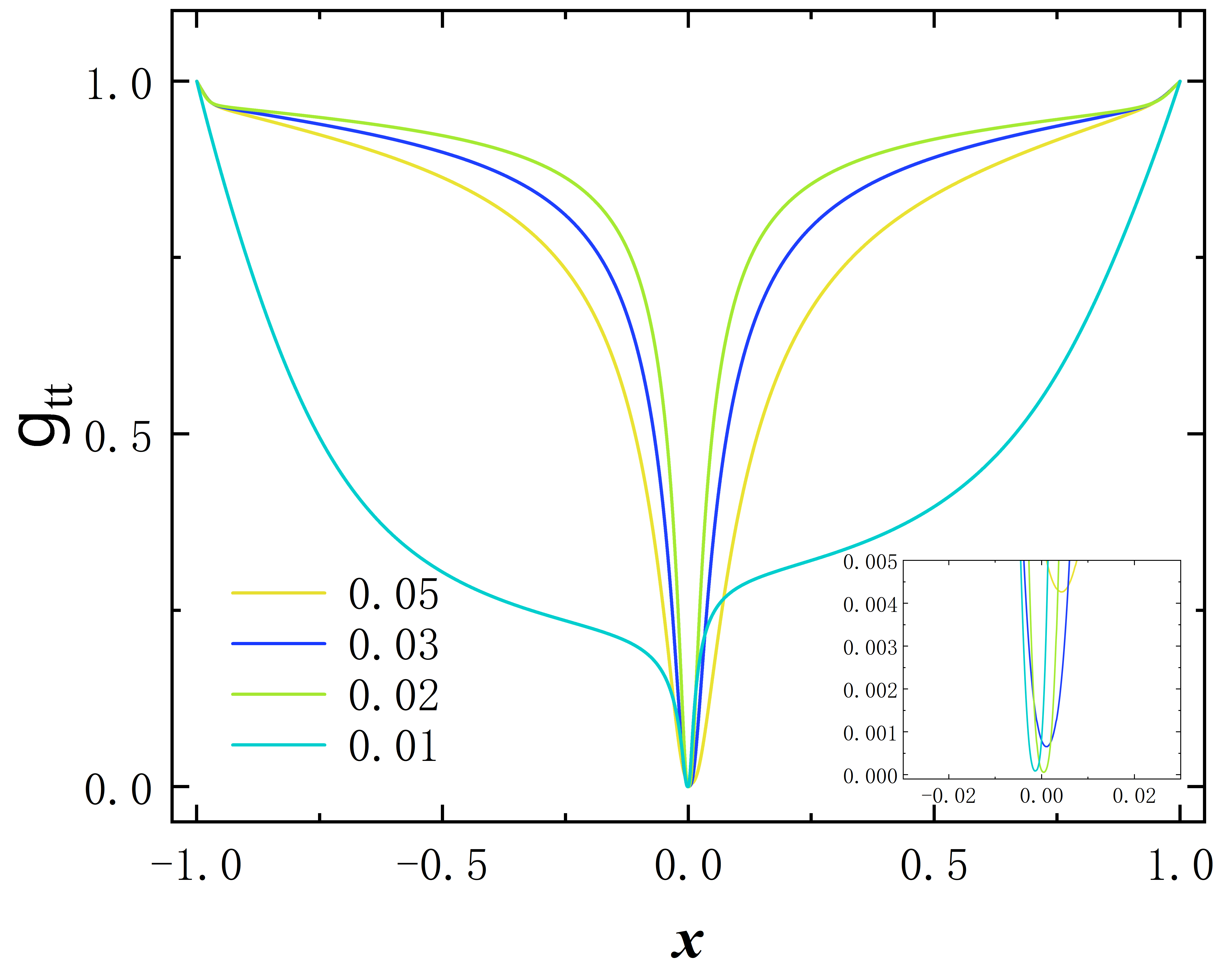}}
\subfigure{\includegraphics[width=0.49\textwidth]{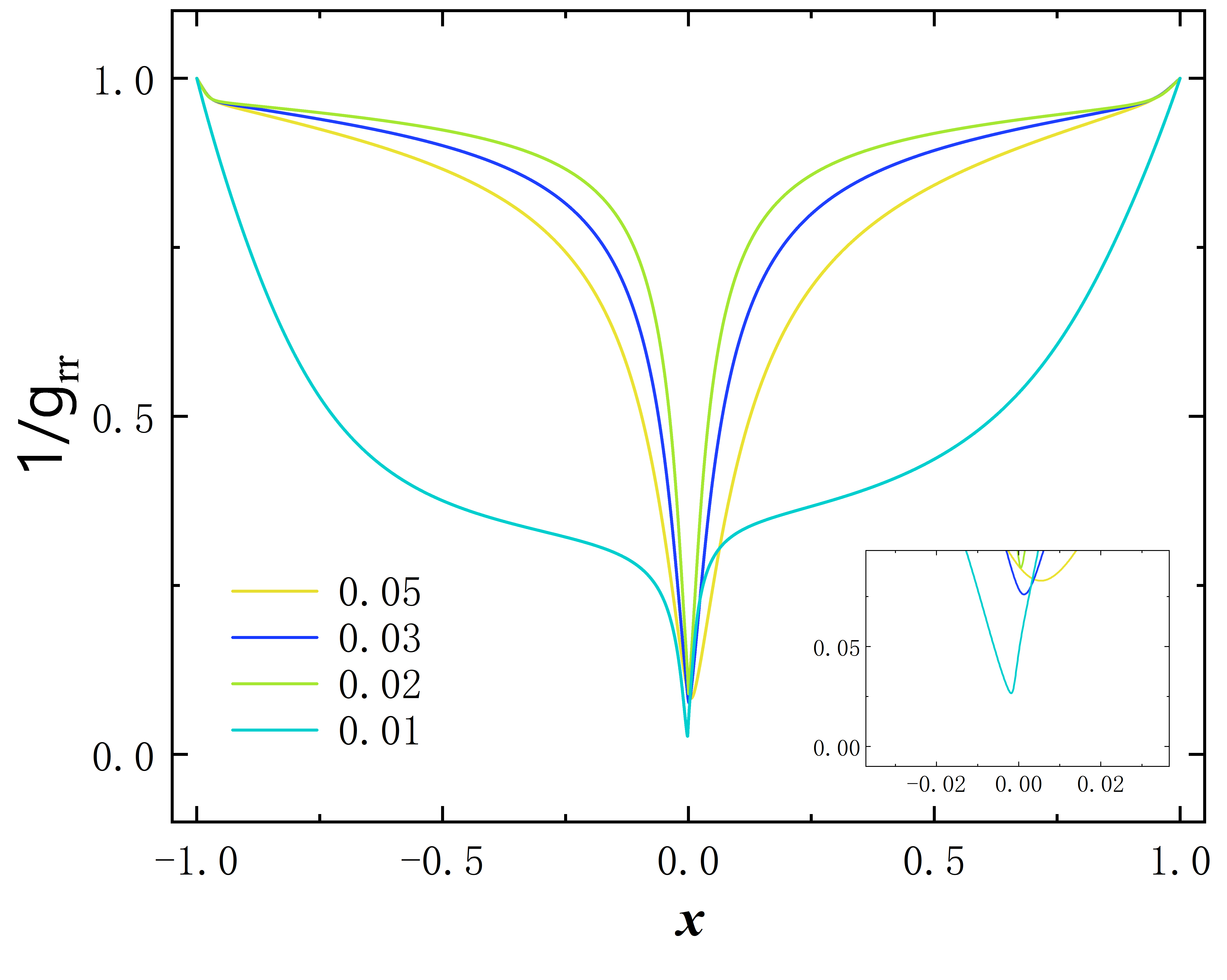}}
  \end{center}
\caption{The left panel is $g_{tt}$ and the right panel is $g_{rr}$, they are functions of $x$. For $r_0$ = $0.05$, $0.03$, $0.02$, the frequency $\omega$ fixed in 0.99. When $r_0$ = $0.01$, the $\omega$ = $0.794$. We use subfigures to show the trend of $g_{tt}$ tending to zero and $g_{rr}$ tending to infinity under four different values of $r_0$.}
\label{phase14}
\end{figure}

\begin{figure}
  \begin{center}
\subfigure{\includegraphics[width=0.49\textwidth]{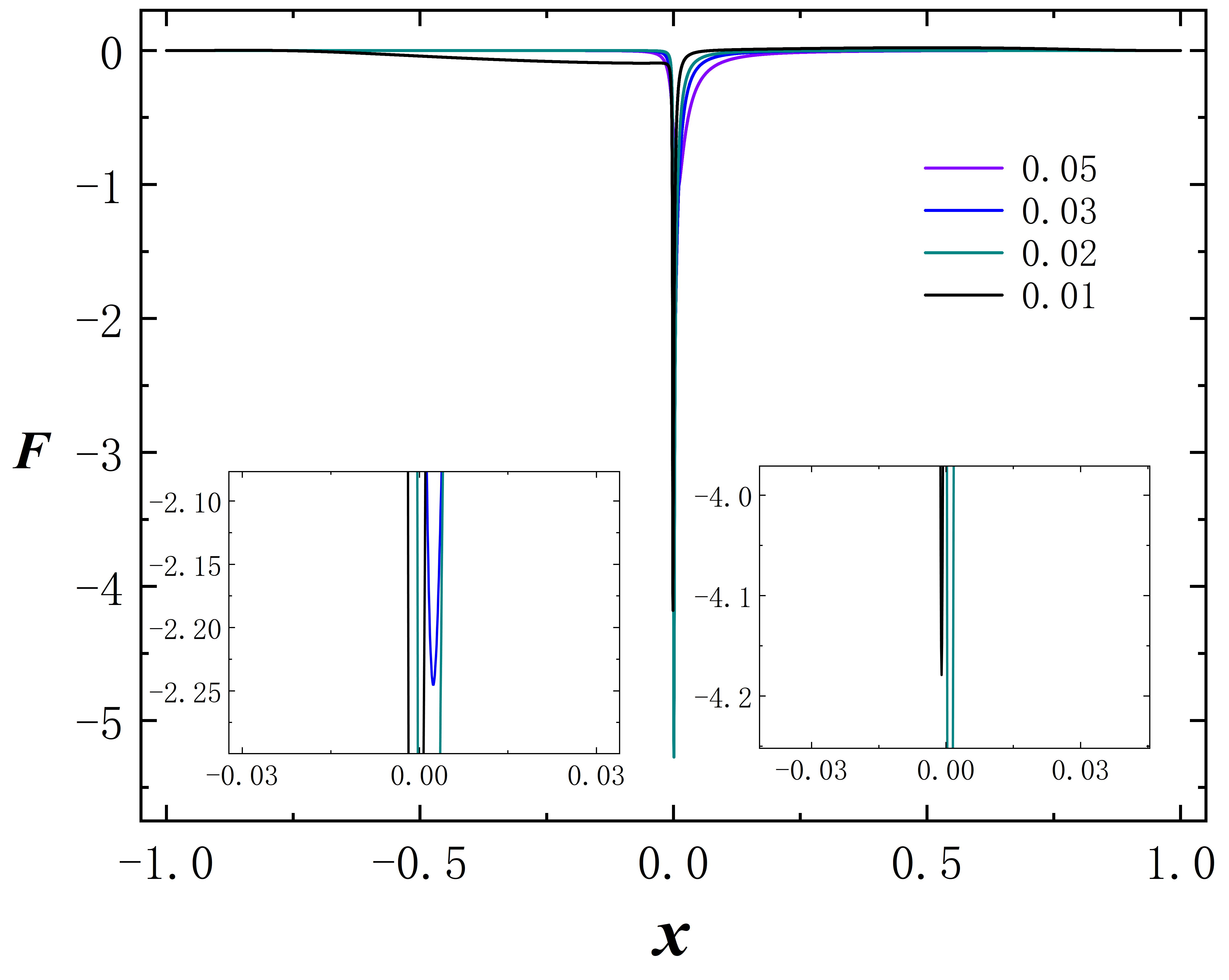}}
\subfigure{\includegraphics[width=0.49\textwidth]{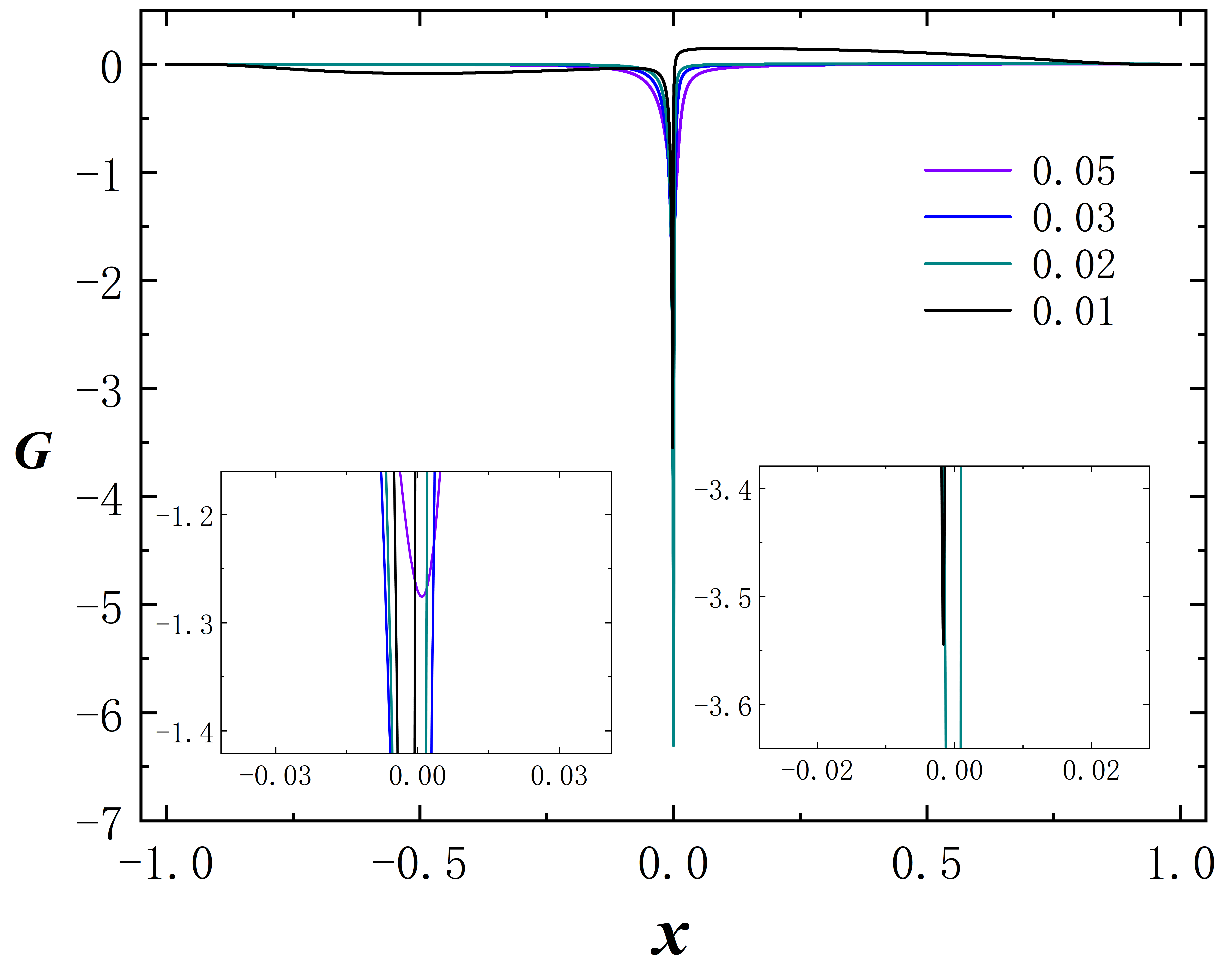}}
  \end{center}
\caption{The left panel is $F$ and the right panel is $G$, they are functions of $x$. For $r_0$ = $0.05$, $0.03$, $0.02$, the frequency $\omega$ fixed in 0.99. When $r_0$ = $0.01$, the $\omega$ = $0.794$. We use subfigures to show the convergence of the fields under four different values of $r_0$.}
\label{phase15}
\end{figure}

We further examined the extremely approximate black hole solution. For small throat size $r_0$, the solution cannot return to the vacuum state. We studied the metrics $g_{tt}$ and $g_{rr}$ and the values of fields $F$ and $G$ at specific frequencies $\omega$, finding similar phenomena and laws as in the previous case, Fig. \ref{phase14}, Fig. \ref{phase15}, for comparison, the parameter selection will be exactly the same as before. From numerical calculations, we infer that as $r_0$ decreases to a certain range, $g_{tt}$ approaches zero, and $g_{rr}$ approaches infinity, indicating the emergence of an extremely approximate black hole solution. As $r_0$ decreases further, the black hole transfers from one side of the wormhole to the other. And the material field distribution concentrates near the horizon and almost diverges.


\begin{figure}
  \begin{center}
\subfigure{\includegraphics[width=0.425\textwidth]{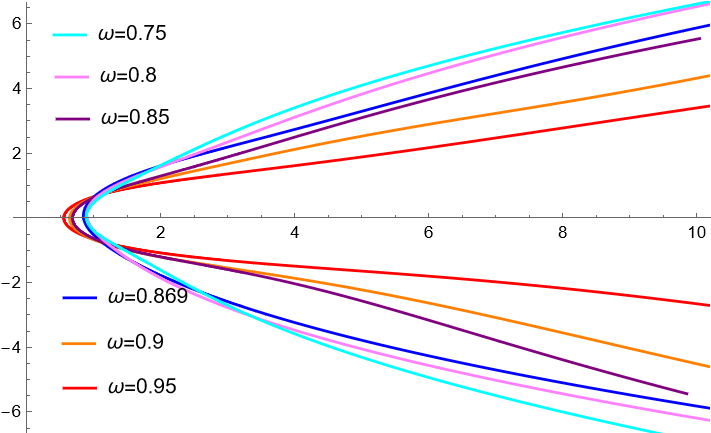}}
\subfigure{\includegraphics[width=0.425\textwidth]{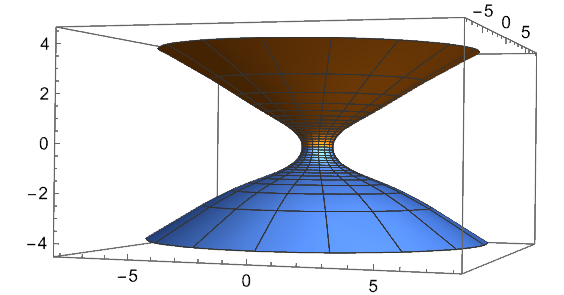}}
\subfigure{\includegraphics[width=0.425\textwidth]{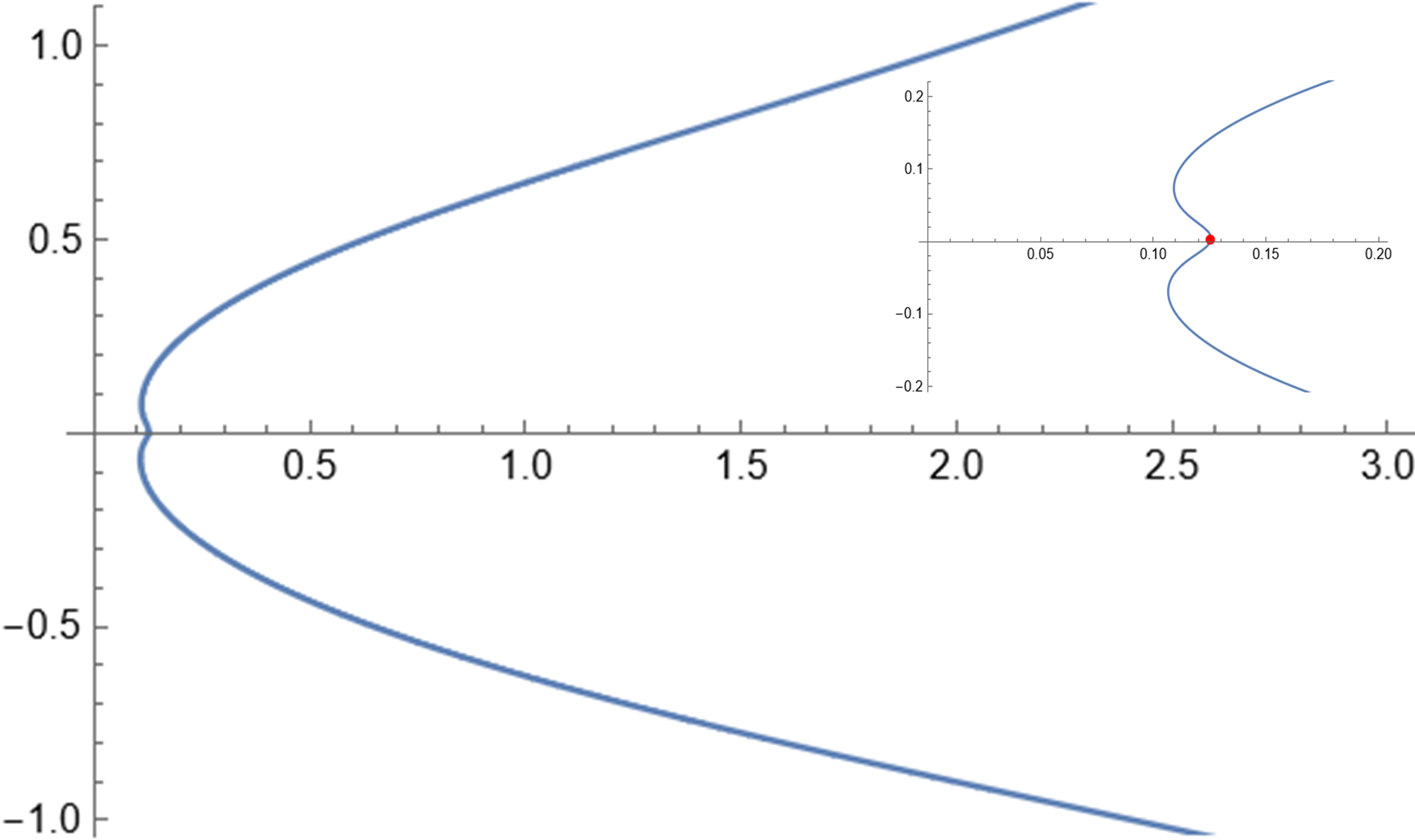}}
\subfigure{\includegraphics[width=0.425\textwidth]{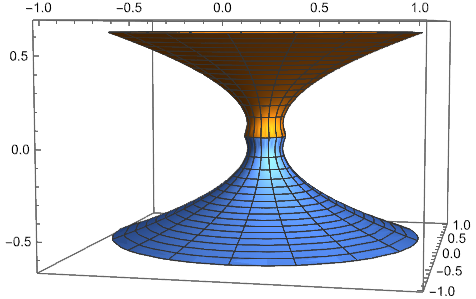}}
\end{center}
\caption{First row: \textit{left}: Two-dimensional view of the isometric embedding of the equatorial plane. \textit{Right}: Isometric embeddings of the equatorial plane of this solution with throat parameter $r_0$ = 0.5 and  $\omega$ = 0.85. Second row: \textit{left}: Two-dimensional view of the isometric embedding of the equatorial plane with $r_0$ = $0.03$, $\omega$ = $0.9$. \textit{Right}: The corresponding 3D embedded image.}
\label{phase16}
\end{figure}

Using the previous conclusions, we can quickly obtain the geometric embedding diagram of the two-dimensional space in this case and then embed the hypersurface into the three-dimensional Euclidean space. In Fig. \ref{phase16}, we show a two-dimensional view of the isometric embedding of the equatorial plane with throat parameter $r_0$ = 0.5 and several values of the frequency $\omega$ at the left panel. The right panel shows the 3D plot corresponding to the frequency $\omega$ = 0.85. The geometry of the wormhole is asymmetric and has only one throat and no equatorial plane. When the throat size $r_0$ is small enough the wormhole has an equatorial plane with a throat on both sides of the equatorial plane, see the picture of the last row ($r_0$ = 0.03, $\omega$ = 0.9).

\textbf{Case3: $F$ has two nodes and $G$ has one node.}

\begin{figure}
\begin{center}
\subfigure{\includegraphics[width=0.49\textwidth]{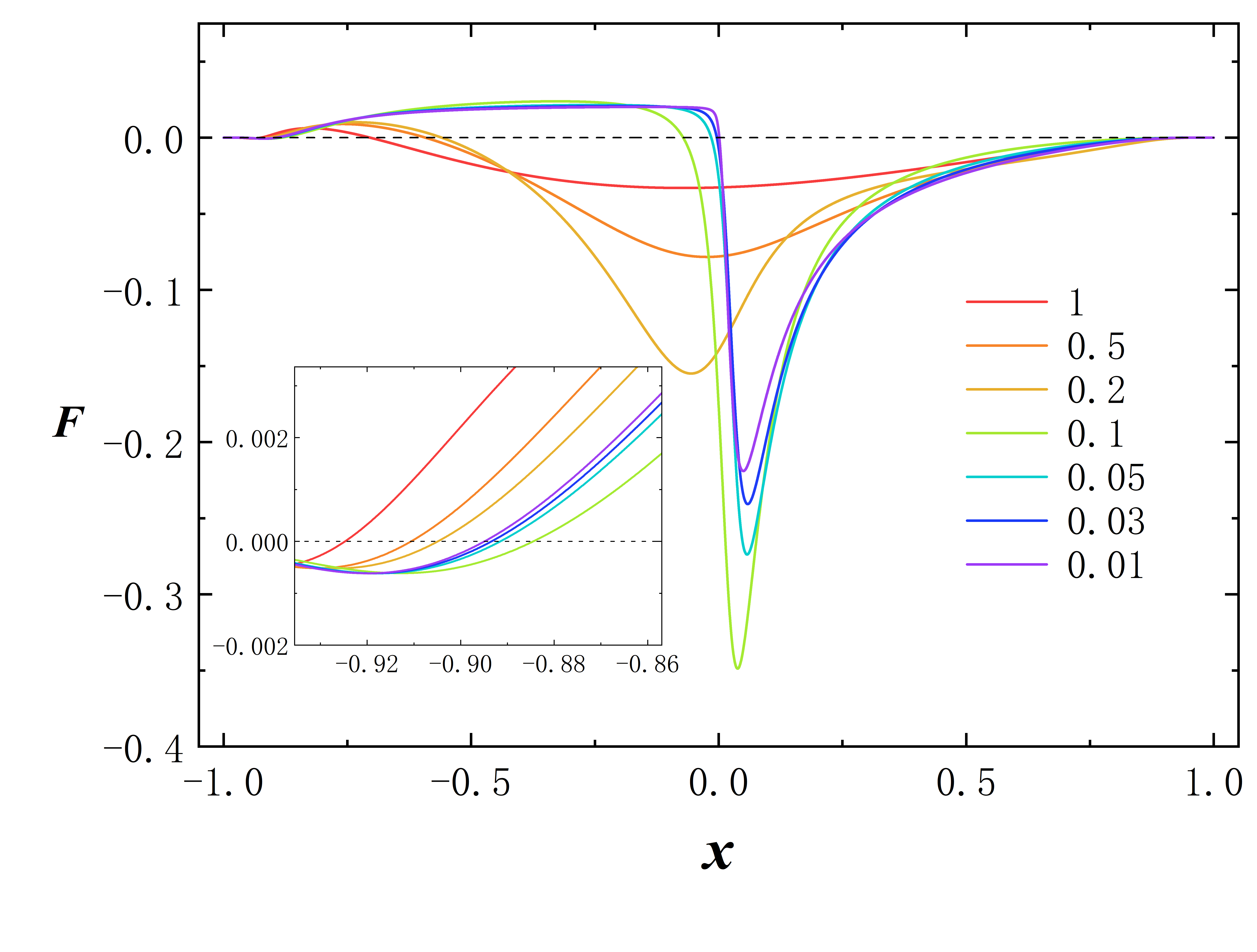}}
\subfigure{\includegraphics[width=0.49\textwidth]{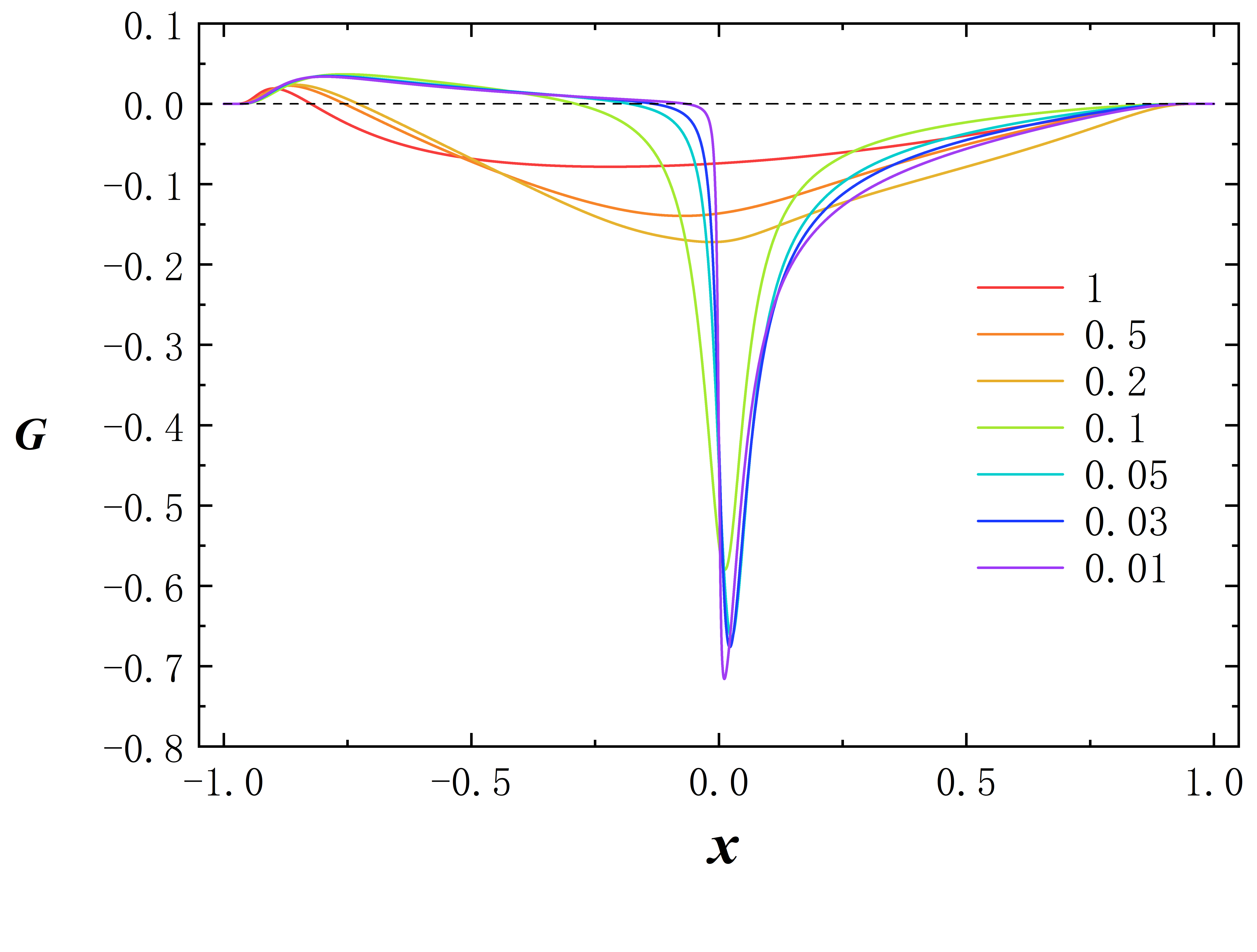}}
\end{center}
\caption{The radial distribution of the dirac fields $F$ and $G$ with several values of $r_0$ for frequency $\omega$ = 0.87.}
\label{phase17}
\end{figure}

Although the number of Dirac field nodes is the same as in the second case, the specific field configuration differs, as indicated by the difference in node positions of the $F$ and $G$ fields. The choice of parameters remains unchanged, Fig. \ref{phase17}

\begin{figure}
  \begin{center}
\subfigure{\includegraphics[width=0.32\textwidth]{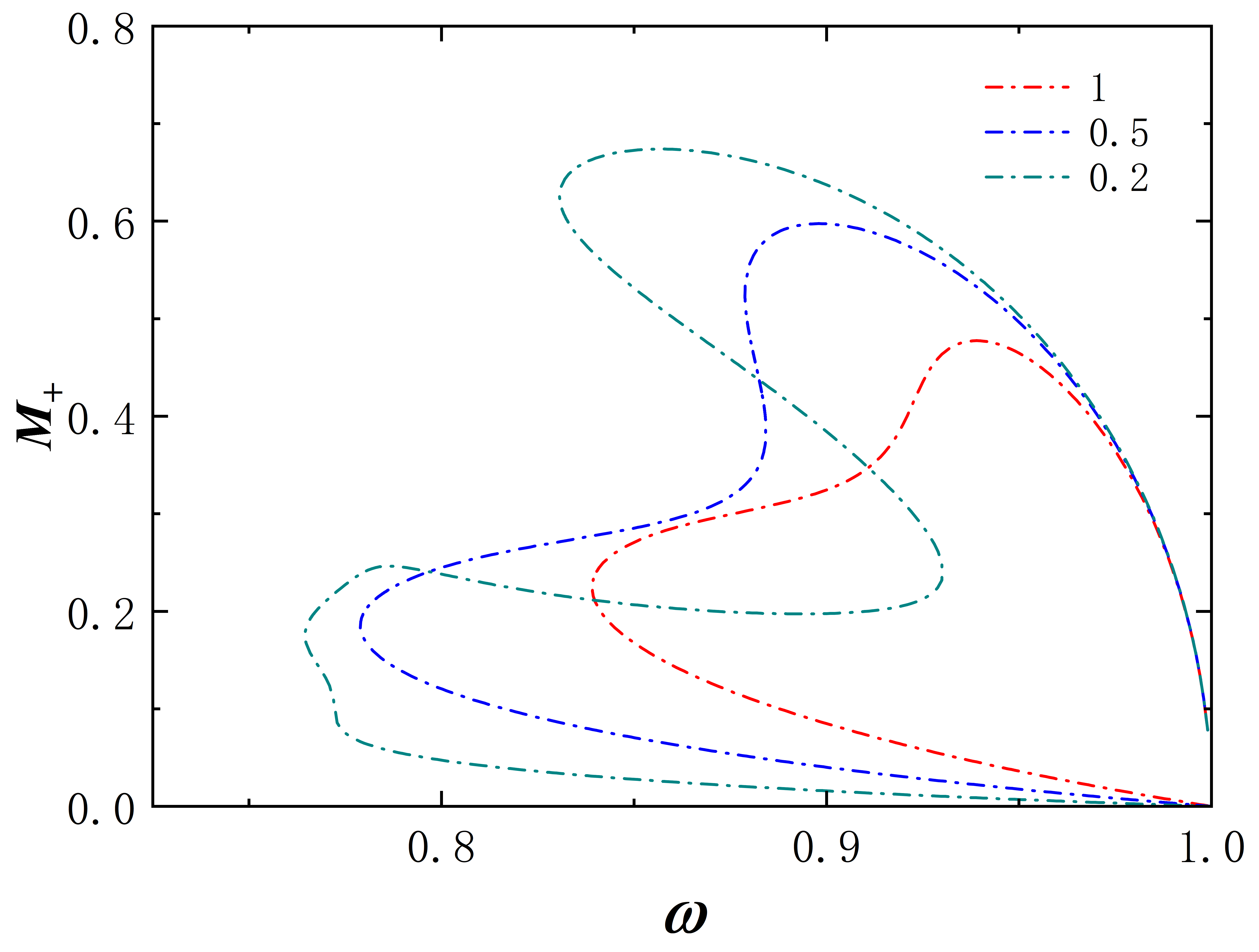}}
\subfigure{\includegraphics[width=0.32\textwidth]{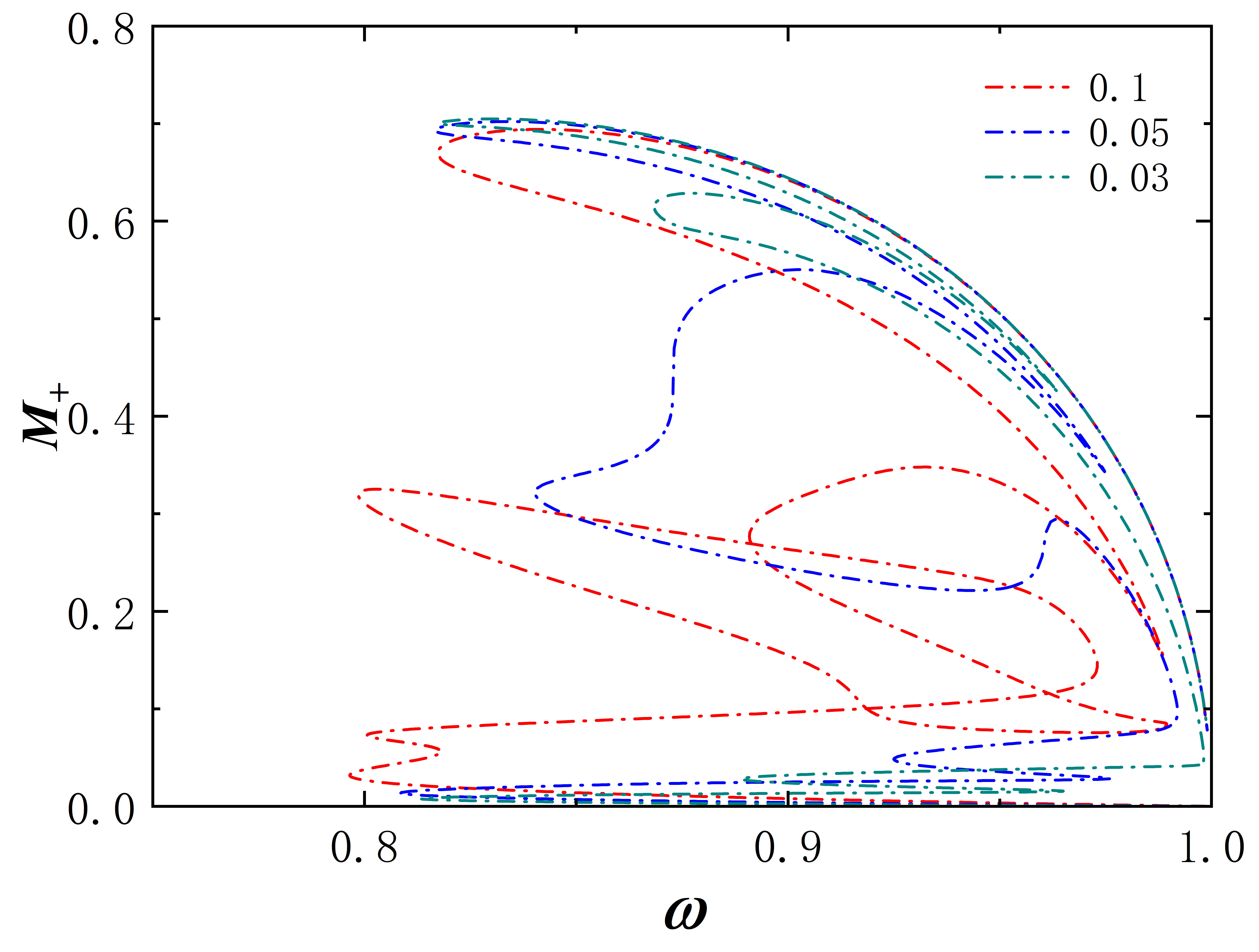}}
\subfigure{\includegraphics[width=0.32\textwidth]{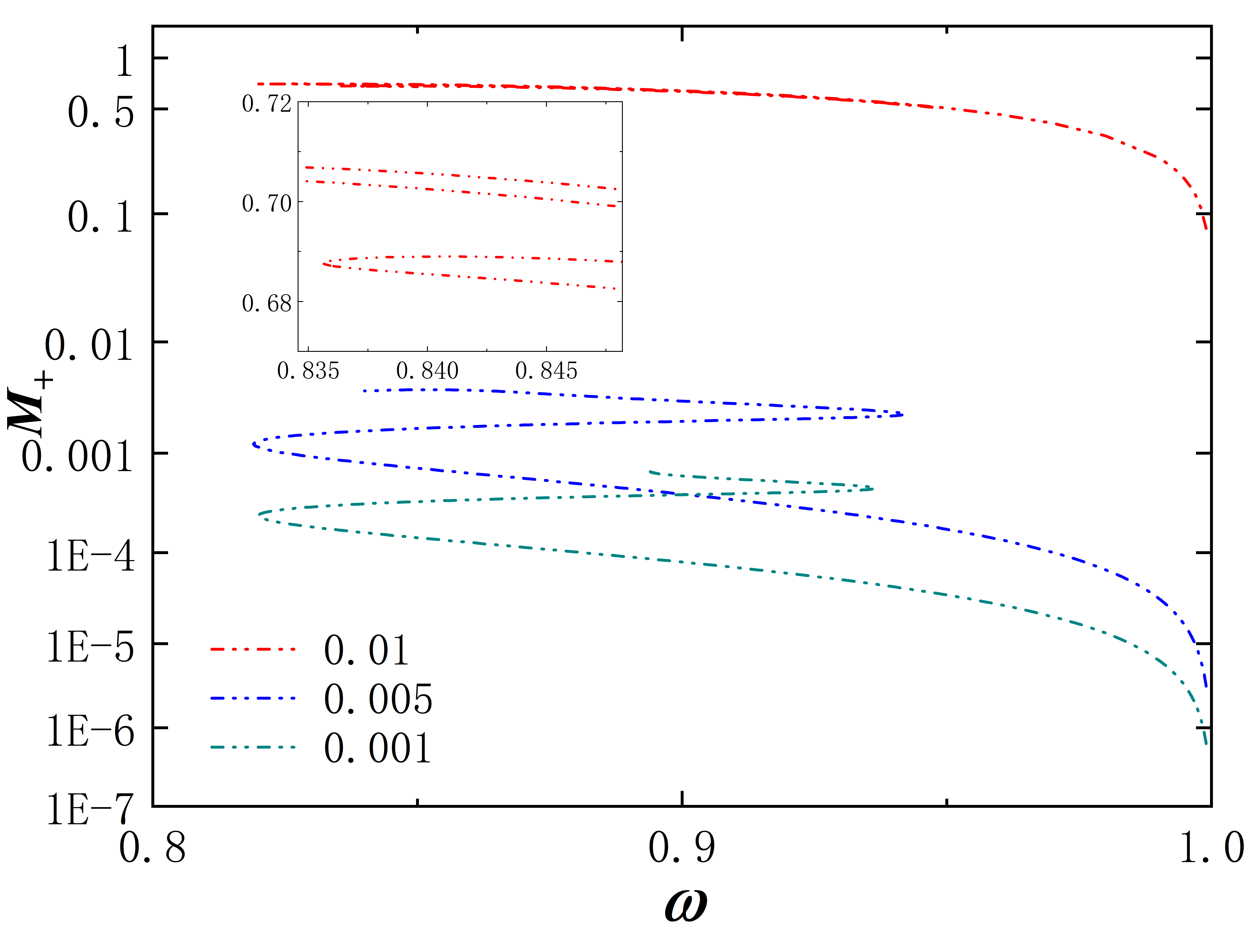}}
\subfigure{\includegraphics[width=0.32\textwidth]{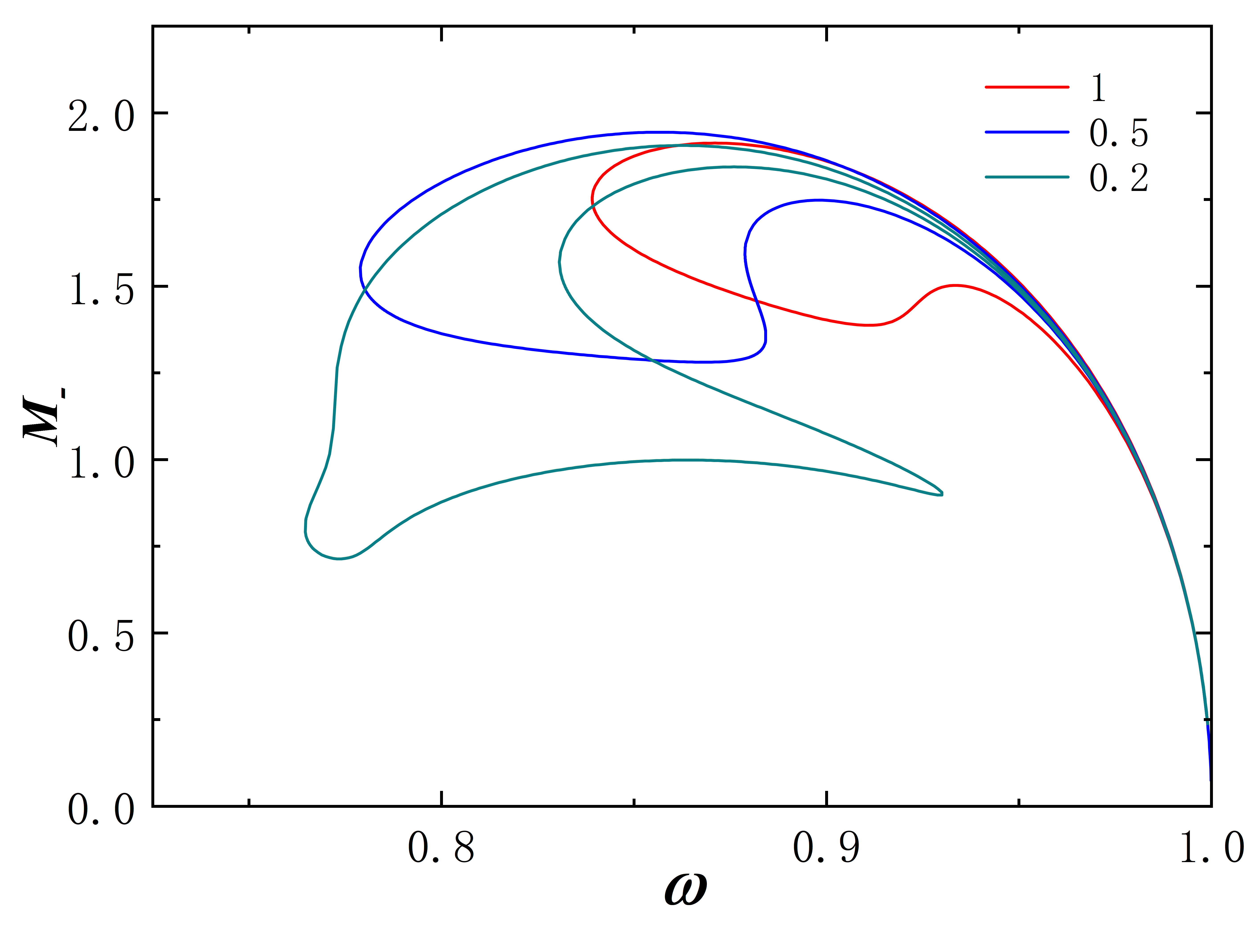}}
\subfigure{\includegraphics[width=0.32\textwidth]{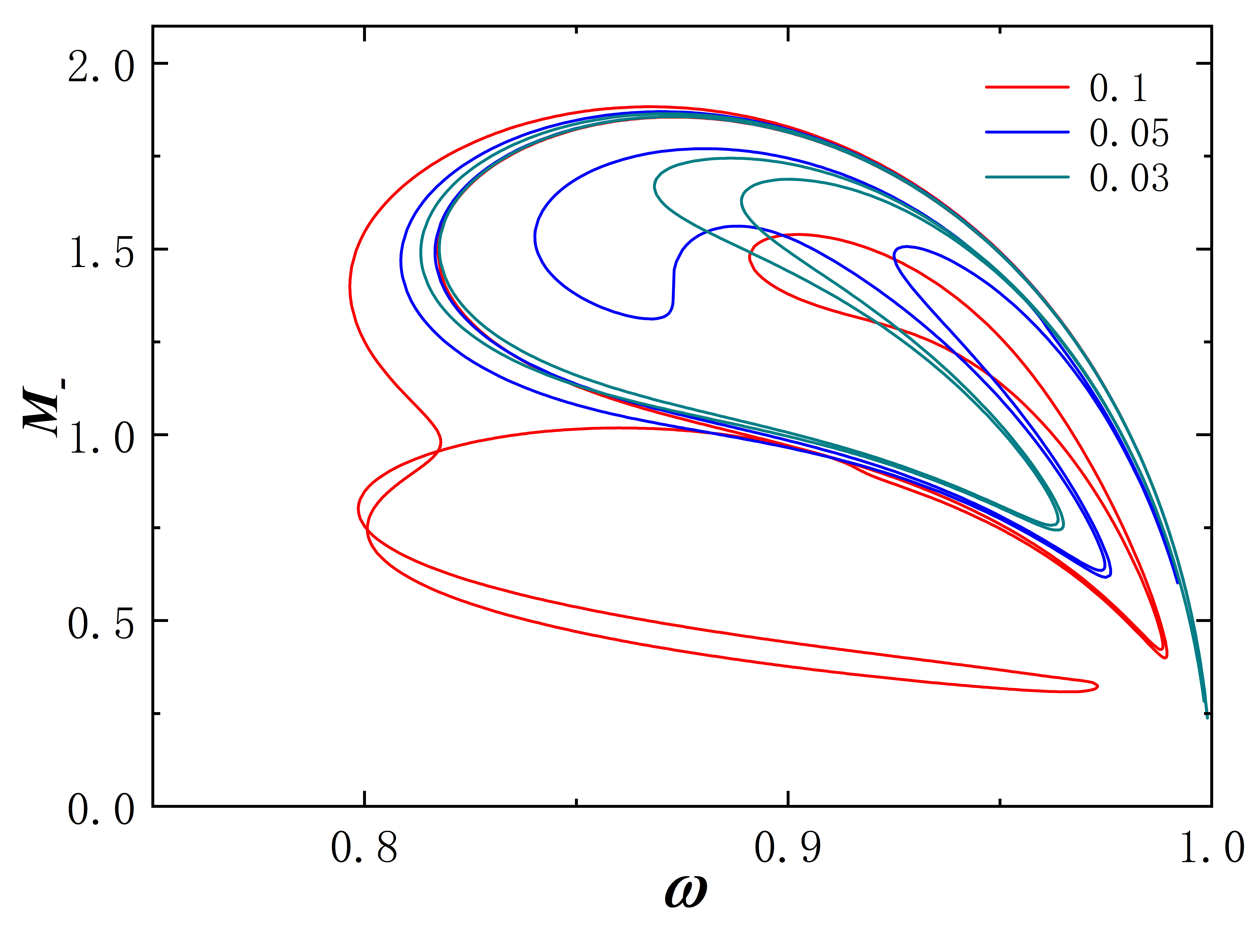}}
\subfigure{\includegraphics[width=0.32\textwidth]{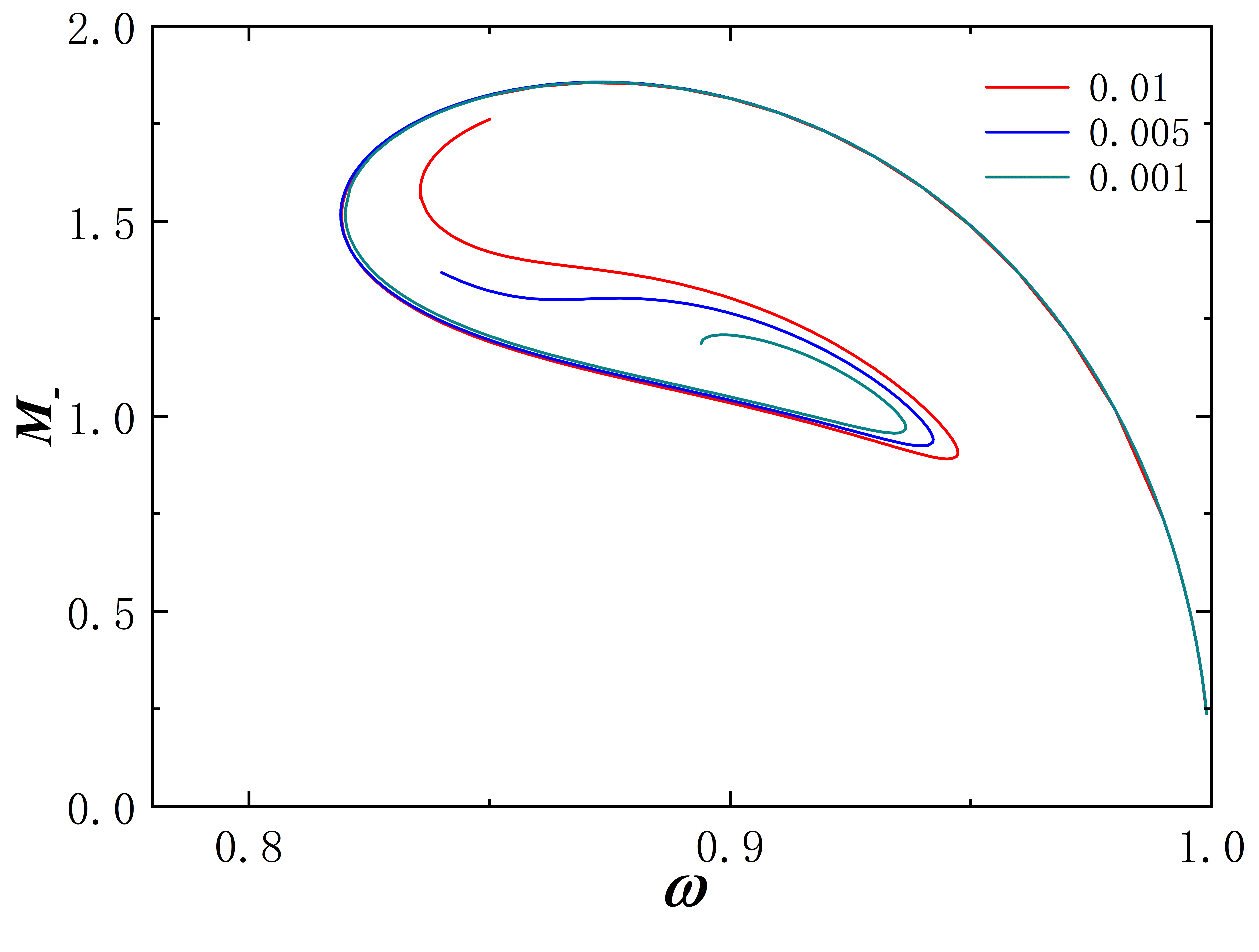}}
  \end{center}
\caption{The mass $M$ as the function of frequency $\omega$ for some values of $r_0$. The dotted line represents $M_+$, the solid line represents $M_-$.}
\label{phase18}
\end{figure}

\begin{figure}
  \begin{center}
\subfigure{\includegraphics[width=0.75\textwidth]{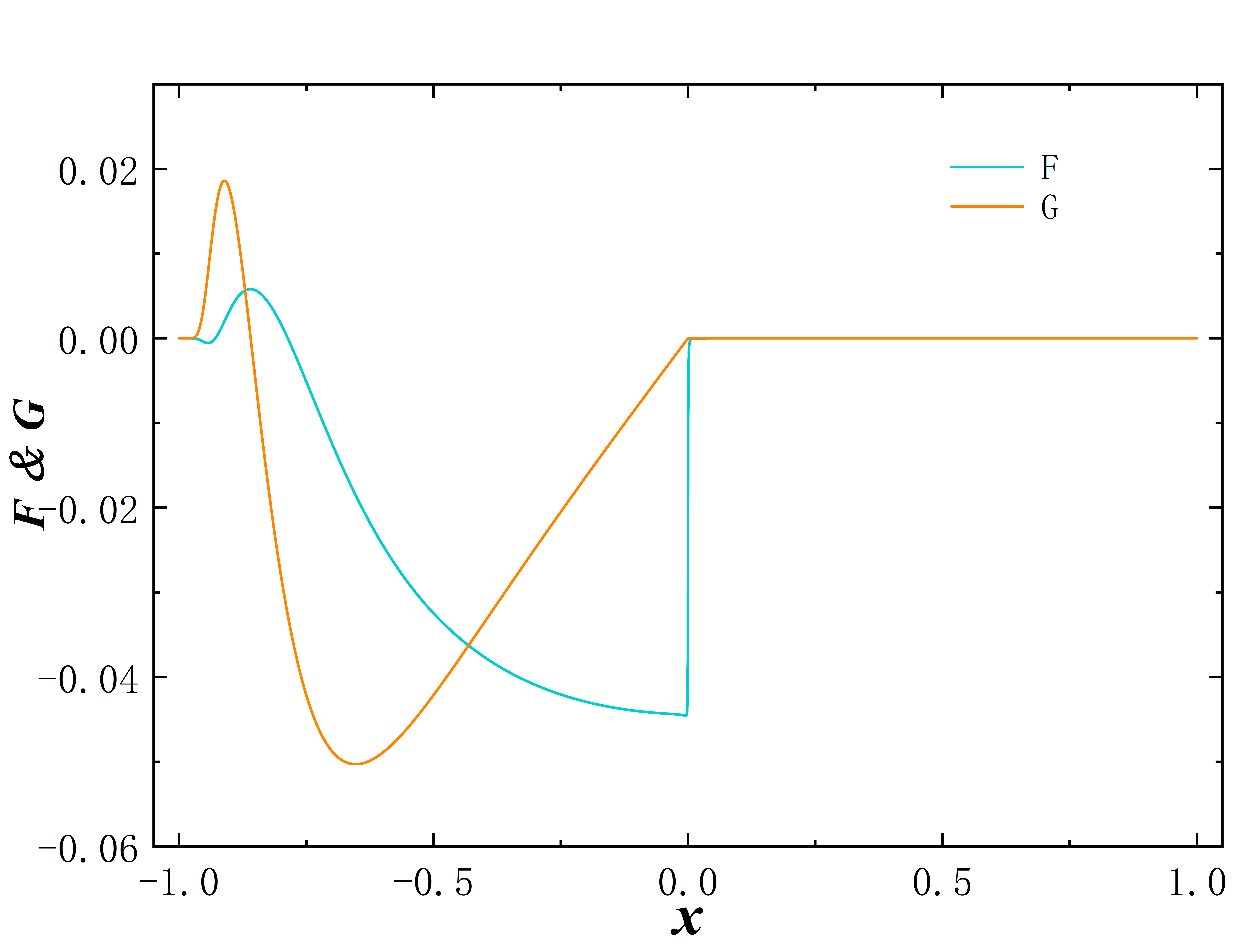}}
\end{center}
\caption{The $F$ and $G$ field with $r_0$ = 0.001 and $\omega$ = 0.85.}
\label{phase19}
\end{figure}

The contribution of ADM mass is similar with the case 1, Fig .\ref{phase18}. We display it in three groups based on parameter selection, as there are new situations not previously encountered. For very small $r_0$ values of 0.005 and 0.001, the value of $M_+$ becomes very small while the value of $M_-$ becomes large. The reason for this is that the field determines the mass distribution. At this time, the $F$ and $G$ fields exhibit almost all their distribution concentrated at $x<0$, with almost no field on the other side. We show it in Fig. \ref{phase19}, with $r_0$ = 0.001 and $\omega$ = 0.85.

\begin{figure}
  \begin{center}
\subfigure{\includegraphics[width=0.32\textwidth]{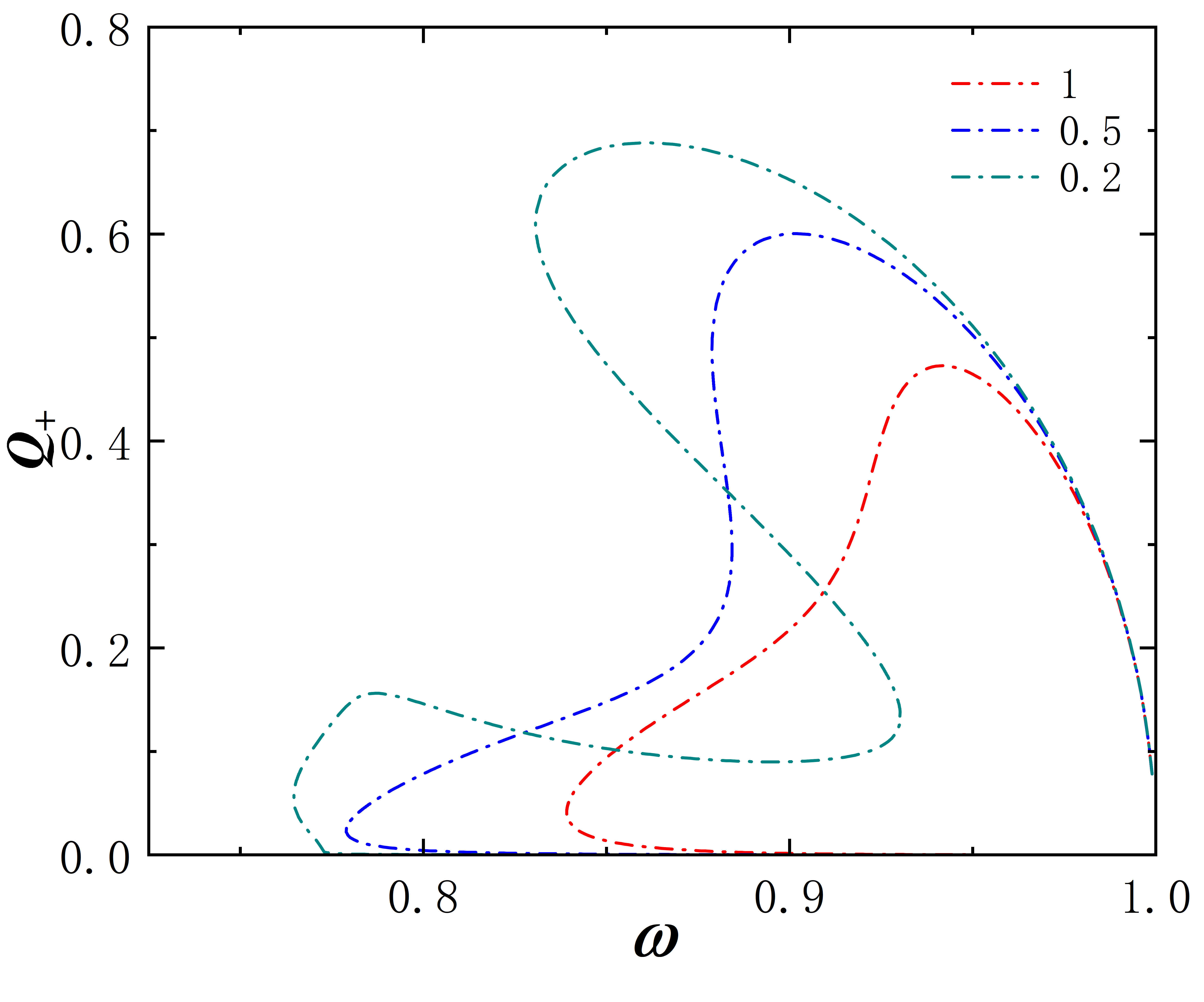}}
\subfigure{\includegraphics[width=0.32\textwidth]{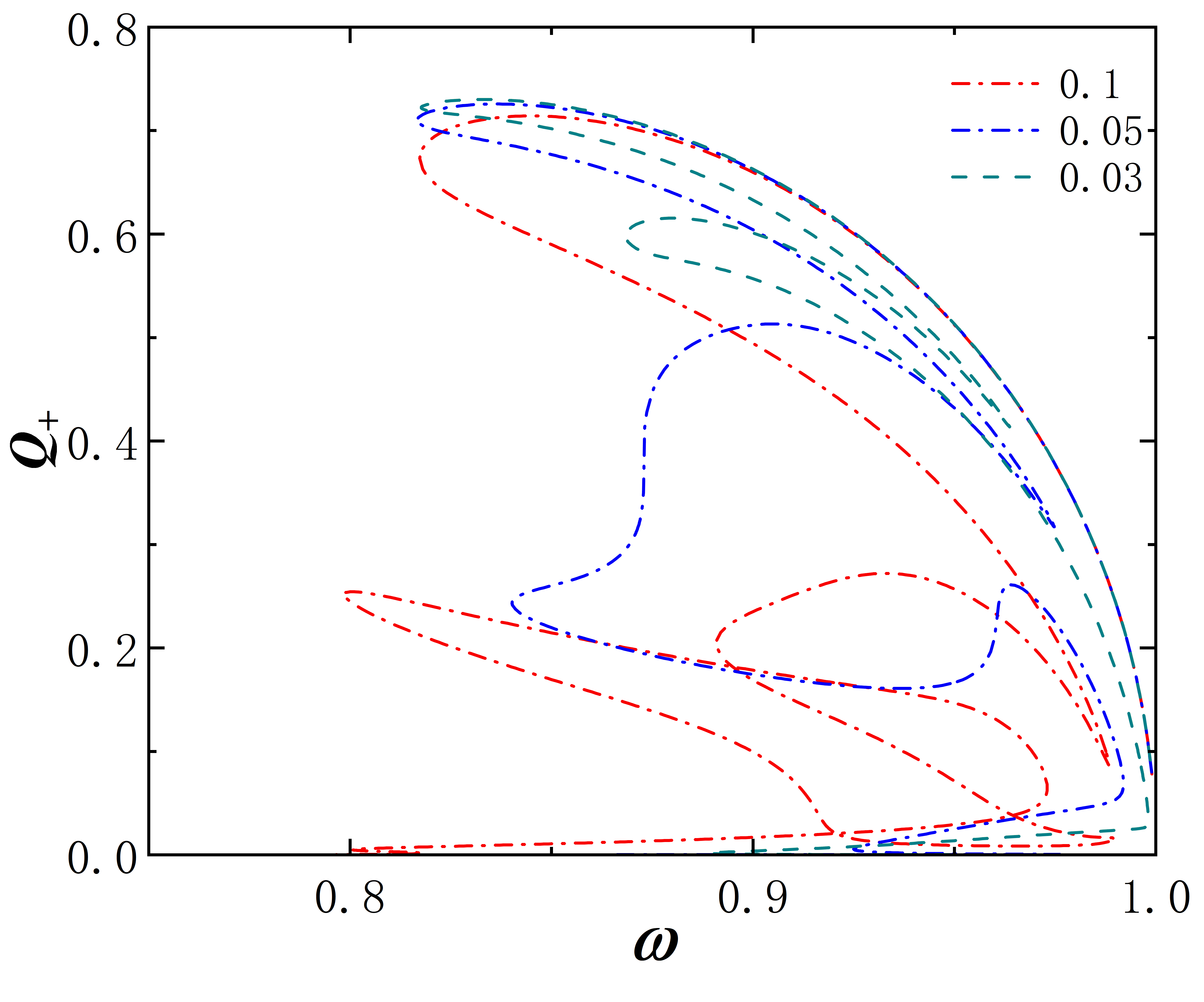}}
\subfigure{\includegraphics[width=0.32\textwidth]{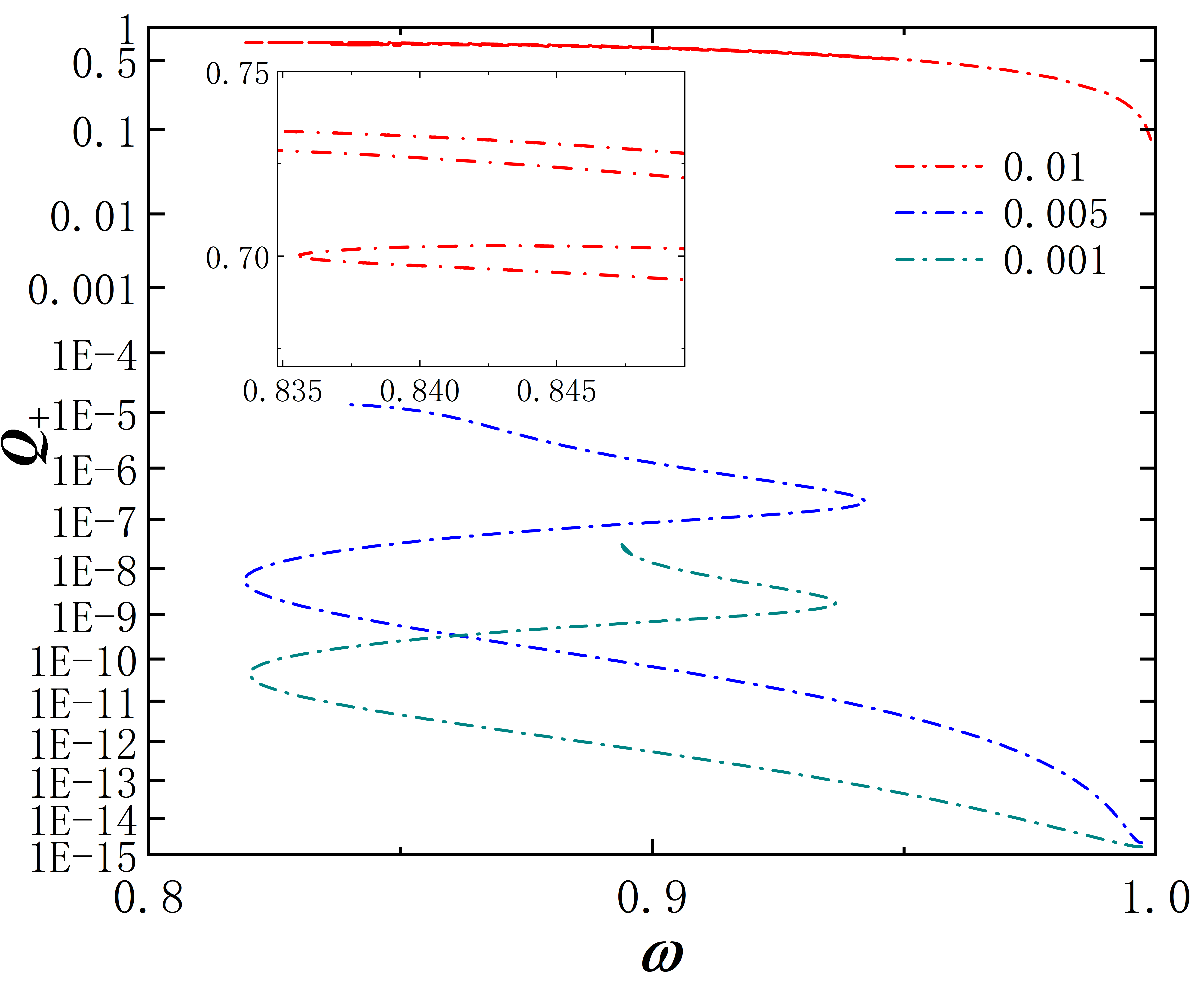}}
\subfigure{\includegraphics[width=0.32\textwidth]{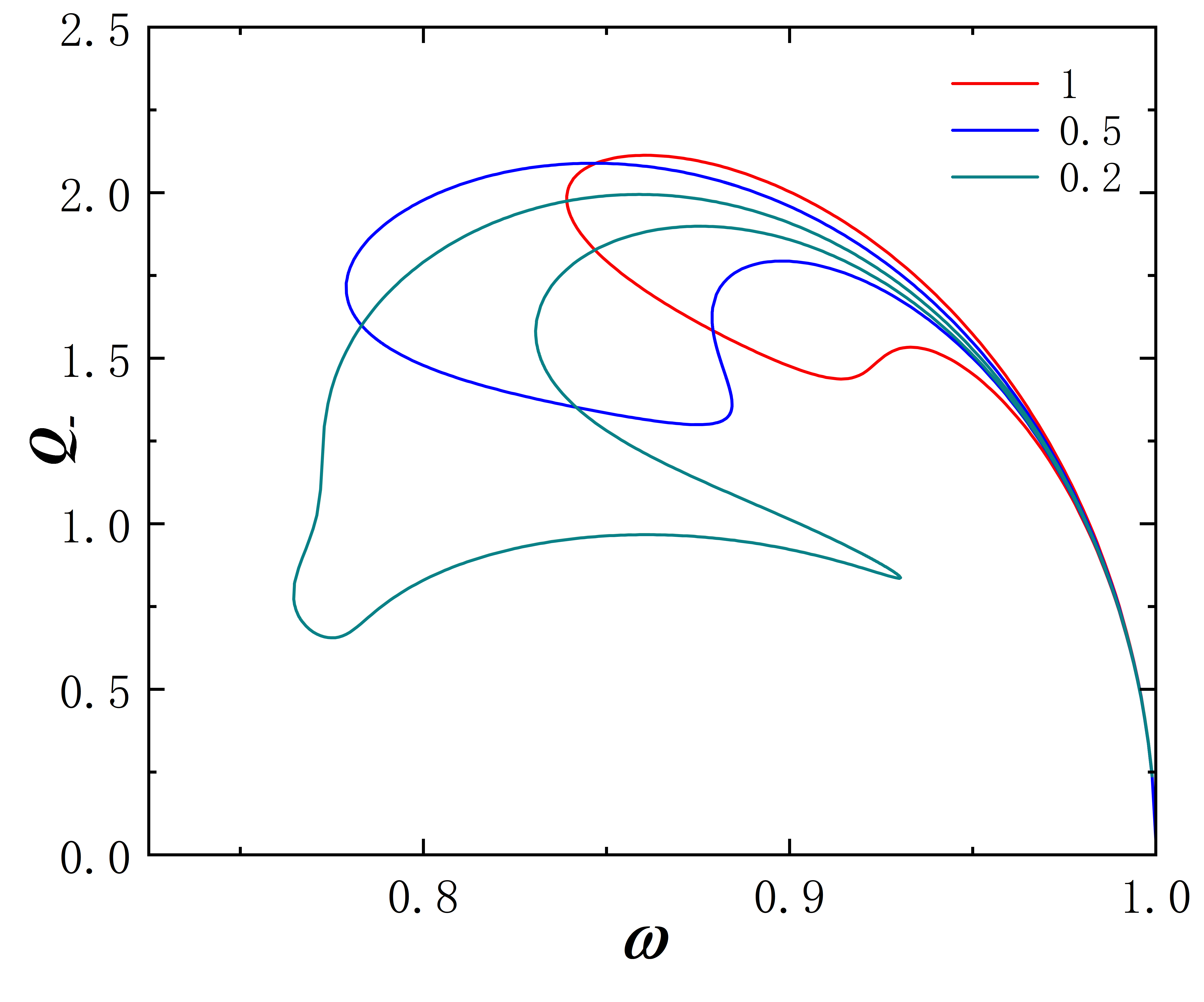}}
\subfigure{\includegraphics[width=0.32\textwidth]{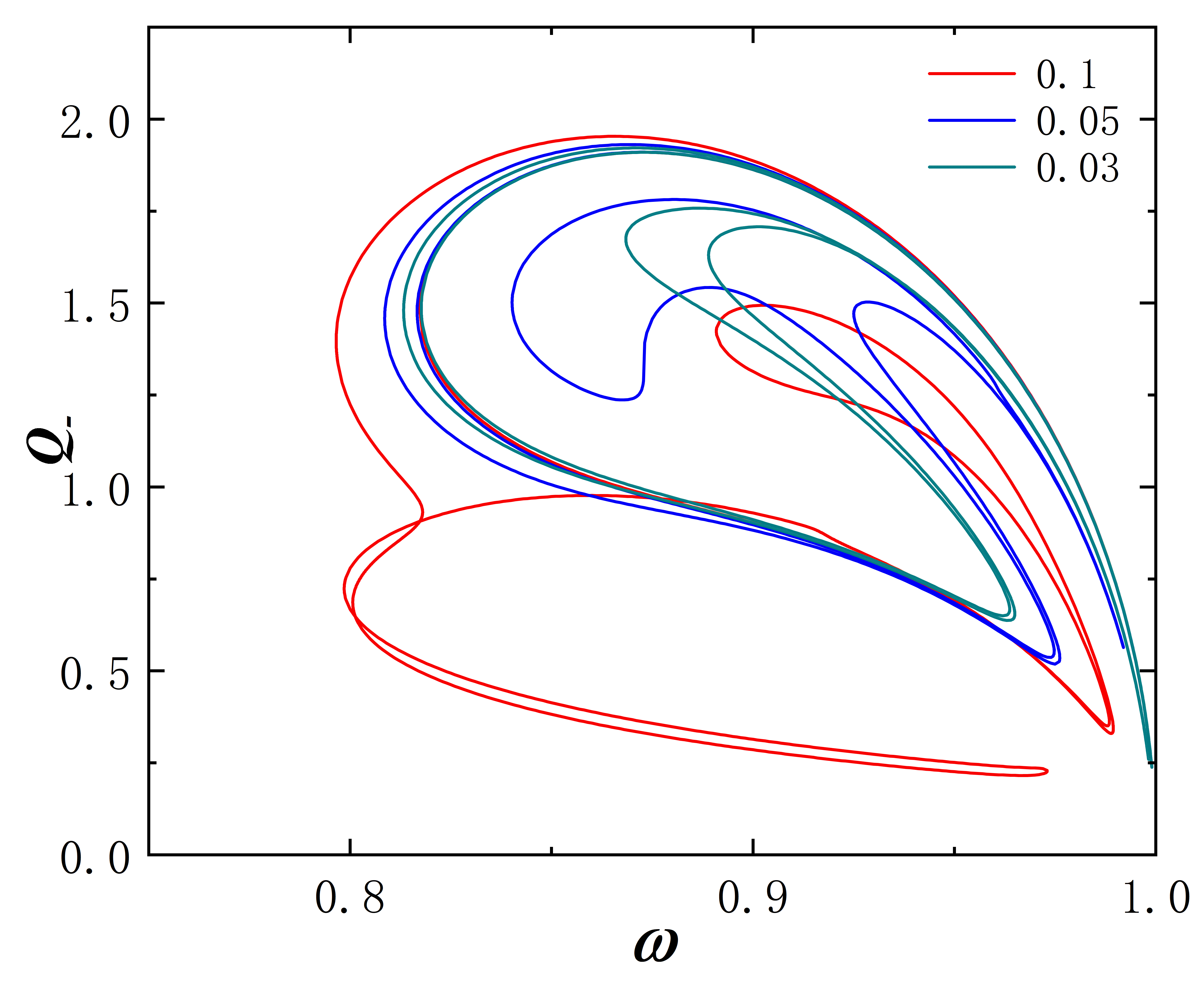}}
\subfigure{\includegraphics[width=0.32\textwidth]{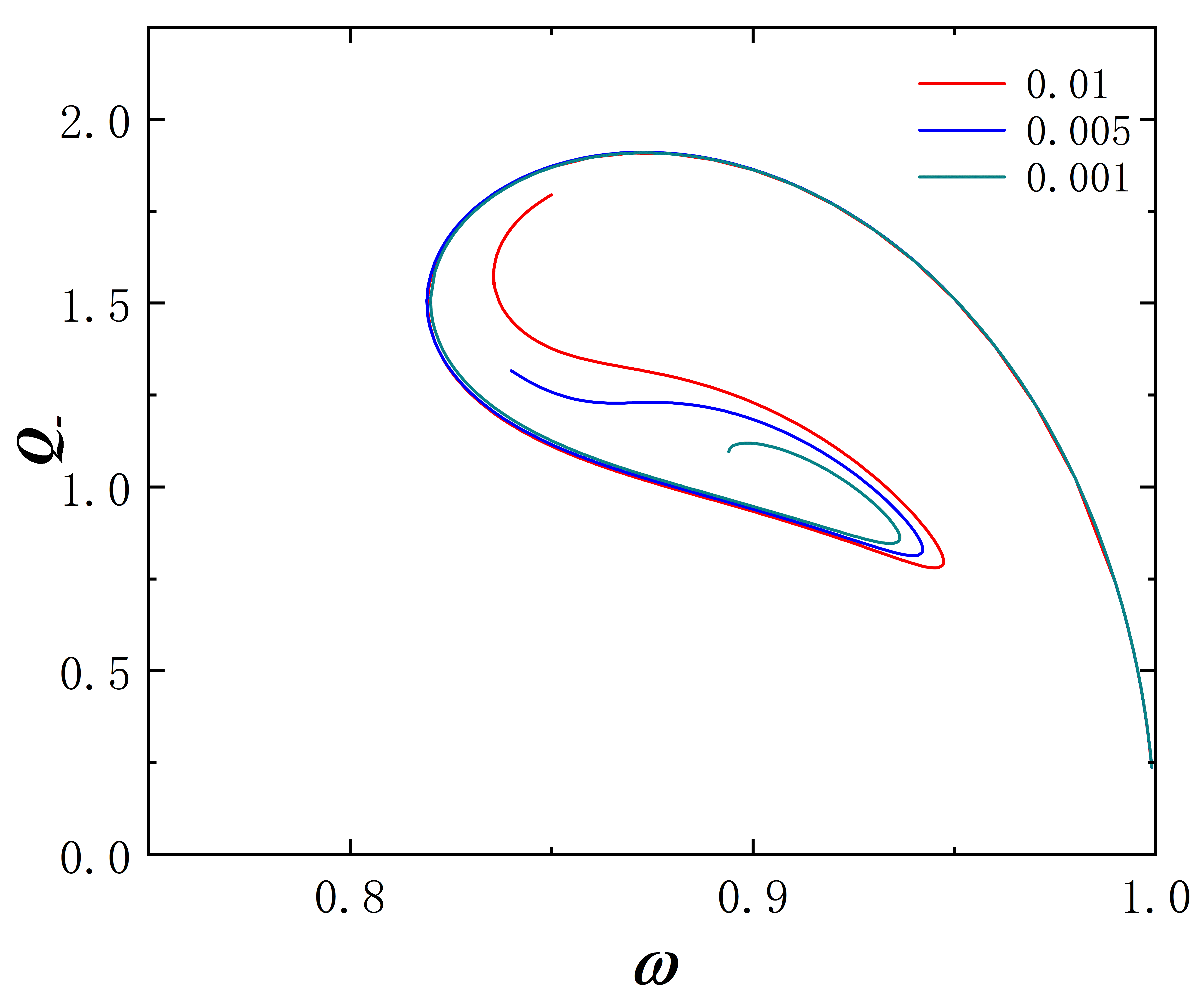}}
  \end{center}
\caption{The Noether charge $Q$ as the function of frequency $\omega$ for some values of $r_0$. The dotted line represents $Q_+$, the solid line represents $Q-$.}
\label{phase20}
\end{figure}

In Fig. \ref{phase20}, we show the Noether charge $Q$. When $r_0$ is relatively small, the field has negligible distribution on the $Q_+$ side, as reflected in the Noether charge $Q$.

\begin{figure}
  \begin{center}
\subfigure{\includegraphics[width=0.45\textwidth]{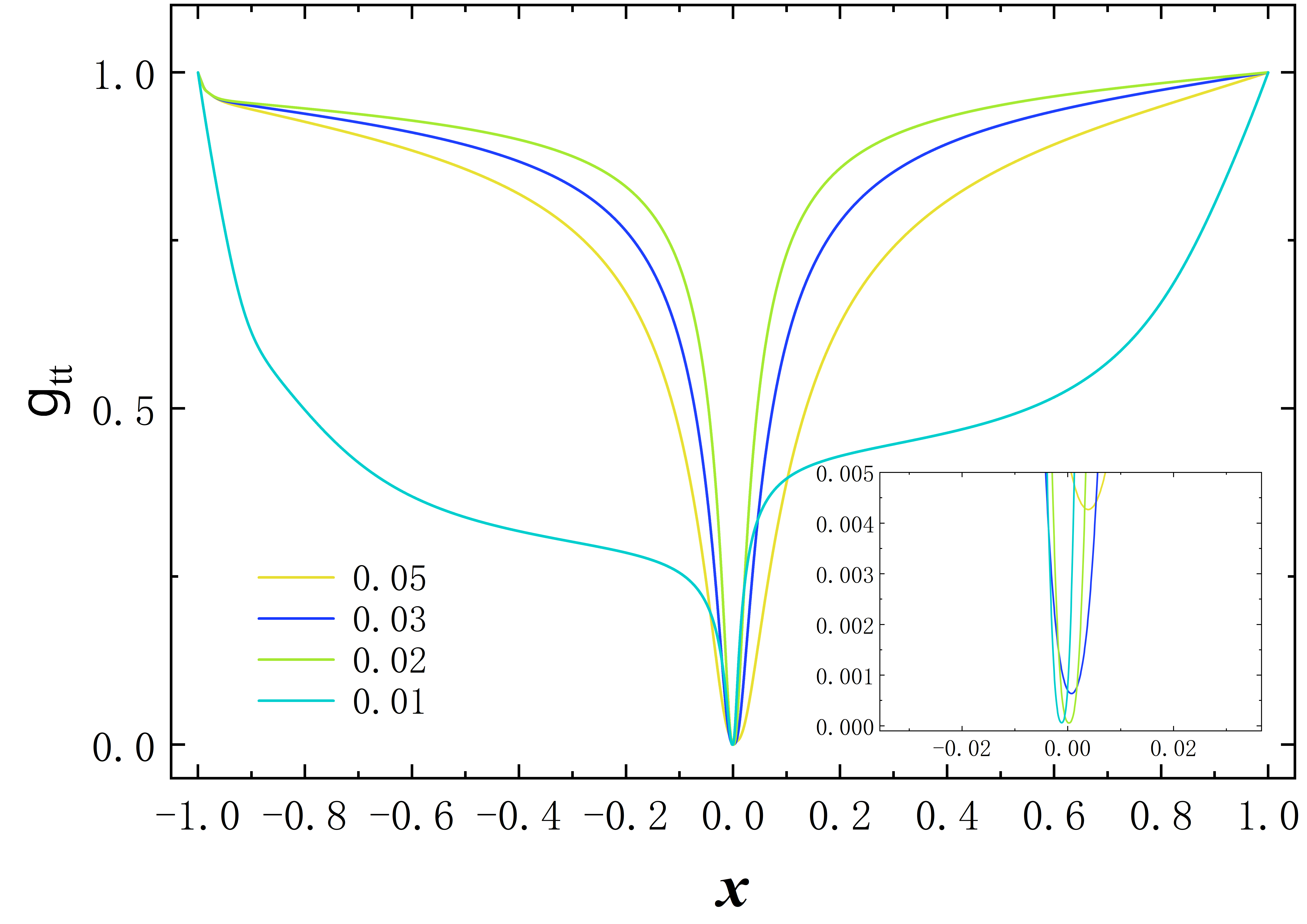}}
\subfigure{\includegraphics[width=0.45\textwidth]{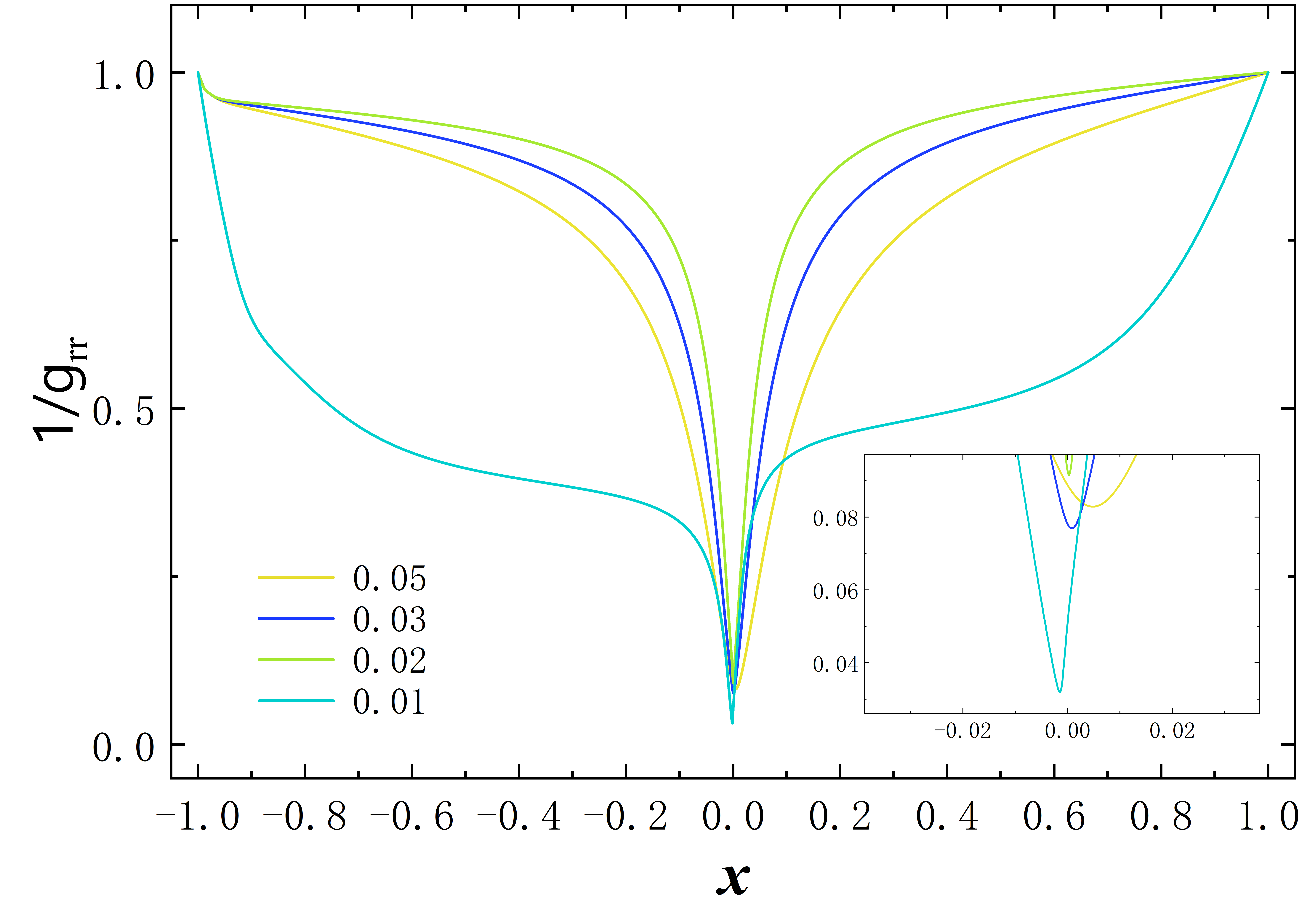}}
  \end{center}
\caption{The left panel is $g_{tt}$ and the right panel is $g_{rr}$, they are functions of $x$. For $r_0$ = $0.05$, $0.03$, $0.02$, the frequency $\omega$ fixed in 0.99. When $r_0$ = $0.01$, the $\omega$ = $0.84$. The subfigures show the trend of $g_{tt}$ tending to zero and $g_{rr}$ tending to infinity under four different values of $r_0$.}
\label{phase21}
\end{figure}

In all cases, when the throat size $r_0$ is small, the last branch of the solution cannot return to the vacuum. We previously interpreted this phenomenon as the emergence of an extremely approximate black hole solution. To further verify it, we show the values of metrics $g_{tt}$ and $g_{rr}$ at a specific $\omega$ in Fig \ref{phase21}. The situation in the field is similar and to save space, we will not show it anymore.

When $r_0$ is 0.05, 0.03, and 0.02 the parameter $\omega$ selection remains the same as before. However, when $r_0$ is 0.01, we set the frequency $\omega$ to 0.84, as this allows the last branch of the solution to be calculated over a wider range of frequencies without losing accuracy. As the throat size $r_0$ decreases, the minimum value of $g_{tt}$ approaches zero, and the maximum value of $g_{rr}$ tends to infinity. Additionally, the phenomenon of the black hole transfer from one side of the wormhole to the other persists. All numerical results confirm our hypothesis.

\begin{figure}[!htbp]
  \begin{center}
\subfigure{\includegraphics[width=0.425\textwidth]{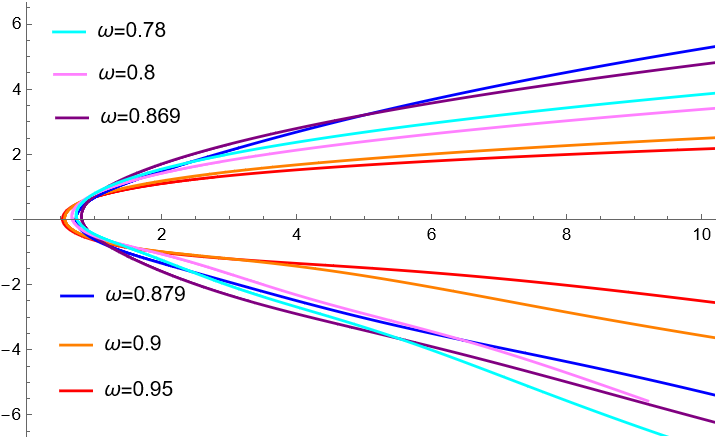}}
\subfigure{\includegraphics[width=0.425\textwidth]{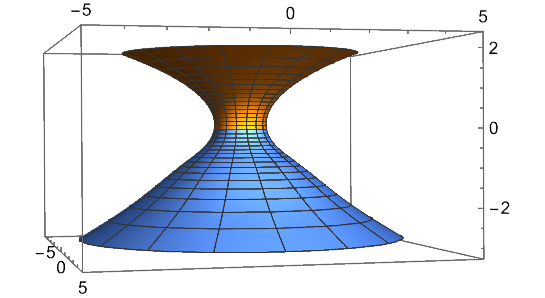}}
\subfigure{\includegraphics[width=0.425\textwidth]{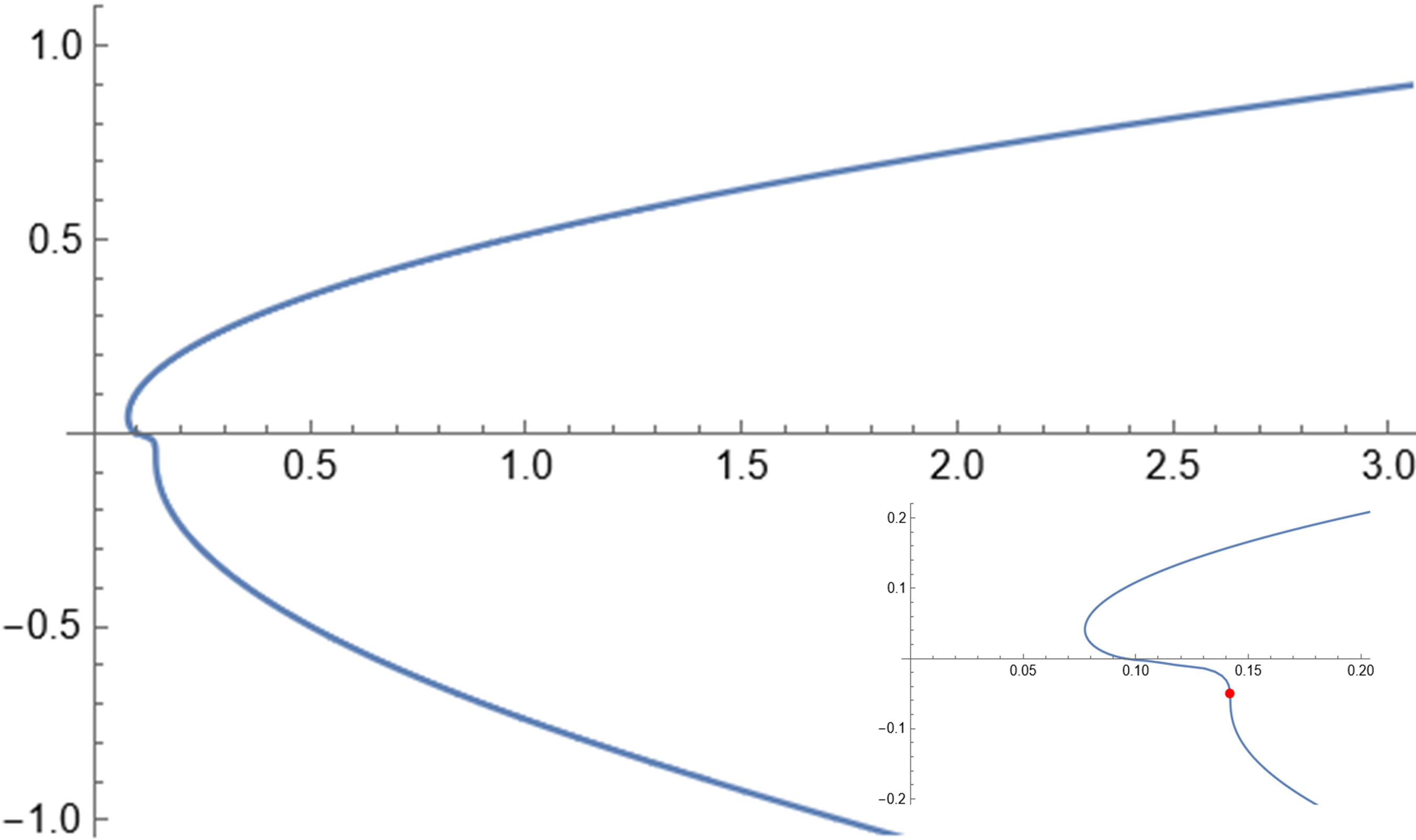}}
\subfigure{\includegraphics[width=0.425\textwidth]{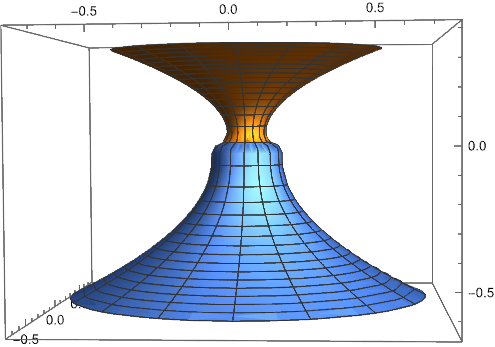}}
\end{center}
\caption{First row: \textit{left}: Two-dimensional view of the isometric embedding of the equatorial plane. \textit{Right}: Isometric embeddings of the solution with throat parameter $r_0$ = 0.5 and  $\omega$ = 0.78. Second row: \textit{left}: Two dimensional view of the isometric embedding with $r_0$ = $0.03$, $\omega$ = $0.9$. \textit{Right}: The corresponding 3D embedded image.}
\label{phase22}
\end{figure}

The result for the scalar charge $\cal D$ of the phantom field is similar to the previous cases. Despite the law remaining constant, the number of $\cal D$ branches matches the number of $M$ and $Q$ branches, which can increase complexity in certain scenarios and we do not show it for the reason of space. The embedded diagram of the wormhole is shown in Fig. \ref{phase22}. In this case, an equatorial plane and a throat appear in the wormhole when the $r_0$ = 0.03, we still choose the frequency $\omega$ = 0.9.

\textbf{The Kretschmann scalar near the throat.}

\begin{figure}[!htbp]
  \begin{center}
\subfigure{\includegraphics[width=0.49\textwidth]{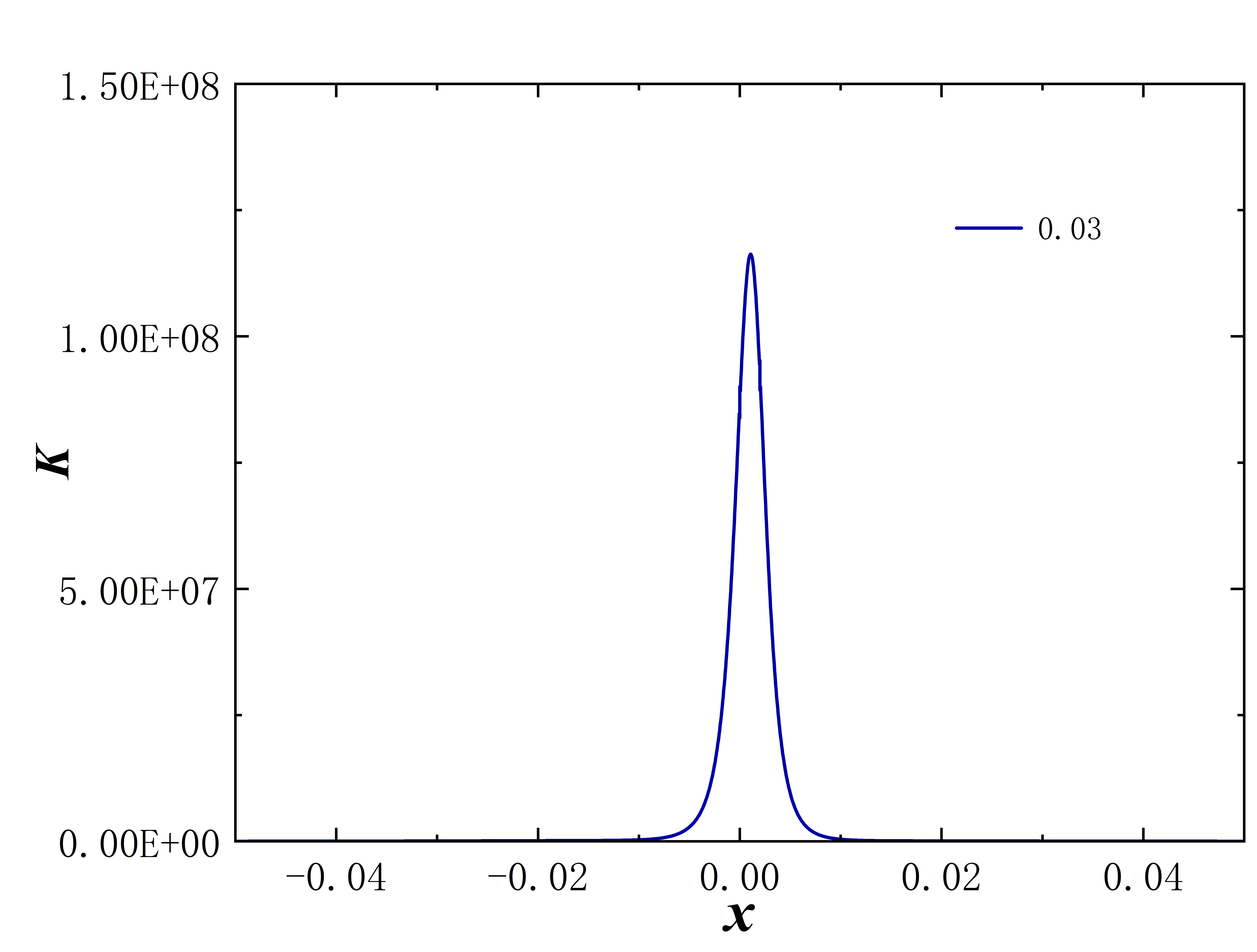}}
\subfigure{\includegraphics[width=0.49\textwidth]{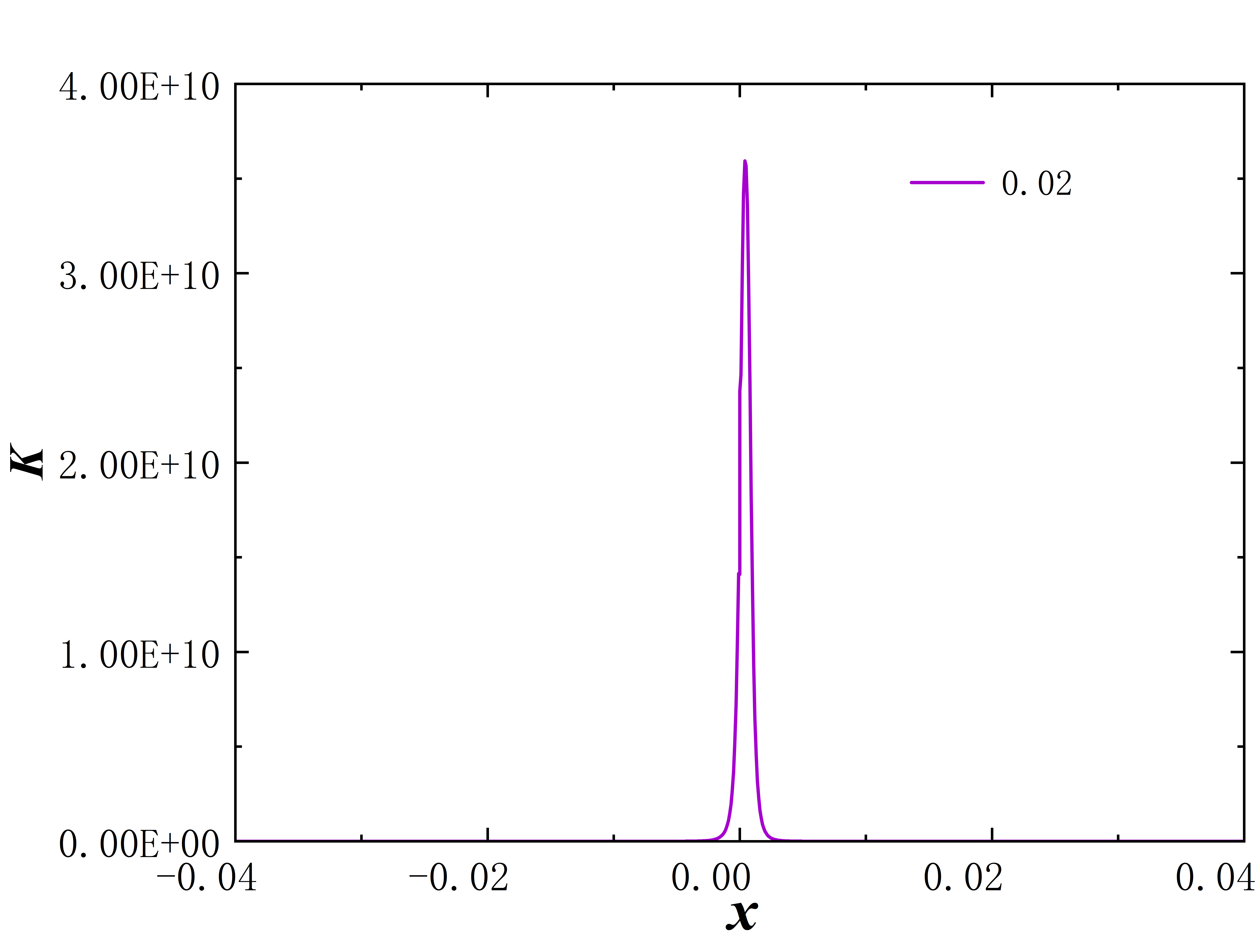}}
\subfigure{\includegraphics[width=0.49\textwidth]{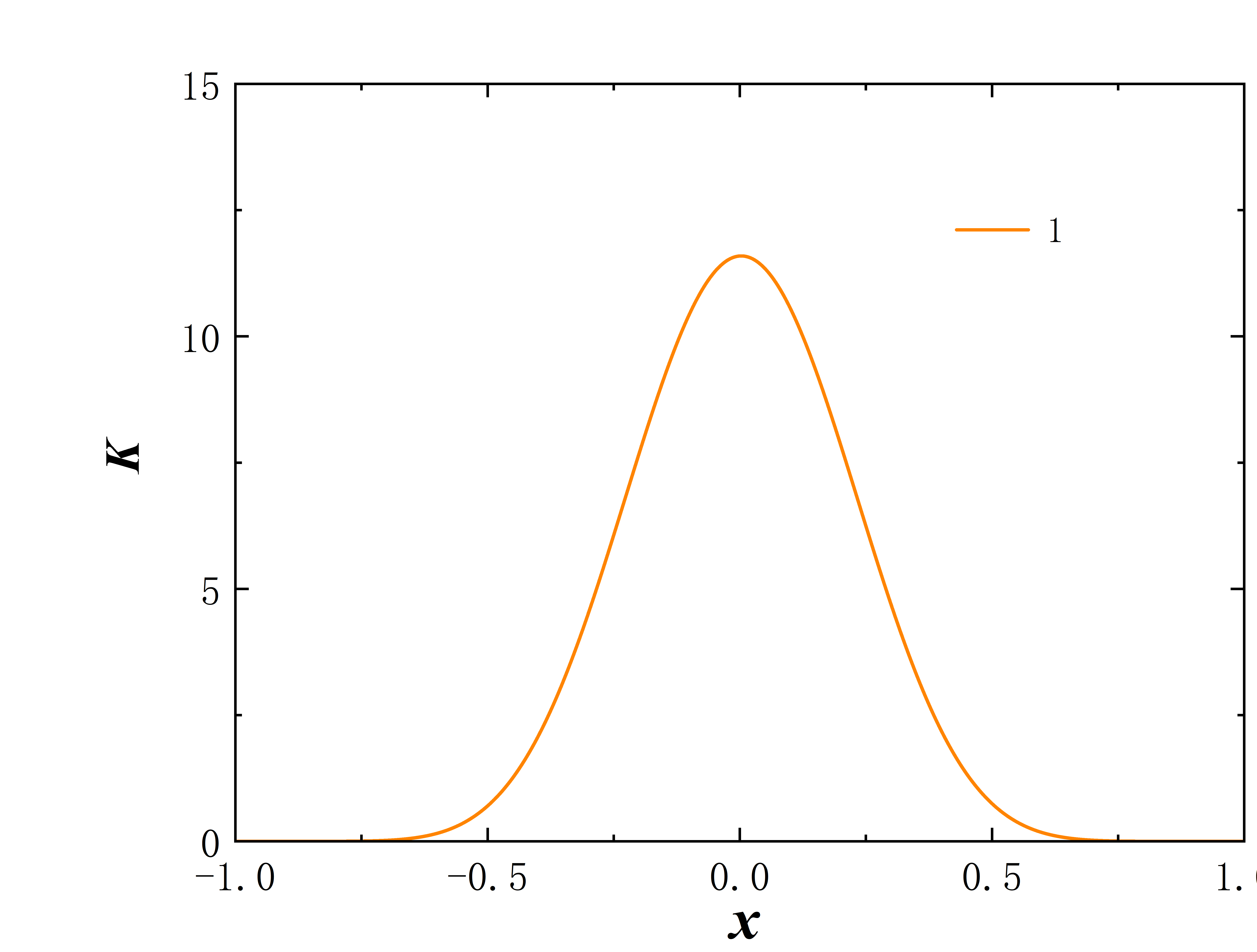}}
\subfigure{\includegraphics[width=0.49\textwidth]{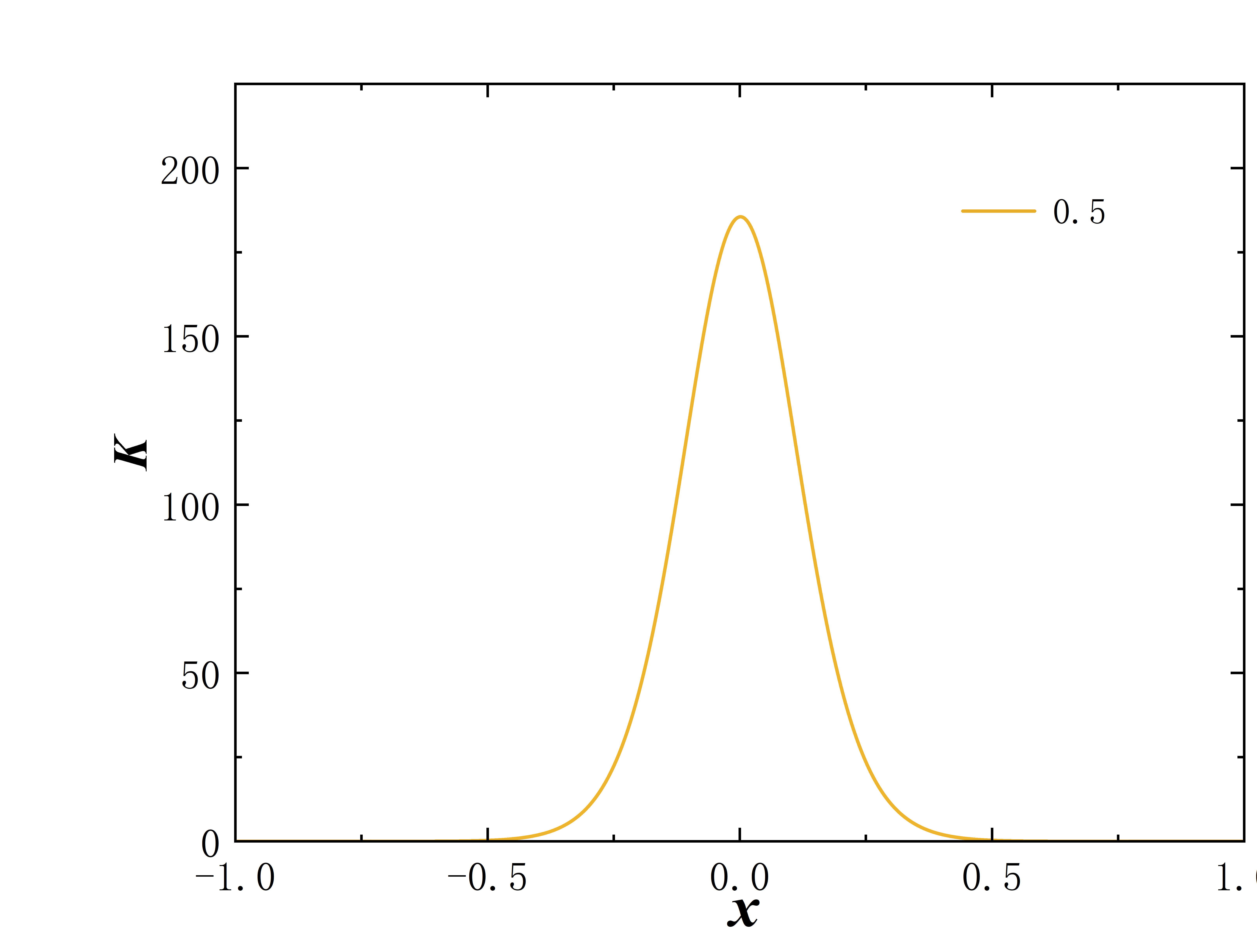}}
\end{center}
\caption{The dark blue line, purple line, brown line, and orange line represent the distribution of the Kretschmann scalar with $r_0$ = 0.03, 0.02, 1, 0.5 respectively.}
\label{phase23}
\end{figure}

Traversable wormholes require no singularities and limited tidal forces at the throat, which imposes constraints on the Kretschmann scalar near the throat. In the above first case, Kretschmann scalar with throat size $r_0$ = 1, 0.5, 0.03, 0.02 are selected for calculation. We limit the frequency $\omega$ = 0.99 at the last branch, where an extremely approximate black hole solution will appear, and observe the distribution of Kretschmann scalar in spatial positions, Fig. \ref{phase23}. The results of the other two cases are similar to it, so they are not shown.

When $r_0$ is large, the value of the Kretschmann scalar is relatively small and does not diverge. At this time, there is no singularity in the metric and no diverge of material field distribution, the wormhole is traversable. With the emergence of extremely approximate black holes, $r_0$ = 0.03 or 0.02, the value of the Kretschmann scalar reaches an astonishing order of magnitude and diverges. This seems to mark the emergence of a singularity and also indicates that the solution corresponds to an untraversable wormhole. In particular, the position $x$ where $K$ diverges coincides with the position of the minimum point of $g_{tt}$, which means a curvature singularity appears at the event horizon. Similar results in the paper \cite{Horowitz:2023xyl}, which in the presence of higher-curvature terms, asymptotically flat extremal rotating black holes have curvature singularities on their horizon.

\section{CONCLUSION}\label{sec5}

In this paper, we investigate the nontrivial topology Dirac star model with a phantom field and derive an asymmetric traversable wormhole solution. By analyzing the ADM mass, Noether's conserved charge, and phantom field scalar charge, we uncover key properties of this solution. In general, the number of branches for mass and conserved charge decreases with increasing throat size $r_0$. As the $r_0$ decreases, the number of branches first increases before decreasing again. The value of $\cal D$ also decreases as the throat size decreases. Notably, we find that when the throat size is below a certain threshold, an extremely approximate black hole solution emerges on one side of the wormhole, it is characterized by the metric $g_{tt}$ tending to zero and $g_{rr}$ tending to infinity, a phenomenon not previously studied. 

In ER = EPR, entangled systems are connected by a wormhole \cite{Maldacena:2013xja,Gharibyan:2013aha,Jafferis:2022crx,Kain:2023ore}. Our work introduces a Dirac field composed of two spin $1/2$ particles in a singlet state, with a phantom field, we obtain a traversable wormhole directly from the Einstein field equations. Numerical results show that the extremely approximate black hole will appear close to one side of the throat and reach the other side as the parameter changes and the wormhole changes from traversable to untraversable. This discovery greatly aids in testing the ER = EPR.

There are some interesting extensions of our work that we plan to investigate in future projects. First, by adding a Maxwell field to this model, can construct a traversable wormhole without the exotic matter. We plan to revisit this to see what results can obtain. Second, in this work, we observed for the first time a direct connection between extremely approximate black holes and traversable wormholes under the Dirac star model with a phantom field. Therefore, we intend to study the traversable wormhole solution with the Proca star model to see if similar interesting properties emerge.

\section*{Acknowledgements}
This work is supported by National Key Research and Development Program of China (Grant No. 2020YFC2201503) and the National Natural Science Foundation of China (Grant No.~12047501 and No.~12275110). Parts of computations were performed on the shared memory system at Institute of computational physics and Complex Systems in Lanzhou University.

\end{document}